\documentclass{book}

\usepackage[english]{babel}

\usepackage{makeidx}         % permette di generare l'indice
\usepackage{multicol}        % usato per l'indice a due colonne

\def\ind#1{\index{#1} #1}
\def\emphind#1{\index{#1} \emph{#1}}

\usepackage{color}

\newcommand{\blu}{\color{blue}}

\def\wfoot#1{}

\def\a{\alpha}
\def\b{\beta}

\def\ga{\gamma}
\def\de{\delta}   %% NON ridefinire come \d !!!!
\def\eps{\varepsilon}
\def\phi{\varphi}
\def\la{\lambda}

\def\s{\sigma}

\def\om{\omega}

\def\vphi{\varphi}

\def\G{{\cal G}}

\def\I{{\cal I}}

\def\L{\mathcal{L}}

\def\R{{\bf R}}
\def\S{{\cal S}}
\def\T{{\rm T}}

\def\Ga{\Gamma}
\def\De{\Delta}

\def\Om{\Omega}

\def\pa{\partial}

\def\d{{\rm d}}       %% derivative

\def\xb{{\bf x}}

\def\o+{\oplus}
\def\xd{{\dot x}}

\def\grad{\nabla}     %% gradient
\def\ss{\subset}

\def\<{\langle}
\def\>{\rangle}

\def\div{\mathrm{div}}

\def\({\left(}
\def\){\right)}
\def\[{\left[}
\def\]{\right]}
\def\=#1{\bar #1}
\def\~#1{\widetilde #1}
\def\wt#1{\widetilde #1}
\def\.#1{\dot #1}
\def\^#1{\widehat #1}
\def\"#1{\ddot #1}

\def\eeq{\end{equation}}
\def\beq{\begin{equation}}

\def\beql#1{\begin{equation} \label{#1}}

\def\eqref#1{(\ref{#1})}

\def\EOR{ \hfill $\odot$ \medskip}
\def\EOE{ \hfill $\diamondsuit$ \medskip}

\def\salta#1{}

\def\VF{VF }

\def\symmref{EMS1,CG,GB,KrV,Olver1,Olver2,Stephani}

\def\HOJ{1}
\def\DETEQDS{2}
\def\WALa{3}
\def\WALb{4}

\def\GS1{5}

\def\GRQa{6}
\def\GRQab{7}

\def\GRQb{8}
\def\GRQbcor{9}

\def\GAEa{10}
\def\GAEb{11}
\def\GAEc{12}

\def\GSstrat{13}

\def\MISa{14}
\def\ALFEI{15}

\def\UNAL{16}

\def\SPAD{17}
\def\GSaa{18}
\def\GSab{19}

\def\SimStrat{20}
\def\GSac{21}
\def\WSTRAT{22}

\def\MISb{23}
\def\MISc{24}
\def\ALFb{25}
\def\MELa{26}
\def\MELb{27}

\def\KOZa{28}
\def\KOZb{29}
\def\KOZc{30}
\def\KOZd{31}
\def\KOZf{32}
\def\KOZg{33}
\def\KOZh{34}

\def\remcorr{\ref{chap:Ito}.5}
\def\remvaria{\ref{chap:Ito}.6}
\def\remJCZ{\ref{chap:Ito}.7}
\def\remLL{\ref{chap:Strato}.8}

\makeindex

\newcommand*{\titleAT}{\begingroup % Create the command for including the title page in the document
\newlength{\drop} % Command for generating a specific amount of whitespace
\drop=0.1\textheight % Define the command as 10% of the total text height

\rule{\textwidth}{1pt}\par % Thick horizontal line
\vspace{2pt}\vspace{-\baselineskip} % Whitespace between lines
\rule{\textwidth}{0.4pt}\par % Thin horizontal line

\vspace{\drop} % Whitespace between the top lines and title
\centering % Center all text
{\Huge Symmetry of stochastic }\\[0.5\baselineskip] % Title line 1
{\Huge non-variational}\\[0.75\baselineskip] % Title line 2
{\Huge differential equations} % Title line 3

\vspace{0.25\drop} % Whitespace between the title and short horizontal line
\rule{0.3\textwidth}{0.4pt}\par % Short horizontal line under the title
\vspace{\drop} % Whitespace between the thin horizontal line and the author name

{\Large \textsc{Giuseppe Gaeta}}\par % Author name

\vfill % Whitespace between the author name and publisher text
{\large \textsc{Dipartimento di Matematica, \\ Universit\`a degli Studi di Milano}}\par % Publisher

\vspace*{\drop} % Whitespace under the publisher text

\rule{\textwidth}{0.4pt}\par % Thin horizontal line
\vspace{2pt}\vspace{-\baselineskip} % Whitespace between lines
\rule{\textwidth}{1pt}\par % Thick horizontal line

\endgroup}

\begin{document}

\pagestyle{empty} % Removes page numbers

\titleAT % This command includes the title page

\frontmatter

%\begin{abstract} \noindent
%I will sketchily illustrate how the theory of symmetry helps in
%determining solutions of (deterministic) differential equations,
%both ODEs and PDEs, staying within the classical theory. I will
%then present a quick discussion of some more and less recent
%attempts to extend this theory to the study of stochastic
%differential equations, and briefly mention some perspective in
%this direction.
%\end{abstract}

%\vfill\eject

%\maketitle

\tableofcontents

\mainmatter

\vfill\eject
%
%\section{Introduction and background} \label{sec:intro}
\chapter{Getting started} \label{chap:intro}

\section{Introduction and Background}

Symmetry based methods to attack (in particular, nonlinear)
deterministic differential equations were introduced long ago, and
have witnessed a flourishing development -- both in the theory and
in concrete applications -- in the last two decades.

Analogous methods for \emph{stochastic} differential equations are
much less widespread and actually at a less advanced stage of
development. Lately there seem to be a growing interest in
applying the symmetry approach to the study of stochastic
differential equations.

The purpose of this text is to provide a short review -- and a
guide to the literature -- of the topic directed to people
familiar with stochastic differential equations (SDEs) but not
with symmetry methods.\footnote{A shorter review has been provided
by E. Bibbona in a talk \cite{Bibbona}, but this has never been
published as an article.}

In order to keep the text short (and also in view of the fact
symmetry considerations, in particular those related to the
Noether theorem, are already routinely present in this specific
field), I decided \emph{not} to treat at all (symmetric)
\emphind{stochastic variational problems}; some bibliographic
notes are however provided on this topic too.
\bigskip

Symmetry attracted mathematicians since the very beginning of
Mathematics; it just suffices to think of the study or regular
polygons and polyhedra by the Greeks (and, of course, their
development of spherical geometry, which they needed to sail
across the Mediterranean \cite{Russo}).

However, for a long time Symmetry was confined to the study of
\emph{geometrical} objects and concepts. The way to apply symmetry
yo the study of analytical objects, and in particular equations,
was paved by Ren\'e Descartes (1596-1650) with his merging of
geometry and analysis in ``G\'eom\'etrie Analytique''. In fact,
Evariste \ind{Galois} (1811-1832) introduced the concept of a
\emph{group} in the study of solutions (or impossibility of
solution in algebraic terms) of algebraic equations.

The modern theory of Symmetry was laid down by Sophus \ind{Lie}
(1842-1899) \cite{Lie1,LieScheffers,LieHistory}. Even in this
case, the motivation behind the work of Lie was not in pure
algebra, but instead in the effort to \emph{solve differential
equations}.

I will assume the reader knows some (very) basic notions about Lie
groups, and illustrate how the theory of symmetry helps in
determining solutions of (deterministic) differential equations,
both ODEs and PDEs, staying within the classical theory (i.e.
disregarding the many -- and rather relevant -- extension of it
developed in the last two decades). I will then present some brief
discussion of more and less recent attempts to extend this theory
to the study of \emph{stochastic differential
equations}.\footnote{It is maybe worth mentioning that symmetry
considerations for Markov processes made their way into the
literature earlier on; see e.g. \cite{Glover,GloMit,Liao}.}

As mentioned above this review is directed primarily to people
familiar with stochastic equations, but not with symmetry methods
for studying differential equations; thus while introducing the
latter -- and actually keeping their discussion to a rather basic
level -- I will assume the reader is familiar with Stochastic
Differential Equations (in this case too we will need only the
very basics of the theory). Here and there I will also point out
some topics which are not central to the development of the theory
but would deserve investigation and promise results.

I will adopt a rather ``applied'' point of view. Many of us are
rather familiar with solving equations with spherical symmetry
(e.g. the central force problem in Mechanics when thinking of
ODEs, or the isotropic wave equation for PDEs), and we all know
that the key step there is to pass to spherical coordinates. These
should be seen as the most familiar example of
\emphind{symmetry-adapted coordinates}, and I will advocate the
idea (which is admittedly a XIX century one) that symmetry
considerations should allow to determine \ind{symmetry-adapted
coordinates} even in cases where these are not immediately obvious.

A more modern and intrinsic approach to symmetry of differential
equations is provided in many books, see e.g. \cite{\symmref}, and
the reader is referred to them for a complete exposition of the
theory and of its applications.\footnote{I should add that the
bibliography on symmetry methods for differential equations is by
now rather extended; thus the selection of these books is just
according to my personal taste and the need to provide a
reasonably short list. Many other good books also exist (some of
these will be quoted below for specific issues), and the reader
can find them more suited to his/her taste.}

I should also mention that I will mainly speak of
\emphind{continuous symmetries}. Discrete ones can also exist but,
apart from cases where they are obvious (typically, reflection
symmetries), they are difficult to detect. The reason is the one
well understood by Lie: for continuous actions one can base
considerations on the tangent space and thus reduce to
\emph{linear} problems, while for fully discrete actions
linearization is not possible.\footnote{It should also be
mentioned -- and stressed -- that we will consider symmetries for
\emph{continuous} equations. Discrete, or differential-difference,
deterministic equations can and have been also studied in terms of
symmetry \cite{LVW,LeWindiscrete1,LeWindiscrete2}, but we will not
touch upon these. As far as I know, no such study has been
performed for discrete, or \index{differential-difference
equations} differential-difference, \emph{stochastic} equations
(this would require to study symmetry of Markov processes
\cite{Glover,GloMit} rather than of SDEs); such a study would
surely be of interest, and is the first of a number of possible
interesting projects we will mention along our discussion.}

Symmetry of stochastic equations is by now not only a mathematical
subject, but also an applied one. I will not speak of concrete
applications, but I would like here to mention that current
research topic include such diverse application as Fluid Mechanics
\cite{ACC,CreDar,Holm,HolmTyr} and Financial Mathematics
\cite{AliPat,EdeGov,Gazizov,Les1,Les2,LesQuiZam,LesZam,Taylorthesis}.

Last but not least, another very important topic will be absent
from my discussion (as specified also in the title). This is
\emphind{symmetry of variational problems}, i.e. the beautiful
theory laid down by Emmy Noether (1882-1935) in 1915 (and
published in 1918) \cite{Noether}, see also \cite{Koss}, and which
played such a great role in the fundamental Physics of the second
half of XX century\footnote{One often forgets that Noether was not
motivated by Classical Mechanics, but by General Relativity;
correspondingly, her original Theorem was of much wider scope than
it is usually taught in the Mechanics courses. A nice discussion
of \ind{Noether theorem} is provided by Olver \cite{Olver1}, while
the work of Noether is discussed at length (and her original work
provided in a reliable translation) together with its influence in
the book by Kosmann-Schwarzbach \cite{Koss}.} -- and earlier on in
Mechanics. This is also developed in the stochastic framework, as
we will briefly recall (mainly to point out some relevant
literature) below; see in particular Remark \remvaria.

\section{Practicalities, notation, plan of the work}

Equations and Propositions are numbered consecutively through the
full body of the work; on the other hand, footnotes, Examples and
Remarks are numbered by Chapter. The end of Remarks and Examples
is marked respectively by the symbols $\odot$ and $\diamondsuit$.
I have tried to avoid too many cross-references among Chapters,
which in some case led to repeated equations.

Some (rather standard) notations are routinely used; thus $\pa_i
\:= (\pa / \pa x^i)$ throughout the work, and summation over
repeated indices is always implied (if not explicitly stated
otherwise).

As stated above, the work is primarily directed to people dealing
with stochastic equations but not familiar with symmetry
techniques. Thus Chapter \ref{chap:det} provides a quick but
self-contained introduction to the latter (limiting of course to
the basic aspects; references are provided for several further
development). On the other hand, we assume reader are familiar
with stochastic differential equations, and will not discuss these
in their general features, going straight to their symmetry
properties. Albeit investigation of symmetry aspects of SDEs
started by considering equations in Stratonovich form, we will
first consider equations in Ito form (Chapter \ref{chap:Ito}) and
only afterwards consider equations in Stratonovich form (Chapter
\ref{chap:Strato}); I think there is no need to justify giving a
prominent role to Ito, i.e. properly defined, equations. We will
deviate from the historical development of the subject also by
putting in the same Chapter old and recent results. On the other
hand we will devote a separate discussion (Chapter \ref{chap:RS})
to a very recent development, i.e. \emph{random symmetries} (of
both types of equations).

Work on symmetry of SDEs concentrated so far to a large extent on
what would be the proper definition of symmetry in this case, and
how symmetries can be actually determined; this is also true of
Chapters \ref{chap:Ito}, \ref{chap:Strato} and \ref{chap:RS} here.
But one, in particular if not familiar with the symmetry approach,
should also wonder what is the \emph{use} of all this. Our final Chapter
\ref{chap:Use} provides a partial answer to this, illustrating
several applications of symmetry considerations.

I would like to stress that the answer here is only partial not
due to laziness by the author, but due to the fact the theory is
under development. Actually, this is just the main motivation
which led me to write these notes, in the hope they can attract
new practitioners to this field and thus contribute somehow to the
development of the field.

\section{Acknowledgements}

This work was stimulated by the workshop ``Stochastics and
Symmetry'' (Milano, October 2015); I would like to thank the
organizers, and in particular Sergio Albeverio, for their kind
invitation.

My older works on symmetry of stochastic equations were performed
in collaboration with prof. N. Rodriguez Quintero (now at
Universidad de Sevilla), while some of the recent results reported
here were obtained with F. Spadaro (now at EPFL Lausanne); I
warmly thank both of them. I would also like to thank L. Peliti
and A. Vulpiani for encouragement; C. Lunini for remarks and
discussions about Section \ref{sec:itovsstrat} and Chapter
\ref{chap:RS}; and J.C. Zambrini for a critical reading of, and
interesting remarks on, the whole manuscript.

\vfill\eject
\chapter{Deterministic equations} \label{chap:det}
\def\nc{\ref{chap:det}}

\section{The geometry of differential equations}
\label{sec:geom}

The key idea in describing the geometrical meaning of a
differential equation (or better is associating a geometrical
object to a differential equation) is to introduce \emph{jet
spaces} (or more precisely \emphind{jet bundles}) \cite{EMS1,CG,GB,Ibrag,KrV,Olver1,Olver2,Ovs,Saunders}.

We will denote the space (the bundle) of independent and dependent
variables as the \emphind{phase space} (\ind{phase bundle}) $M$;
note that in the case of ODEs we are actually referring to the
complete phase space, rather than the reduced one (only dependent
variables) which is often used in the case of autonomous dynamical
systems. Here I will consider the case where the DEs (differential
equations) at hand are defined in $\R^q$, with dependent variables
taking also value in $\R^p$, and do not have boundary conditions,
so that the geometry is the simplest possible one. (Hence I will
speak of ``spaces'' rather than ``manifolds'' or ``bundles''.) I
will also assume, for ease of language and discussion, that the
DEs under consideration do not involve non-algebraic functions of
the derivatives.

The jet space can be thought of as the space of dependent
($u^1,...,u^p$) and independent ($x^1,...,x^q$) variables,
together with the partial derivatives of the $u$ with respect to
the $x$. In principle one can -- and for certain questions (e.g.
in the case of PDEs for some side conditions) should -- consider
the infinite order jet space \cite{KrV,Olver1,Olver2}, i.e.
partial derivatives of all orders. But if we are dealing with a DE
of order $n$, then for most questions it will suffice to consider
the jet space of order $n$, $J^n M$ i.e. to consider partial
derivatives of order $k \le n$ only. These will be denoted as
$u^a_J$, where $J = (J_1,...,J_q)$ (here $j_i \ge 0$) is a
\emphind{multi-index} of order $|J| = j_1 + ... + j_q$, and $u^a_J
:= \pa u^a / (\pa x_1^{j_1} ... \pa x_q^{j_q} )$.

A differential equation (or system thereof) $\De$ is then a
standard equation in $J^n M$, and hence it describes as usual a
manifold in it; this is also called the \emphind{solution
manifold} for $\De$, and denoted as $S_\De \ss J^n M$. This is a
geometrical object, and we can now apply geometrical tools to
study it. E.g., we can consider maps or vector fields which leave
them invariant.\footnote{For obvious reasons, one is primarily
interested in maps defined in the ``physical'' phase manifold and
then lifted to the Jet manifold; as for the way of performing this
lift, see below.}

There is of course a problem: the variables $u^a_J$ represents
derivatives of the $u^a$ w.r.t. the $x^i$, hence they are not
really independent variables -- albeit our description considered
them as such. In order to take this fact into account, the jet
space should be equipped with an additional structure, the
\emphind{contact structure}. This is associated to the name of
Elie \ind{Cartan} (1869-1951) \cite{Cartan1,Cartan2,Cartan3};
\ind{jet bundles} are associated with the name of his pupil
Charles \ind{Ehresmann} (1905-1979) \cite{Ehr}.

The information carried by the contact structure is that e.g.
$u^a_i$ is the derivative of $u^a$ w.r.t. $x^i$. This can be
expressed by introducing the one-forms \beq \om^a \ := \ \d u^a \ -
\ \sum_{i=1}^q \, u^a_i \, \d x^i \eeq and looking at their kernel
(or annihilator), i.e. the set of vector fields $X$ such that
$\iota_X \om = 0$.

We will thus consider general (smooth) vector fields in $M$, but
as for vector fields in $J^n M$ only those compatible with the
contact structure will respect the special nature of the $u^a_J$
variables.

Note that if we think of a vector field (VF) as an infinitesimal
transformation of the $x$ and $u$ variables, once this is defined
the transformations of the derivatives are also implicitly
defined. This rather trivial observation can be made precise; the
procedure of extending a \VF in $M$ to a \VF in $J^n M$ by
requiring the preservation of the contact structure is also called
\emphind{prolongation}. Correspondingly, there is a
\emphind{prolongation formula} providing the components of the
prolonged \VF in terms of the components of the original one.

This is better expressed in recursive form, but we need first to
introduce some notation. We will write \beq \pa_i \ := \
\frac{\pa}{\pa x^i} \ , \ \ \pa_a \ := \ \frac{\pa}{\pa u^a} \ ; \
\ \pa^a_J \ := \ \frac{\pa}{\pa u^a_J} \ . \eeq Thus a \VF in $M$
will be written in components as \beq \label{eq:\VF} X \ = \ \xi^i
(x,u) \, \pa_i \ + \ \phi^a (x,u) \, \pa_a \ . \eeq Here and in
the following, the Einstein summation convention will be routinely
employed.

We will also write $D_i$ for the substantial derivative w.r.t.
$x^i$, i.e. \beq D_i \ = \ \pa_i \ + \ u^a_i \pa_a \ + \ u^a_{ij}
\pa^a_j \ + \ ... \ , \eeq and denote by $u^{[k]}$ the set of all
the derivatives of the $u$ of order $k$.

A vector field in $J^n M$ will then be written (with $\psi^a_0 \equiv \phi^a$) as \beq
\label{eq:P\VF} Y \ = \ \xi^i (x,u,...,u^{[n]}) \, \pa_i \ + \
\sum_{|J|=0}^n \psi^a_J (x,u,...,u^{[n]}) \, \pa^a_J \ . \eeq This
is the prolongation of a vector field in $M$ if and only if \beq
\xi^i (x,u,...,u^{[n]}) \ = \ \xi^i (x,u) \ , \ \ \psi^a_J
(x,u,...,u^{[n]}) \ = \ \psi^a_J (x,u,...,u^{[|J|]} )  \eeq and
moreover the components satisfy the \emphind{prolongation formula}
\beq \label{eq:prolform} \psi^a_{J,i} \ = \ D_i \psi^a_J \ - \
u^a_{J,k} \, D_i \xi^k (x,u) \ . \eeq We will refer again to
standard books \cite{\symmref} for its derivation.

A \VF $X$ defined in $M$ is then an infinitesimal \index{Lie-point
symmetry} Lie-point \emph{symmetry} (more precisely, the generator
of a one-parameter local group of symmetries) if its prolongation,
also written $X^{(n)}$, satisfies \beq \label{eq:symmcond} X^{(n)}
\ : \ S_\De \ \to \ \T S_\De \ . \eeq

This definition is consistent and precise; its drawbacks is that
it only concerns the manifold $S_\De$, while when considering a DE
we would usually be interested in its \emphind{solutions}. It
turns out an equivalent characterization of symmetries of DE is to
map solutions into (generally, different) solutions.

In order to understand this, we should characterize geometrically
solutions to a given DE $\De$. A function $u = u(x)$ can also be
seen as a section of the bundle $(M,\pi_0,X)$, i.e. of the phase
manifold (or just space) seen as a bundle over the manifold (or
just space) of independent variables. This is just the set of
points \beq \ga_f \ = \ \{ (x,u) : u = f(x) \} \in M \ . \eeq This
section is naturally lifted to sections $\ga_f^{(k)}$ in Jet
spaces, simply by computing derivatives. E.g., at first order we
have \beq \ga_f^{(1)} \ = \ \{ (x,u,u_x) \ : \ u = f(x) \ , \ u_x
= f'(x) \} \ . \eeq
With this construction, a function $u=f(x)$ is a solution to $\De$
if and only if \beq \ga^{(n)}_f \ss S_\De \ss J^n M \ . \eeq The
details of the prolongation operation guarantee that $X^{(n)}$
will transform (locally) sections into sections, hence (if it is
tangent to $S_\De$) solutions into solutions.

We conclude that indeed \emph{a symmetry maps solutions into
(generally, different) solutions}. In the case a solution is
mapped into itself, we say it is an \index{invariant solutions}
\emph{invariant solution}.

It is immediately apparent from this discussion that a first use
of symmetry can be that of \emph{generating new solutions from
known ones}.\footnote{E.g., the fundamental solution of the
\ind{heat equation} can be generated in this way starting from the
trivial (constant) solution; see \cite{Olver1} and Example
\ref{chap:det}.14 below.}

This is surely of interest, but it is not the only way in which
knowing the symmetry of a differential equation can help in
determining (all or some of) its solutions, as we will discuss
below; see sects.\ref{sec:ODEs} and \ref{sec:PDEs}.

Finally, we should mention that in some cases one considers
\emphind{generalized symmetries}, corresponding to the action of
generalized vector fields. These are vector fields acting in $M$
but with coefficients which depend also on derivatives of the $u$
with respect to the $x$. Thus they are properly defined only in
the space of sections of the bundle, or in \ind{infinite order jet
spaces}. We will not enter in the details of how one makes sense
of these, referring the reader to the literature \cite{\symmref},
but some results in Sect. \ref{sec:Hojman} and
Sect. \ref{sec:Misawa} will be stated in terms of these. In the
rest of the paper we will just consider standard vector fields.

\section{Determining the symmetry of a differential equation}
\label{sec:determining}
\def\cs{\ref{sec:determining}}

Before discussing how symmetry are used, we should briefly discuss
how one can determine the (continuous) symmetries of a given
differential equation.\footnote{As mentioned above, determination
of discrete symmetries  -- when not trivial -- is in general a
harder problem, and a non-algorithmic one (but see
\cite{GaeRoddiscr,Hydpap1,Hydpap2,Hydon} for constructive albeit
non-general methods).}

The first step is to consider a general \VF of the form
\eqref{eq:\VF}, with $p$ and $q$ (i.e. the number of dependent and
independent variables) as suitable for the equation (or system
thereof) under consideration; one should then apply the
prolongation formula \eqref{eq:prolform} and thus obtain the
prolonged vector field $Y = X^{(n)}$ corresponding to $X$, see
\eqref{eq:P\VF}; here of course contact is made with
\eqref{eq:\VF} by choosing $\psi^a_0 = \phi^a$. In this way we
obtain a vector field which depends on the unknown functions
$\xi^i (x,u)$, $\phi^a (x,u)$.

We should then determine what are the conditions on these function
which guarantee that \eqref{eq:symmcond} is satisfied. As $S_\De$
is identified as the zero-level set of the function $\De : J^n M
\to \R^s$ (here $s$ is the number of equations to be satisfied,
i.e. the dimension of the system $\De$), we just have to apply $Y$
(considered as a differential operator) on $\De$ (considered as a
set of functions) and require that $Y (\De) = 0$ whenever $\De =
0$.

Note that we could as well require that $X (\De ) = 0$ in general;
this will however be too strong a requirement (it corresponds to
requiring that \emph{all} the level set -- not just the zero one
--  of $\De$ are invariant), and in this case one also speaks of
\index{strong symmetry} \emph{strong symmetries}.

\medskip\noindent
{\bf Remark \nc.1.} Actually a rather strict relation exists between
general symmetries of a given equation and strong symmetries of
some equivalent equation, as shown by Carinena, Del Olmo and
Winternitz \cite{CDW}; the reader is referred to their work for
details. \EOR
\bigskip

When writing down the condition $Y(\De) = 0$ on $\De=0$, we should
recall that the $(x,u,u_J)$ should be considered as independent
variables; moreover all the dependencies on the $u_J$ with nonzero
$J$ are completely explicit. Thus the vanishing of $Y(\De^k)$
amounts to the vanishing \emph{separately} of the coefficients of
different monomials in the $u_J$. In this way we have a (usually,
rather large) set of equations, known as the (symmetry)
\emphind{determining equations},  to be satisfied by the unknown
functions $\xi,\phi$.

These are coupled PDEs, but the relevant point is that they are
\emph{linear}; this of course descends from the fact we are
considering the infinitesimal action of vector fields, i.e.
\emph{linearized} transformations, and thus touches to the core of
Lie theory.

Despite being in large number, these equations do have a
hierarchical structure: those corresponding to monomials with high
order derivatives of the $\xi, \phi$ functions will be rather
simple and are easily solved. One then has an {\it ansatz} for the
$\xi,\phi$, and further equations get simpler. Proceeding
systematically one is often able to solve apparently very complex
systems. When systems are too large -- or the scientist too lazy
-- one can also resort to computer programs written in symbolic
manipulation languages (e.g. the package {\tt symmgrp}
\cite{symmgrp}; see also \cite{Her2} and \cite{willy} for other software).

In fact, the key point here is that the solution of the
\ind{determining equations}, and hence the determination of the
continuous Lie-point symmetries, is \emph{completely
algorithmic}.\footnote{The exception to this is for
\emphind{dynamical systems}, i.e. systems of first order ODEs (the
reason is obvious: in this case we lack the above-mentioned
hierarchical structure in the determining equations), discussed in
Sect. \ref{sec:dynamicalsystems} below. As suggested by Stephani
\cite{Stephani}, for these it is useful to use the theory of
characteristics in reverse, i.e. transform the dynamical system
into an equivalent first order quasilinear PDE. A different
approach goes through combining symmetry analysis with
perturbation theory \cite{CGnf1,CGnf2,CG}, and in particular
\index{Poincar\'e} \index{Birkhoff} (Poincar\'e-Birkhoff)
\emphind{normal forms} \cite{ArnGMDE,Elp,Iooss}; this should be
compared with the approach to normal forms for stochastic
differential equations, as also briefly mentioned later on (Sect.
\ref{sec:randiff}).}

Needless to say, it may also be rather complex computationally, so
that before the introduction of symbolic manipulation computer
languages their solution could be just too difficult in practice,
albeit possible in principle.

Detailed examples of actual computations with solution of the \ind{determining
equations} for relevant differential equations are given in any
book on symmetries of DEs, e.g. \cite{\symmref}, and the reader is
referred to them.

It is also worth mentioning that the concept of symmetry of a
differential equation has been extended (from the Lie-point
framework considered here) in several directions; this is not the
place to discuss such extensions, for which the reader is -- as
usual -- referred to the literature.\footnote{Readers interested
in certain real-world applications can also be alerted about the
extension of the symmetry approach to difference or
\ind{differential-difference equations}, see e.g.
\cite{LVW,LeWindiscrete1,LeWindiscrete2,Yamilov}.}

We will now suppose to have determined the symmetries of our
differential equation $\De$, and turn to the problem of how to use
it to obtain information on the solutions of $\De$.

As we will see in a moment, the key idea is the same for ODEs and
PDEs, and amounts to the use of \emphind{symmetry-adapted
coordinates}; but the scope of the application of symmetry methods
is rather different in the two cases.

From now on, for the sake of discussion, I will restrict to the
scalar case (a single equation for a single dependent variable);
the general case is considered e.g. in the books mentioned
above.\footnote{Also, several of our examples will concern linear
equations; this is just for ease of discussion and get simple
computations, but the theory is primarily concerned with
\emph{nonlinear} equations.}

\medskip\noindent
{\bf Example \nc.1.} Let us consider the (linear, homogeneous)
equation \beq \label{eq:ex1ode} x^2 \, \frac{d^2 u}{d x^2} \ + \ x
\, \frac{d u}{d x} \ + \ u \ = \ 0 \eeq for $u = u(x)$. We
consider vector fields of the form
$$ X \ = \ \xi (x,u) \, \pa_x \ + \ \vphi (x,u) \, \pa_u \ ; $$
their second prolongation will be written in the form
\begin{eqnarray*}  X^{(2)} \ = \ Y &=& \xi (x,u) \, \frac{\pa}{\pa x} \ + \ \vphi (x,u) \, \frac{\pa}{\pa u} \\
& & \ + \ \psi (x,u,u_x) \, \frac {\pa}{\pa u_x} \ + \ \chi (x,u,u_x,
u_{xx} ) \, \frac{\pa}{\pa u_{xx}} \ ; \end{eqnarray*} here the functions
$\psi$ and $\chi$ are explicitly computed (in terms of the unknown
functions $\xi$ and $\vphi$) via the prolongation formula
\eqref{eq:prolform}. We obtain
$$ \psi (x,u,u_x) \ = \ \vphi_x + u_x \vphi_u - u_x \xi_x - u_x^2
\xi_u \ , $$ and a more involved formula (which we do not report
here) for $\chi$.

Let us first look for \emph{strong} symmetries of the equation
\eqref{eq:ex1ode}. By applying $Y$ on the equation $\De$ we obtain
an expression of the form \begin{eqnarray} Y (\De) &=& \a_0 (x,u)
\ + \ \a_1 (x,u) \, u_x \ + \ \a_2 (x,u) \, u_x^2 \ + \ \a_3 (x,u)
\, u_x^3 \nonumber \\ & & \ \ \ + \ \b_1 (x,u) \, u_{xx} \ + \
\b_2 (x,u) \, u_x \, u_{xx} \ , \end{eqnarray} where $\a_i$ and
$\b_j$ are explicit functions of the unknown $\xi$, $\vphi$. As
the dependencies on $u_x$ and $u_{xx}$ are explicit, we require
the vanishing of the $\a_i$ and $\b_j$. With straightforward
algebra, we obtain that necessarily \beq \xi (x,u) \ = \ c_1 \, x
\ , \ \ \vphi (x,u) \ = \ \phi (x) \eeq where $\phi (x)$ is an
arbitrary solution to \eqref{eq:ex1ode} (as discussed in
Sect. \ref{sec:Bluman} below, this corresponds to the fact
\eqref{eq:ex1ode} is linear, and we will consider these symmetries
as trivial\footnote{They are associated to the (linear)
superposition principle for solutions to linear equations.});
actually as \eqref{eq:ex1ode} can be solved exactly, we have
$$ \phi (x) \ = \ k_1 \, \sin (\log |x| ) \ + \ k_2 \, \cos (\log
|x| ) \ . $$ Thus the only nontrivial symmetry is in this case
generated by $$ X \ = \ x \, \pa_x \ , $$ which corresponds to a
scale transformation $$ x \ \to \ \la \ x \ ; $$ needless to say,
this symmetry was immediately apparent from the form of our
equation  \eqref{eq:ex1ode}.

Let us now look for \emph{standard} (as opposed to strong)
symmetries of the same equation \eqref{eq:ex1ode}. In this case we
have to restrict $Y(\De)$ to the solution manifold $S(\De)$, which we do by
substituting for $u_{xx}$ according to \eqref{eq:ex1ode}; we
obtain now \beq
\[ Y (\De) \]_{S_\De} \ = \ \ga_0 (x,u) \ + \ \ga_1 (x,u) \, u_x \ + \ \ga_2 (x,u)
\, u_x^2 \ + \ \ga_3 (x,u) \, u_x^3 \ , \eeq where $\ga_i$ are
explicit functions of the unknown $\xi$, $\vphi$. As the
dependencies on $u_x$ are explicit, we require the vanishing of
the $\ga_i$. The computation is made easier by starting from
$\ga_3 = 0$ and $\ga_2 = 0$, which allows to make also the
dependencies on $u$ fully explicit. Proceeding by standard
computations, we obtain in the end that there are six symmetry
generators, i.e.
$$ \begin{array}{l}
X_1 \ = \ x \, \pa_x \ , \\
X_2 \ = \ u \, \pa_u \ , \\
X_3 \ = \ x \, u \, \sin (\log |x|) \, \pa_x \ + \ u^2 \, \cos
(\log |x|) \, \pa_u \ , \\
X_4 \ = \ x \, u \, \cos (\log |x|) \, \pa_x \ - \ u^2 \, \sin
(\log |x|) \, \pa_u \ , \\
X_5 \ = \ 2 \, x \, \sin (\log |x|) \, \cos (\log |x|) \, \pa_x \
+ \ u \, \cos (2 \log|x|) \, \pa_u \ , \\
X_6 \ = \ x \, \cos (2 \log|x|) \, \pa_x \ - \ 2 \, u \, \sin(\log
|x| ) \, \cos (\log |x|) \, \pa_u \ , \end{array} $$
apart from symmetries of the form $$ X_\phi \ = \ \phi (x) \, \pa_u $$ with $\phi$ a solution to \eqref{eq:ex1ode} (see above for these). \EOE

\medskip\noindent
{\bf Example \nc.2.} Let us consider the ODE \beq\label{eq:SGred}
(d^2 u / d x^2 ) \ = \ - \ \sin (u) \eeq for $u = u(x)$; this
describes uniform (in space: here $x$ actually represents time)
solutions to the sine-Gordon equation \eqref{eq:SG} to be met
later on. In this case there is only one symmetry generator, which
is the obvious one, $ X = \pa_x$; this is actually a strong
symmetry.\footnote{In this and the following examples we do not
provide details of the computations (which would be
space-consuming) but the reader is invited to perform them in
order to get accustomed with the approach.} \EOE

\medskip\noindent
{\bf Example \nc.3.} Consider \cite{Stephani} the (second order,
nonlinear) ODE \beq \frac{d^2 u}{d x^2} \ = \ (x - u) \ \( \frac{d
u}{d x} \)^3 \ . \eeq This has the maximal number of symmetry
generators for a second order ODE, i.e. eight, and hence an
eight-parameter Lie symmetry group.\footnote{This is also the case
for $d^2 u / d x^2 = 0$, in which case one gets the (eight
parameters) group of projective transformations.}

The generators are
\begin{eqnarray*}
X_1 &=&
 \frac{1}{2} \[ \left(1-2 u^2+4 x u-2 x^2 \right)
   \sin (u)+2 (x-u) \cos (u)\] \pa_x \ + \ [(x-u) \cos (u)] \pa_u \ , \\
X_2 &=& \frac{1}{2} \[\left(2 u^2-4 x u+2 x^2-1\right)
   \cos (u)+2 (x-u) \sin (u)\] \pa_x \ + \ [(x-u) \sin (u)] \pa_u \ , \\
X_3 &=& (x-u) \pa_x \ , \ \ \ X_4  \ = \ \pa_x \ + \ \pa_u \ , \\
X_5 &=& - \[  \cos (u) (\cos (u)+(u-x) \sin (u)) \] \pa_x \ - \  \cos^2(u) \pa_u \ , \\
X_6 &=& \[ \frac{1}{2} ((x-u) \cos (2 u)+\sin (2 u)) \] \pa_x \ +
\ \sin (u) \cos (u) \pa_u \ , \\
X_7 &=& \cos (u) \pa_x \ , \ \ \ X_8 \ = \ \sin (u) \pa_x \ .
\end{eqnarray*}
{} \EOE

\medskip\noindent
{\bf Example \nc.4.} The standard example for determination of
symmetries of a PDE is the heat equation; this is discussed in
virtually any book on symmetries of differential equations (see
e.g. those mentioned above). We will refrain from considering it,
referring the reader e.g. to \cite{Olver1}. \EOE

\medskip\noindent
{\bf Example \nc.5.} We will instead consider the KdV equation
\cite{CalDeg,Dunajski} \beq \label{eq:KdV} u_t \ + \ u_{xxx} \ + \
6 \, u \, u_x \ = \ 0 \ . \eeq We write vector fields in the form
$$ X \ = \ \xi (x,t,u) \, \pa_x \ + \ \tau (x,t,u) \, \pa_t \ + \
\vphi (x,t,u) \, \pa_u \ , $$ and proceed according to the general
method. With some standard computations we get that there are four
symmetry generators, given by $$ \begin{array}{ll}
X_1 \ = \ \pa_x \ , & \ \ X_2 \ = \ \pa_t \ ; \\
X_3 \ = \ x \, \pa_x \ + \  3 \, t \, \pa_t \ - \ 2 \, u \, \pa_u
\ , & \ \ X_4 \ = \ 6 \, t \, \pa_x \ + \ \pa_u \ .
\end{array} $$
{} \EOE

\section{Symmetry and ODEs}
\label{sec:ODEs}
\def\cs{\ref{sec:ODEs}}

\subsection{Reduction}
\label{sec:reduction}

If we have an ODE $\De$ of order $n$ and this admits a Lie-point
symmetry, the equation can be \emph{reduced} to an equation of
order $n-1$. The solutions to the original and to the reduced
equations are in correspondence through a \emphind{quadrature},
i.e. an integration;   this of course introduces an integration
constant, hence the correspondence is certainly not one-to-one.

It should be noted that in the case of multiple symmetries we do
not always have as many reductions as symmetries: this depends on
the Lie algebraic structure of the symmetry algebra -- i.e. of the
Lie algebra of the symmetry vector fields
\cite{\symmref}.\footnote{This is also related to a recent
development, i.e. so called \emphind{twisted symmetries} of
differential equations \cite{Muriel1,Muriel2,Muriel3}; for these
see also \cite{twisted1,twisted2}. In this context one should also
mention \emphind{solvable structures} \cite{BP1,Hartl,ShP}.}

In this case one can describe the reduction procedure and the
reason why it works in rather simple terms (the notation and
discussion are simplified by dealing with scalar equations, as we
do here).

We will start by considering an ODE of order $n$, which we write
quite generally as \beq \De \ := \ \ F(x,u;u_x,u_{xx},...,u^{[n]}
) \ = \ 0 \ ; \eeq here $F$ is a smooth function of its arguments.

Suppose we have determined a vector field \eqref{eq:\VF} which is
a symmetry of $\De$, and assume moreover (for ease of discussion)
it is actually a strong symmetry, i.e. $X(\De ) = 0$. We will then
change variables via an invertible map \beq \label{eq:map} (x,u)
\to (y,z) \eeq (here $y$ should be thought of as the independent
variable, $z$ as the dependent one), so that in the new variables
$X$ is written as \beq \label{eq:XhatODEs} \^X \ = \
\frac{\pa}{\pa z} \ . \eeq (The notation $\^X$ should not cause
confusion: $X$ and $\^X$ are the same geometrical object, but
expressed in different systems of coordinates; we introduce a
different notation since the prolongation depends on the
coordinates and on which variable is considered as the independent one,
and we will need to prolong the vector field.)

By doing this we also have to write the equation in the new
variables, which yields \beq \De \ := \ \
G(y,z;z_y,z_{yy},...,z^{[n]} ) \ = \ 0 \ . \eeq The detailed form
of $G$ will depend on the change of variables, but as both $X$ and
$S_\De$ are geometrical object, the tangency condition
\eqref{eq:symmcond} does not depend on the coordinates we are
using. That is, we know that necessarily $\^X^{(n)} (G) = 0$; on
the other hand, the prolongation formula \eqref{eq:prolform}
guarantees that with $\^X$ as in \eqref{eq:XhatODEs}, we just have
$\^X^{(2)} = \^X = \pa_z$.

But if this the case, it just means $G$ does not depend on $z$,
\beq G(y,z;z_y,z_{yy},...,z^{[n]} ) \ = \ H (
y;z_y,z_{yy},...,z^{[n]} ) \ . \eeq We can then perform a new
change of coordinates (involving only the dependent coordinate)
\beq \label{eq:w} w \ := \ z_y \ . \eeq In these coordinates, we
write the equation as \beq H(y,w;w_y,...,w^{[n-1]} ) \ = \ 0 \ .
\eeq That is, we have reduced the equation to one of lower order.

Suppose now we are able to determine a solution $w = h(y)$ to the
reduced equation. This identifies solutions $z = g(y)$ to the
original equation (in ``intermediate'' coordinates) simply by
integrating \eqref{eq:w}, \beq z (y) \ = \ \int w(y) \, d y \ ;
\eeq note a constant of integration will appear here.

We can finally go back to the original coordinates; this is done
simply by inverting the map \eqref{eq:map}.

It should be noted that the reduced equation could still be too
hard to solve; the method can only guarantee that we are reduced
to a problem of lower order, i.e. hopefully simpler than the
original one.\footnote{It should be stressed that this is
\emph{not} the only possible strategy; in several situations, it
is actually convenient to \emph{increase} the dimension of the
system, embedding a nonlinear problem into a linear one. This is
done e.g. in solving the \ind{Calogero system}
\cite{CalInt,CalDeg,OlP}, or more generally in the
\ind{Kazhdan-Kostant-Sternberg approach} \cite{Mar2015,KKS}. Also,
one can try not to reduce (or increase) the order of the system,
but take it into a more convenient form, e.g. to deal with an
\emph{autonomous} system \cite{CurOli,DonOli,OliSpe}.}

We should also mention that other applications of the symmetry
approach to ODEs are also possible; among many, we would like to
recall here the study and determination of \emphind{nonlinear
superposition principle} \cite{CdL4,CdL3,CGM1,CGM2,CdL1,CdL2},
similar to the familiar one for the \ind{Riccati equation}
\cite{CarRam,Kras,Reid}. This is related to the well known
\ind{Wei-Norman method} \cite{CMN,CKWN1,CKWN2,WN1,WN2}; the
stochastic counterpart of this, including \ind{nonlinear
superposition principle}, has been studied in \cite{LSstoch}.

As mentioned above, standard symmetry techniques often fails in
the (relevant!) case of \ind{dynamical systems}. In this
framework, one is often interested in the situation near a known
solution, and this is investigated by perturbation techniques;
most of these are based in a way or another on the approach
pioneered by \ind{Poincar\'e} and also known as the method of
\emphind{normal forms}. The interplay between \ind{symmetry and
perturbations} (which is, as well known, of paramount importance
in Quantum Mechanics \cite{Dirac,LLQM,Mess,Tin,Weyl}) has also
been investigated in the literature, both in general and for the
specific case of the \ind{normal forms} approach
\cite{ArnGMDE,CG,Elp,Iooss}, see e.g.
\cite{GWsplit3,GWsplit1,GWsplit2}.\footnote{See also the
proceedings of the series of conferences on ``Symmetry and
Perturbation Theory'' (SPT) \cite{SPTproc}; and related volumes
\cite{SPTAAM}.}

\medskip\noindent
{\bf Example \nc.6.} Let us consider the equation
$$ x^2 \, u'' \ + \ x \, u' \ + \ u \ = \ 0 $$ seen in Example \nc.1
above; in this case the (linear) equation is easily solved, and
the solutions are
$$ u(x) \ = \ c_1 \, \sin(\log|x|) \ + \ c_2 \, \cos (
\log|x|) \ . $$ We will use symmetry to obtain these solutions,
and we deal with the symmetry generated by
$$ X \ = \ x \, \pa_x \ . $$ The associated change of variables $(x,u) \to (y,z)$,
putting the \VF in the form $X = \pa_z $, is
$$ x \ = \ e^z \ , \ \ u \ = \ y \ ; \ \ \[ \mathrm{with \ inverse} \ \ \ y = u \ , \ \ z \ = \
\log (x) \] \ . $$ With these, and recalling $y$ is the new
independent variable, $z$ the new dependent one, we have
$$ \begin{array}{l} du = d y , \ d x = e^z \, d z ; \\ du/dx = 1/(e^z dz/dy) \ ; \\
d^2 u/ d x^2 = - (e^{-2z}/ z_y) (1 + z_{yy}/z_y^2) \ , \end{array}
$$ and hence the equation is rewritten as
$$ - \ \( \frac{1}{z_y} \ + \ \frac{z_{yy}}{z_y^3} \) \ + \
\frac{1}{z_y} \ + \ y \ = \ 0 \ ; $$ passing now to the variable
$w = z_y$, we get the first order separable equation
$$ \frac{w_y}{w^3} \ = \ y \ . $$
This is immediately solved, yielding
$$ w(y) \ = \ \pm \ \frac{1}{\sqrt{k_1 - y^2 }} \ . $$
Going back to the variable $z$, we have to solve
$$ \frac{dz}{dy} \ = \ \pm \ \frac{1}{\sqrt{k_1 - y^2 }} \ , $$
which provides
$$ z(y) \ = \ \pm \ \arctan \[ \frac{y \ \sqrt{k_1 - y^2}}{y^2 -
k_1}\] \ + \ k_2 \ . $$ Inverting the original change of
coordinates, i.e. going back to the $(x,u)$ coordinates, this
reads
$$ \log[x] \ = \ \pm \arctan \[ \frac{u \ \sqrt{k_1 - u^2}}{u^2 -
k_1}\] \ + \ k_2 \ , $$ which is inverted to give
\begin{eqnarray*} u &=& \pm \ \frac{\sqrt{k_1} \ \tan (k_2 - \log
x )}{\sqrt{1 + \tan^2 (k_2 - \log x )}} \\
&=& \pm \ \sqrt{k_1} \ \sin (k_2 - \log x ) \\
&=& \pm \ \sqrt{k_1} \[ \sin (k_2) \, \cos (\log x) \ - \ \cos
(k_2) \, \sin (\log x) \] \\
&=& c_1 \, \sin (\log x ) \ + \ c_2 \, \cos (\log x )  \ .
\end{eqnarray*}
{} \EOE

\medskip\noindent
{\bf Example \nc.7.} Let us consider the equation
$$ u'' \ = \ (x - u) \ (u')^3 $$
of Example \nc.3 above. Among its eight
symmetries, we consider
$$ X_3 \ = \ (x - u) \, \pa_x  $$
and the associated change of variables
$$ t \ = \ u \ , \ \ s \ = \ \log (x - u) \ ; \ \ \[ u = t \ , \ \
x = t + e^s \] \ . $$ This entails
\begin{eqnarray*}
\frac{du}{dx} &=& \frac{1}{1 + e^s s_t} \\
\frac{d^2 u}{d x^2} &=& - \ \frac{e^s}{(1 + e^s s_t)^3} \ \(
s_{tt} + s_t^2 \) \ . \end{eqnarray*}
 In this way the equation reads
$$ \frac{d^2 s}{d t^2} \ + \ \( \frac{d s}{d t} \)^2 \ + 1 \ = \ 0
\ ; $$ this can be solved directly, yielding
$$ s(t) \ = \ \^k_2 \ + \ \log \[ \cos (t - k_1 ) \] \ ; $$
or we can complete our general procedure by passing to the
dependent variable $w := (d s / d t)$, in terms of which we get a
first order (separable) equation,
$$ dw/dt \ + \ w^2 \ + \ 1 \ = \ 0 \ . $$
This gives
$$ w(t) \ = \ - \tan (t - k_1) \ = \ \tan (k_1 - t) $$
and therefore we get $s(t)$ as above.

Going back to the original variables, and writing $\^k_2 = \log
(k_2)$, this reads
$$ \log (x-u) \ = \ \log (k_2) \ + \ \log \[ \cos (k_1 - u ) \] \ , $$
namely
$$ (x-u) \ = \ k_2 \ \cos (k_1 - u ) \ , $$ which provides the
solution (in implicit form) to our original equation, as can be
checked by explicit computation. \EOE

\medskip\noindent
{\bf Example \nc.8.} In this and the following example, taken from
Olver \cite{Olver1}, we will actually consider classes of (second
order) ODEs. We start by considering a general autonomous second
order equation, \beq F (u,u_x,u_{xx} ) \ = \ 0 \ . \eeq This is
invariant under $X = \pa_x$, i.e. translations of the independent
variable. To let this fit into our scheme, we must therefore
invert the role of dependent and independent variables; our change
of coordinates will be
$$ x = z \ , \ u = y \ ; $$ with this we get
$$ \frac{d u}{d x} \ = \ \frac{1}{z_y} \ , \ \ \frac{d^2 u}{d x^2}
\ = \ - \, \frac{z_{yy}}{z_y}^2 \ \frac{d z}{d y} \ = \ - \,
\frac{z_{yy}}{z_y^2} \, u_x \ = \ - \, \frac{z_{yy}}{z_y^3} \ .
$$ The original equation reads therefore
\beq F \( y , \frac{1}{z_y} , - \frac{z_{yy}}{z_y^3} \) \ := \ H
(y , z_y , z_{yy} ) \ = \ 0 \ . \eeq with the new change of
dependent variable $z_y = w$ this reads \beq H (y,w,w_y) \ = \ 0 \
. \eeq
{} \EOE

\medskip\noindent
{\bf Example \nc.9.} Consider a general linear homogeneous second
order equation, \beq \frac{d^2 u}{d x^2} \ + \ p(x) \ \frac{d u}{d
x} \ + \ q(x) \ u \ = \ 0 \ . \eeq Being linear, this is invariant
under scale transformations in $u$, and these are generated by
$$ X \ = \ u \ \pa_u \ . $$ The associated change of variables is
$$ x \ = \ y \ , \ \ u \ = \ e^z \ ; \ \
\[ \mathrm{with \ inverse} \ y = x \ , \ \ z = \log
(u) \] \ . $$ In these variables we have $X = \pa_z$; moreover,
$$ u_x = e^z z_y \ , \ \ u_{xx} = e^w (z_{yy} + z_y^2) \ . $$
Therefore the equation reads now
$$ e^w \ \[ z_{yy} \ + \ z_y^2 \ + \ p(y) \, z_y \ + \ q (y) \] \
= \ 0 \ . $$ We can eliminate the factor $e^w$ (which is never
zero); by the usual change of dependent variable $w = z_y$ we
further rewrite the equation as \beq w_y \ + \ w^2 \ + \ p(y) \, w
\ + \ q(y) \ = \ 0 \ ; \eeq note this is a Riccati equation
\cite{Kras,Reid}. \EOE

\subsection{Conserved quantities}
\label{sec:Hojman}

As well known, Noether theory \cite{Koss,Noether,Olver1} provides
the connection between symmetry and conservation laws (for PDEs)
or directly \ind{conserved quantities} (for ODEs), in the case of
\emph{variational} systems. However, even for non-variational
systems of ODEs there are conserved quantities associated to
(some) symmetries, as first noticed by Hojman \cite{Hoj}.

This fact played an important role in early investigation of
symmetries of stochastic differential equations (see in particular
the works by Misawa and by Albeverio and Fei, to be discussed in
Sect. \ref{sec:Misawa} below), and it is thus worth discussing it
in some detail in the present context.

Following Hojman (and reverting to consider $t$ as the independent variable, $x^i$ as the dependent ones), we consider a system of second order ODEs for $x (t) \in V$, with $V$ an $n$-dimensional manifold; as the result is
local, we can just consider $V = \R^n$. Our system will be written, at least locally, in the form \beq \label{eq:Hoj1} \ddot{x}^i \
= \ F^i (x,\xd , t) \ , \eeq with $F$ a smooth function (which
will be required to satisfy some additional condition). We will
consider symmetry (generalized) vector fields \cite{Olver1} of the
form \beq \label{eq:HojX} X \ = \ \phi^i (x,\xd,t) \
\frac{\pa}{\pa x^i} \ ; \eeq this might be the evolutionary
representative \cite{Olver1} of a vector field $$ X_0 \ = \ \tau
(x,t) \, \pa_t \ + \ \xi^i (q,x) \, \pa_i \ , $$ but this is not
necessarily the case. \footnote{Note that the evolutionary
representative of $X_0$ is $X_v = (\xi^i - \tau \xd^i) \pa_i $,
which sets a severe limitation on the form of the $Q^i$ in $X$ for
the latter to be in fact an evolutionary representative.}

Note that the statement that $X$ is a symmetry of \eqref{eq:Hoj1}
means in this case that its components satisfy \beq D_t^2 ( \phi^i
) \ = \ \phi^j \, \frac{\pa F^i}{\pa x^j} \ + \ (D_t \phi^j) \,
\frac{\pa F^i}{\pa \xd^j} \ . \eeq This can be checked by direct
explicit computation, and the same holds for the following
proposition \cite{Hoj}.

\medskip\noindent
{\bf Proposition \HOJ.} {\it Let $X$ as in \eqref{eq:HojX} be a
symmetry of eq.\eqref{eq:Hoj1}, and let $F$ in \eqref{eq:Hoj1}
satisfy \beq \label{eq:HojF} \frac{\pa F^i}{\pa \xd^i} \ = \ - \
D_t [ \log (\la )] \eeq for some smooth function $\la = \la (x)$.
Then the quantity \beq J_\la \ := \ \frac{1}{\la} \ \[ \frac{\pa
(\la \phi^i )}{\pa x^i} \ + \ \frac{\pa [\la \, (D_t \phi^i)]}{\pa
\xd^i} \] \eeq is conserved under the flow of \eqref{eq:Hoj1}. }
\bigskip

\medskip\noindent
{\bf Remark \nc.2.} It should be noted that for $\la$ constant, so that \eqref{eq:HojF} reduces to
\beq \label{eq:HojF1} \frac{\pa F^i}{\pa \xd^i} \ = \ 0 \ , \eeq
the conserved quantity identified by Proposition {\HOJ} is just \beq
J_0 \ := \ \frac{\pa \phi^i}{\pa x^i} \ + \ \frac{\pa (D_t
\phi^i)}{\pa \xd^i} \ . \eeq
{} \EOR

\medskip\noindent
{\bf Example \nc.10.} Consider a harmonic oscillator (which of
course would admit a variational description) in polar coordinates
$(r,\theta)$ \cite{Hoj}, so that \eqref{eq:Hoj1} is now
\begin{eqnarray*} \ddot{r} &=& - \, \om^2 \, r \ + \ r \,
\dot{\theta}^2 \ , \\
\ddot{\theta} &=& - \, \frac{2}{r} \, \dot{r} \, \dot{\theta} \ .
\end{eqnarray*}
The (generalized) vector field
$$ X \ = \ r^3 \, \dot{\theta} \ \pa_r $$ is a (generalized) symmetry for these
equations. On the other hand, eq.\eqref{eq:HojF} is satisfied by
choosing $\la = r^2$. In this case one obtains
$$ J_\la \ = \ 6 \, r^2 \ \dot{\theta} \ , $$
which is proportional to the angular momentum (which is of course itself
conserved). \EOE

\section{Dynamical systems} \label{sec:dynamicalsystems}

I declared above that I would consider \emph{scalar} equations; an
exception is however in order, i.e. to consider \emphind{dynamical
systems}. By this we mean systems of first order ODEs of the
form\footnote{Note that, at difference with other sections but
conforming to the general use in the literature, here we denote
the independent variable as $t$ and the dependent ones as $x^i$.}
\beq \label{eq:DSna} \frac{d x^i}{d t} \ = \ f^i (x,t) \ . \eeq
These may represent equations in $M_0 = \R^n$, or the $x^i$
($i=1,...,n$) can be more generally local coordinates on a
$n$-dimensional manifold $M_0$, the restricted phase manifold; we
will also consider $M = M_0 \times \R$ (the second factor
representing time), the phase manifold.\footnote{In the dynamical
systems literature these are sometimes referred to as,
respectively, the phase manifold and the augmented (or extended)
phase manifold.}

Note that (unless we consider problem focusing on the zeroes of
$f$, as e.g. in \ind{normal forms} theory
\cite{ArnGMDE,Elp,Iooss}) we can always reduce to considering
\emph{autonomous} dynamical systems, i.e. to the case of \beq
\label{eq:DSa} \frac{d x^i}{d t} \ = \ f^i (x) \eeq simply by
adding a new variable $x^0$ with evolution equation $d x^0 / dt =
1$.

Moreover, in applications one is quite often concerned from the
beginning with autonomous systems.

Here we will just give some basic result, also in order to ease
comparison with the results for \emph{stochastic} dynamical
systems to be considered later on. A specific discussion of symmetries for
dynamical systems is provided e.g. in \cite{CG} (see in particular
Chapter III there), to which the reader is referred for further
detail.

\subsection{Symmetry of dynamical systems} \label{sec:symmDS}

To the dynamical system \eqref{eq:DSa} is naturally associated a
vector field in $M_0$, i.e. \beq X_f \ = \ f^i (x) \, \pa_i \ ;
\eeq note that for a general dynamical system \eqref{eq:DSna} we
need to consider vector fields defined in $M$, i.e. $$ X_f \ = \
f^i (x,t) \, \pa_i \ . $$ We can also associate to the same
dynamical system \eqref{eq:DSna} or \eqref{eq:DSa} a full
dynamical vector field (always defined in $M$), taking into
account also the flow of time; this reads \beq Z_f \ = \ \pa_t \ +
\ f^i (x,t) \, \pa_i \ . \eeq

When looking for symmetries of a dynamical system, one should
consider general vector fields of the form \beq \label{eq:symmDS}
X \ = \ \tau (x,t) \, \pa_t \ + \ \vphi^i (x,t) \, \pa_i \ . \eeq
The general procedure, described in Sect. \ref{sec:determining},
would produce an under-determined system of $n$ equations for the
$n+1$ functions $(\tau; \vphi^1,...,\vphi^n)$; for \eqref{eq:DSa}
these read \cite{CG} \beq \label{eq:deDSgen} \frac{\pa
\vphi^i}{\pa t} \ + \ f^j \, \frac{\pa \vphi^i}{\pa x^j} \ - \
\vphi^j \, \frac{\pa f^i}{\pa x^j} \ = \ \( \frac{\pa \tau}{\pa t}
\ + \ f^j \ \frac{\pa \tau}{\pa x^j} \) \ f^i \ . \eeq

In this context, it is quite natural to look for more restricted
classes of symmetries: that is,
\begin{itemize} \item[(i)] those
which act on $t$ just by a reparametrization (that is, with $\tau
= \tau (t)$ only), also designed as \index{fiber-preserving
symmetries} \emph{fiber-preserving} symmetries; \item[(ii)] those
which do not act on $t$ (automorphisms, possibly time-dependent,
of $M_0$: that is, with $\tau = 0$), also designed as
\index{time-preserving symmetries} \emph{time-preserving}
symmetries; \item[(iii)] or even which neither act nor depend on
$t$ (automorphisms of $M_0$: that is, with $\tau =0$, $\vphi^i_t =
0$); these are also designed as \index{LPTI symmetries} \emph{Lie
point time-independent} (LPTI) symmetries.
\end{itemize}

In discussing these, and more generally symmetries of dynamical
systems, it is useful to introduce the \emphind{Lie-Poisson
bracket} of two ($C^\infty$) functions defined on $M_0$ (which is
again a $C^\infty$ function on $M$). This reads \beq
\label{eq:LPB} \{ f , g \} \ := \ (f \cdot \grad) \, g \ - \ (g
\cdot \grad) \, f \ ; \eeq in components, we have \beq \{ f , g
\}^i \ = \ f^j \, \frac{\pa g^i}{\pa x^j} \ - \ g^j \, \frac{\pa
f^i}{\pa x^j} \ . \eeq The bracket is obviously antisymmetric, $\{
g , f \}  =  - \{ f , g \}$, and satisfies the \ind{Jacobi identity}.

This bracket has an immediate relation with the commutator of the
vector fields $X_f$ and $X_g$ associated to the functions $f$ and
$g$. In fact, \begin{eqnarray*} [ X_f , X_g ] &=& [ f^j \pa_j ,
g^m \pa_m ] \ = \ [f^j (\pa_j g^i) - g^j (\pa_j f^i)] \, \pa_i \\
&=& \{ f , g \}^i \ \pa_i \ = \ X_{\{ f , g \} } \ .
\end{eqnarray*}
In other words,
$$ \{ f , g \} \ = \ h \ \Longleftrightarrow \ [X_f , X_g ] \ = \
X_h \ . $$ With this notation, we have

\medskip\noindent
{\bf Proposition \DETEQDS.} {\it The general \ind{determining equations}
\eqref{eq:deDSgen} for symmetries of an autonomous dynamical
system \eqref{eq:DSa} read \beq \vphi^i_t \ + \ \{f , \vphi\}^i \
= \ (Z_f \tau) \, f^i \ . \eeq For symmetries of class $(i)$ above
these reduce to \beq \label{eq:deDSi} \vphi^i_t \ + \ \{f ,
\vphi\}^i \ = \ \tau_t \, f^i \ ; \eeq for class $(ii)$ we just
have \beq \label{eq:deDSii} \vphi^i_t \ + \ \{f , \vphi\}^i \ = \
0 \ ; \eeq and for LPTI, i.e. class $(iii)$, symmetries we get
\beq \label{eq:deDSiii} \{f , \vphi\}^i \ = \ 0 \ . \eeq}
\bigskip

Note that \eqref{eq:deDSi} can be reduced to the form
\eqref{eq:deDSii} by defining $$ \psi \ = \ \vphi \ - \ \tau \, f
\ ; $$ in fact with this \eqref{eq:deDSi} just reads \beq \psi^i_t
\ + \ \{f , \psi\}^i \ = \ 0 \ . \eeq

In the following, we will denote by $\G_f$ the set of
time-preserving symmetries for $f$, that is of vector fields $X =
\vphi^i (x,t) \pa_t$ satisfying \eqref{eq:deDSii}.

\subsection{Constants of motion, and the module structure}
\label{sec:moduleSDS}

Assume now that the dynamical system \eqref{eq:DSa} admits a
conserved quantity (or constant of motion, or first integral)
$\a$, i.e. a smooth function $\a : M \to \R$ such that $Z_f (\a) =
0$; we will denote by $\I_f$ the set of these functions (clearly
sums and products of such functions still give functions in
$\I_f$). In the case of $\a$ not depending on time, this also
reads $X_f (\a) = 0$. In this case, if $X$ is a (time-preserving)
symmetry for \eqref{eq:DSa}, then $\wt{X} = \a X$ is also a
symmetry for the same dynamical system. In fact,
\begin{eqnarray*}
\wt{\vphi}^i_t \ + \ \{ f , \wt{\vphi} \}^i &=& \pa_t (\a \,
\vphi^i) \ + \ \{ f , \a \vphi \}^i \\ &=& \a_t \, \vphi^i \ + \
\a \, \vphi^i_t \ + \ \a
\, \{ f , \vphi \}^i \ + \ f^j \pa_j (\a ) \\
&=& \a \ \( \vphi^i_t \ + \ \{ f , g \}^i \) \ + \ Z_f (\a ) \ = \
0 \ . \end{eqnarray*} We conclude that the set $\G_f$ of
(time-preserving) symmetry generators for a given dynamical system
has, beside the structure of algebra, also the structure of a
\emphind{module} over $\I_f$. More precisely, we have the
following \cite{Wal330}

\medskip\noindent
{\bf Proposition \WALa.} {\it The set $\G_f$ is a finitely
generated module over $\I_f$.}

\medskip\noindent
{\bf Remark \nc.3.} In the presence of nontrivial $I_f$, the
set $\G_f$ will be infinite dimensional as a Lie algebra, and
finite dimensional as a Lie module. \EOR

\medskip\noindent
{\bf Remark \nc.4.} Our discussion, and Proposition {\WALa}, also hold
for the set $\Ga_f$ of general symmetries for \eqref{eq:DSna},
i.e. of vector fields of the form \eqref{eq:symmDS} satisfying
\eqref{eq:deDSgen}. In fact, consider the vector field $\wt{X} =
\a X$ as above, and assume $X$ is a symmetry for \eqref{eq:DSna}.
Then \eqref{eq:deDSgen} yields
\begin{eqnarray*}
& & \frac{\pa \wt\vphi^i}{\pa t} \ + \ f^j \, \frac{\pa
\wt\vphi^i}{\pa x^j} \ - \ \wt\vphi^j \, \frac{\pa f^i}{\pa x^j} \
- \ \( \frac{\pa \wt\tau}{\pa t} \ + \ f^j \ \frac{\pa
\wt\tau}{\pa x^j} \) \ f^i \\
&=& \a \ \[ \frac{\pa \vphi^i}{\pa t} \ + \ f^j \, \frac{\pa
\vphi^i}{\pa x^j} \ - \ \vphi^j \, \frac{\pa f^i}{\pa x^j} \ - \
\( \frac{\pa \tau}{\pa t} \ + \ f^j \ \frac{\pa \tau}{\pa x^j} \)
\ f^i \] \\
& & \ + \ \a_t \, \vphi^i \ + \ \vphi^i \, f^j \pa_j (\a) \ - \
\tau \, \( \a_t + f^j \pa_j (\a) \) \, f^i \\
&=& \( \vphi^i \, - \, \tau \, f^i \) \ Z_f (\a ) \ = \ 0 \ ;
\end{eqnarray*}
the term in square brackets was cancelled since it is zero by the
assumption that $X$ is a symmetry. \EOR

\medskip\noindent
{\bf Example \nc.11.} Consider the simple system
\begin{eqnarray*}
dx/dt &=& \[ 1 + \exp[ - (x^2 + y^2)] \] \, y \\
dy/dt &=& - \ \[ 1 + \exp[ - (x^2 + y^2)] \] \, x \ .
\end{eqnarray*}
This obviously admits $X_0 = y \pa_x - x \pa_y$ (i.e. rotations)
as a symmetry, and $\rho = (x^2 + y^2)$ as a conserved quantity.
In fact, it is easy to check that any vector field of the form
$$ X_k \ = \ \rho^k \ X_0 $$ is also a symmetry. \EOE

\subsection{Orbital symmetries} \label{sec:orbital}

It is well known that studying dynamical systems it is often
fruitful to focus on \emphind{trajectories} (also called
\emphind{solution orbits}) rather than on full solutions (which
include the law of displacement along trajectories, i.e. the time
parametrization of solution curves).

Thus it appears one could have some advantage in considering,
beside full symmetries (which map solutions into solutions) and
invariant full solutions, also so called \emphind{orbital
symmetries} and \emphind{invariant trajectories}. This approach
was pursued in particular by Walcher \cite{Wal332,Wal333} (see
also \cite{CGWLie1}), and here we will give some basic notions
about it.

First of all we should characterize dynamical systems of the form
\eqref{eq:DSa} which have the same solution orbits, i.e. introduce
an equivalence relation in the set of systems of this form. It is
clear that \eqref{eq:DSa} will have the same trajectories as any
system of the form (where $\mu (x) \not= 0$ whenever $f(x) \not= 0$)
\beq dx^i / d t \ = \ \mu (x) \ f^i (x) \ ;
\eeq note that one could also consider a multiplier $\mu$
depending on time as well, but this would lead us into the realm
of time-dependent dynamical systems \eqref{eq:DSna}.

Now we recall that we defined symmetries of an equation $\De$ as
vector fields $X$ such that their prolongation (in this case,
$X^{(1)}$) leaves the solution manifold $S_\De$ invariant. Dealing
with equivalence classes of vector fields, we should require that
$X^{(1)}$ maps $S_\De$ into a possibly different manifold
$S_{\wt{\De}}$ which is the solution manifold for a possibly
different equation $\wt{\De}$, in the same equivalence class as
$\De$.

By standard computations, we obtain that (the first prolongation
of) a generic vector field
$$ X \ = \ \tau (x,t) \, \pa_t \ + \ \vphi^i (x,t) \, \pa_i $$
maps the dynamical system \eqref{eq:DSa} into a new dynamical
system \beq d x^i / d t \ = \ g^i (x,t) \eeq with \beq g^i \ = \
f^i \ - \ \eps \ \[ \vphi^i_t \ + \ \{ \vphi , f \}^i \ - \ (D_t
\tau ) \, f^i \] \ . \eeq For what we have seen above, the new
system is in the same equivalence class as the old one if and only
if $g$ is proportional to $f$ through a scalar function, i.e. if
and only if there is a function $\mu$ such that \beq \vphi^i_t \ +
\ \{ \vphi , f \}^i \ - \ (D_t \tau ) \, f^i \ = \ \mu (x,t) \ f^i
\ . \eeq As the last term in the l.h.s. is surely of the required
form (that is, proportional to $f$ and thus can be absorbed into
the definition of $\mu$), we are reduced to study the equation
\beq \vphi^i_t \ + \ \{ \vphi , f \}^i \ = \ \mu (x,t) \ f^i \ .
\eeq Note that this is the same as the equation determining
standard (time-preserving) symmetries, except that in that case we
required $\mu \equiv 0$.

We would like to recall an equivalent (local) characterization of
the equivalence relation we are considering here; this is taken
from \cite{CGWLie1} (see also \cite{CG94,CGmcm} for somewhat related
results).

\medskip\noindent
{\bf Proposition \WALb.} {\it Two dynamical systems are locally
orbit-equivalent if and only if they admit the same first
integrals, and hence the same invariant sets, near any
non-stationary point.}

\section{Symmetry and PDEs}
\label{sec:PDEs}
\def\cs{\ref{sec:PDEs}}

In the case of PDEs we will, for the sake of definiteness,
consider an equation with two independent variables $(x,t)$ and a
dependent one, $u$. Again to keep things simple we will focus on
the case of a second order equation, say \beq \De \ := \ \
F(x,t,u;u_x,u_t,u_{xx},u_{xt},u_{tt} ) \ = \ 0 \ . \eeq The vector
fields can in this framework be written as \beq \label{eq:VFpde} X
\ = \ \xi (x,t,u) \, \pa_x \ + \ \tau (x,t,u) \, \pa_t \ + \ \phi
(x,t,u) \, \pa_u \ ; \eeq we refrain from writing down explicitly
its second order prolongation $Y = X^{(2)}$ (see e.g.
\cite{GB,Olver1} for the explicit expression).

The approach in the case of PDEs is in a way at the opposite as
the one for ODEs. That is, if we have determined a vector field of
the form \eqref{eq:VFpde} which is a symmetry for $\De$, we will
perform a change of coordinates \beq (x,t;u) \ \to \ (y , s ; v )
\eeq (here $v$ should be thought as the new dependent variable)
such that in the new coordinate the \VF $X$ reads
as\footnote{Again the notation $\wt{X}$ should not cause
confusion, this is the same geometrical object as $X$ expressed in
different coordinates.} \beq \wt{X} \ = \ \frac{\pa}{\pa y} \ ;
\eeq that is, it should be all along one of the \emph{independent}
coordinates (as opposed to the ODE case, where it was set to be
along the \emph{dependent} one).

In the new coordinates, the equation will be written as \beq
\wt{\De} \ := \ \ G (y,s,v;v_y,v_s,v_{yy},v_{ys},v_{ss} ) \ = \ 0
\ . \eeq Again the tangency condition \eqref{eq:symmcond} holds
independently of the used coordinates, and again prolongation
formula guarantees that $\wt{X}^{(2)} = \wt{X}$.

Now our goal will \emph{not} be to obtain a general reduction of
the equation, but instead to obtain a (reduced) equation which
determines the \emphind{invariant solutions} to the original
equation (by this we mean $X$-invariant, of course).

In the new coordinates, this is just obtained by imposing $v_y =
0$, i.e. $v = v(s)$. Note that the reduced equation will have
(one) less independent variables than the original one. Needless
to say, this is specially good when we started from a PDE with two
independent variables, but it is useful in general. On the other
hand, this reduced equation will \emph{not} have solutions in
correspondence with general solution to the original equation:
only the invariant solutions will be common to the two equations
(note also that in this case, contrary to the ODE case, we do not
need to solve any \index{reconstruction} ``reconstruction problem'').

In this sense, symmetry is just providing a way to make educated
guesses (or educated \emph{ansatzes}) about the functional
dependence of some classes of special solutions.\footnote{It will
of course provide much more, but we do not want to enter in the
details of the theory \cite{\symmref}.}

\medskip\noindent
{\bf Remark \nc.5.} It should be mentioned that one can also
determine invariant solutions under symmetry groups which are
\emph{not} symmetries of the equation, see e.g. the approach by
Levi and Winternitz in terms of \index{conditional symmetries}
``conditional symmetries'' \cite{LeWinCond,OlvVor} (these are also related to the \index{non-classical method} ``non-classical method'' \cite{ClarKru,OlvRos}; see \cite{PuS} for a discussion of such relation) or more generally the so called
\index{partial symmetries} ``partial symmetries'' \cite{CGpart}. The latter also admit an asymptotic formulation \cite{Gasy,GaeMan1,GaeMan2}. \EOR

\medskip\noindent
{\bf Remark \nc.6.} Here we only consider standard Lie-point
symmetries, i.e. those corresponding to standard vector fields.
These have been generalized in many ways (which cannot be
discussed here), including \ind{generalized symmetries}
\cite{\symmref}, the \ind{non-classical method}
\cite{Clarkson,ClarKru,ClarMan,OlvRos} \ind{potential symmetries}
\cite{Blumanpot,BRK,PuS,Sacpot,Zhda}, and nonlocal ones
\cite{AGI,AnBlu96,KrV2} (also in relation \cite{CF,CFM1,CFM2} to
so called \ind{solvable structures} \cite{BP1,Hartl,ShP}); in
these cases the transformations considered are not generated by
standard vector fields (the names correspond to the kind of
transformations considered).

More recently so called \index{twisted symmetries} ``twisted symmetries'' have been introduced by Muriel and Romero \cite{Muriel1,Muriel2,Muriel3},
see also \cite{CF,twisted1,twisted2}; in this case one considers
standard vector fields but a deformation of the prolongation
operation. The interested reader is referred to the literature for
more details. As far as I know, no corresponding extensions exist
for stochastic equations. \EOR

\medskip\noindent
{\bf Example \nc.12.} As mentioned above, symmetries can be used
to look for \emphind{invariant solutions} to a given PDE in terms
of a simpler (reduced) equation; in particular this might be an
ODE.

We will consider the \ind{KdV equation} \eqref{eq:KdV}; its
symmetries were computed in Example \nc.4 above. In particular we
will look at solutions invariant under $X = X_2 - v X_1$; these
are obviously functions of the form
$$ u(x,t) \ = \ \eta (x - v t) \ := \ \eta (z) \ , $$ i.e. travelling waves (with
speed $v$). Inserting this ansatz into the KdV equation, we get a
reduced (ordinary) equations, which reads \beq - v \, \eta_z \ + \
\eta_{zzz} \ + \ 6 \, \eta \, \eta_z \ = \ 0 \ . \eeq

In this way we achieved our task (reduction to an equation with
less independent variables). We also note this is immediately
integrated once, yielding (the $k_i$ will be integration
constants)
$$ - v \, \eta \ + \ \eta_{zz} \ + \ 3 \, \eta^2 \ + \ k_1 \ = \ 0 \
. $$ Multiplying this by $\eta_z$ and integrating again, we get
$$ - \, \frac{v}{2} \, \eta^2 \ + \ \frac12 \, \eta_z^2 \ + \
\eta^3 \ + \ k_1 \, \eta \ + \ k_2 \ = \ 0 \ . $$ This equation is
solved in terms of special (elliptic) functions; see e.g.
\cite{Olver1} (example 3.4) for details. For $k_1 = k_2 = 0$ we
get
$$ \eta (z) \ = \ \frac{v}{2} \ \frac{1}{\cosh^2 (\b )} \ , $$
where $\b =  - (\sqrt{v}/2) (k_3  \pm  x )$. This is of course the
well known \ind{one-soliton solution}; the simplest writing is
obtained by choosing $v=2$ and $k_3 = 0$ (the sign in $\b$ is then
inessential, due to parity of $\cosh^2 (x)$), in which case we get
$$ \eta \ = \ \frac{1}{\cosh^2 (x) } \ . $$
{} \EOE

\medskip\noindent
{\bf Example \nc.13.} As a second example of this procedure,
consider the \ind{sine-Gordon equation} \beq \label{eq:SG} u_{tt}
\ - \ u_{xx} \ = \ - \ \sin ( u) \ . \eeq This is autonomous,
hence surely invariant under both of
$$ X_1 \ = \ \pa_t \ \ , \ \ \ X_2 \ = \ \pa_x $$
and any linear combination thereof (reduction under $X_2$ gives --
upon a slight change of notation -- eq.\eqref{eq:SGred},
considered in Example \nc.2 above). Looking for solutions
invariant under $ X = X_2 - v X_1 $ amounts to looking for
travelling waves with speed $v$. Writing
$$ u(x,t) \ = \ \eta (x - v t) \ := \ \eta (z) $$ the sine-Gordon
equation is reduced to \beq \frac{d^2 \eta}{d z^2} \ = \
\frac{1}{1 - v^2} \ \sin (\eta ) \ , \eeq i.e. to the motion of a
particle of unit mass in an effective potential \beq W(\eta ) \ :=
\ \frac{1}{1 - v^2} \ \cos (\eta ) \ . \eeq Note that this is
qualitatively different depending on $v^2 < 1$ or $v^2 > 1$; in
particular it turns out that nontrivial solutions will not be able
to satisfy the natural boundary conditions (inherited from a
\ind{finite energy condition}) in the case $v^2 > 1$
\cite{CGspeed1,CGspeed2}. \EOE

\medskip\noindent
{\bf Example \nc.14.} The one-dimensional \index{heat equation}
heat (or diffusion) equation \beq \label{eq:heat} u_t \ = \ u_{xx}
\eeq has a symmetry algebra spanned (beside the infinite factor
related to its linear character, see Sect. \ref{sec:Bluman}) by six
vector fields:
\begin{eqnarray*}
X_1 &=& \pa_x \ , \ \ X_2 \ = \ \pa_t \ , \ \ X_3 \ = \ u \, \pa_u
\ , \ \ X_4 \ = \ x \, \pa_x \ + \ 2 t \, \pa_t \ ; \\
X_5 &=& 2 t \, \pa_x \ - \ x \, u \, \pa_u \ , \ \ X_6 \ = \ 4 \,
t^2 \, \pa_t \ + \ 4 \, t \, x \, \pa_x \ - \ (x^2 + 2 t) \, u \,
\pa_u \ ; \end{eqnarray*} note the first four generators
correspond to rather obvious (translation or scaling) symmetries,
$X_5$ is related to \ind{Galilean boosts} to moving coordinate
frames (see \cite{Olver1}, example 2.41), while $X_6$ is
nontrivial.

There is no doubt that $u = c$ is a (highly trivial!) solution to
the heat equation; on the other hand if we act on this by $X_6$,
it gets transformed into the one-parameter family of nontrivial
solutions
$$ u(x,t;s) \ = \ \frac{c}{\sqrt{1 + 4 s t}} \ \exp \[ - s \, \frac{x^2}{1
+ 4 s t} \] \ , $$ where $s$ is the group parameter.

Choosing $s = \pi c^2$ we get
$$ u(x,t) \ = \ \frac{c}{\sqrt{1 + 4 \pi c^2 t}} \ \exp \[ - \pi c^2 \, \frac{x^2}{1
+ 4 \pi c^2 t} \] \ ; $$ by a time translation (generator $X_2$)
of an amount
$$ \de t \ = \ \( \frac{1-c}{c} \) \, t \ - \ \frac{1}{4 \, c^2 \,
\pi}
$$ this is transformed into the fundamental solution
$$ u(x,t) \ = \ \frac{1}{\sqrt{4 \, \pi \ t}} \ \exp \[ - \,
\frac{x^2}{4 t} \] \ . $$

Thus by (two) symmetry transformations we have mapped the trivial
solution $u = c$ into the fundamental solution of the heat
equation. \EOE

\medskip\noindent
{\bf Example \nc.15.} The \ind{sine-Gordon equation} \eqref{eq:SG}
is invariant under the \ind{Lorentz group} (with the units used in
\eqref{eq:SG}, the limit speed is $c=1$). Thus we can determine
static solutions $u (x,t) = \eta (x)$ (as suggested by our
notation, these are obtained as in Example by setting $v=0$) and
then transform them into moving solution by a \ind{Lorentz boost}.
More generally, this applies to any Lorentz-invariant equation. \EOE

\section{Symmetry and linearization}
\label{sec:Bluman}
\def\cs{\ref{sec:Bluman}}

It is interesting to note that, as shown by Bluman and Kumei
\cite{Bluman,KB} (see also \cite{BLK3,BLK1,BLK2}), the
(algorithmic) symmetry analysis is also able to detect if a
nonlinear equation can be linearized by a change of
coordinates.\footnote{In the Calogero classification of integrable
systems \cite{Calwhat,CalDeg}, these would be \ind{C-integrable}; it is
remarkable that \ind{S-integrable} equations are also characterized by
(generalized) symmetry. We cannot touch upon this topic here, and
the reader is just referred e.g. to (chapter 5 in) the book by
Olver \cite{Olver1}, or to \cite{CalDeg,Mar2015,Dunajski}.}

The reason can be made quite clear intuitively: a linear equation
has a rather large symmetry algebra, corresponding to the (linear)
\emphind{superposition principle}. That is, we know \emph{apriori}
that any transformation adding to the independent variable an
\emph{arbitrary} solution of the equation itself, or actually a linear
combination of solutions, will map solutions into solutions.

On the other hand, such a property cannot be destroyed by a change
of coordinates. Thus if an equation is linear in some coordinates
but is expressed in a different system of coordinates (in which it
is nonlinear), there will be trace of it being linearizable. In
fact, the trace will show up in the symmetry computation in that
the \ind{determining equations} will admit, beside other solutions,
vector fields of the form \beq X_\a \ = \ \a (x,t) \ \pa_u \ ,
\eeq with $\a (x,t)$ an arbitrary solution of some \emph{linear}
equation. Then the nonlinear equation under study can be
transformed exactly to this linear equation (the symmetry also
gives a hint about how to obtain this, i.e. about the required
transformation).\footnote{This question has also been studied in the context of perturbation theory for dynamical systems \cite{BCGM,GaeMarmo}.}

\medskip\noindent
{\bf Example \nc.16.} The \ind{Burgers equation}
$$ v_t \ = \ v_{xx} \ + \ 2 \, v \, v_x \ = \ D_x \, (v_x + v^2) $$
can be rewritten in ``potential form'' setting (note this will
introduce an integration to go back to the original coordinates)
$v := u_x $; it reads then
$$ u_t \ = \ u_{xx} \ + \ u_x^2 \ . $$
This equation admits six symmetries (which correspond to those of
the heat equations) plus a family of the form mentioned above,
i.e. symmetry {\VF}s of the form
$$ X \ = \ \a (x,t) \, e^{- u} \, \pa_u \ . $$
To eliminate the factor $e^{-u}$, it suffices to set $u = \log w$,
i.e. $w = e^u$. With this, we get
$$ X_\a \ = \ \a (x,t) \ \pa_w \ . $$
Actually, writing the equation in terms of $w$ we get, in view of
$$ u_t = \frac{1}{w} \, w_t \ , \ \ u_x = \frac1w \, w_x \ , \ \
u_{xx} = \frac{w_{xx} w - w_x^2}{w^2} \ , $$ just the \ind{heat
equation}. Thus the symmetry analysis led us to ``discover'' the
well known \ind{Hopf-Cole transformation}
\cite{BlumanCole,Olver1}. \EOE

\vfill\eject

\chapter{Ito stochastic equations}
\label{chap:Ito}
\def\nc{\ref{chap:Ito}}

\section{Symmetry, SDEs, diffusion equations}
\label{sec:SDEs}
\def\cs{\ref{sec:SDEs}}

It is now time to come to the central issue of interest in this
paper, i.e. consider \emph{stochastic} differential equations
(SDEs) \cite{LArnold,BelDal,Evans,Freedman,Guerra,Ikeda,ItoMcKean, Kunita,McKean,Oksendal,Stroock,Kampen}
and their symmetries. Albeit the early works on this matter
\cite{AlbFei,Mis1,Mis2,Mis3,Mis4,Mis5} considered Stratonovich
equations\footnote{These will be considered later on, in Chapter
\ref{chap:Strato}.}, I will first (and mainly) focus on SDEs in
Ito form, \beq \label{eq:Ito} d x^i \ = \ f^i (x,t) \, d t \ + \
\s^i_{\ k} (x,t) \, d w^k \ ; \eeq note that I will only consider
\emph{ordinary} (as opposed to partial\footnote{For an attempt at
considering symmetry of partial SDEs, see \cite{Melnick}.}) SDEs.

In \eqref{eq:Ito}, as customary, $f^i$ and $\s^i_{\ j}$ are smooth
functions, $\s$ a nonzero matrix, and the $w^k$ are independent
homogeneous standard Wiener processes, satisfying \beq \langle
|w^i (t) - w^j (t) |^2 \rangle \ = \ \de^{ij} \, \de (t-s) \ .
\eeq

It is well known that to the Ito equation \eqref{eq:Ito} is
associated a diffusion (Fokker-Planck or
Chapman-Kolmogorov\footnote{In the following we will always refer
to this as the Fokker-Planck (FP) equation, conforming to the
Physics community notation.}) equation, which reads
\beq\label{eq:FP} u_t \ + \ A^{ij} \, u_{ij} \ + \ B^i \, u_i \ +
\ C \, u \ = \ 0 \ , \eeq where $u_i = (\pa u / \pa x^i)$ (and
similarly for $u_{ij}$), while the coefficients $A,B,C$ are
functions of the independent variables $(\xb,t)$, given by
\begin{eqnarray*}
A^{ij} &=& - \frac12 \ (\s \s^T)^{ij} \ , \\
B^i \, &=& f^i \ - \ \pa_j \, (\s \s^T)^{ij} \ , \\
C \ &=& (\pa_i \cdot f^i) \ - \ \frac12 \, \pa^2_{ij} \, (\s
\s^T)^{ij} \ . \end{eqnarray*}

\medskip\noindent
{\bf Remark \nc.1.} Here I will consider only (infinitesimal
generators of) continuous symmetries. Discrete symmetries of
stochastic equations are considered e.g. in the Appendix to
\cite{GRQ2}. \EOR

\medskip\noindent
{\bf Remark \nc.2.} As the Fokker-Planck equation \eqref{eq:FP} is
linear, it will have the symmetries of the form $X_\a$ with $\a$
an arbitrary solution of \eqref{eq:FP} itself, see
Sect. \ref{sec:Bluman}. We will consider these as trivial
symmetries in our context. \EOR

\medskip\noindent
{\bf Remark \nc.3.} A more careful analysis would uncover a
delicate question: when dealing with the Fokker-Planck equation in
a probabilistic context, we are interested only in solutions
$u(x,t)$ which satisfy the normalization condition $$
\int_{-\infty}^{+ \infty} u (x,t) \, dx \ = \ 1 \ \ \ \ \ \
\forall t \ . $$ This condition does \emph{not} correspond to a
linear space, and hence the linear superposition principle does
\emph{not} apply here. Correspondingly, the symmetries $X_\a$
mentioned in Remark \nc.2 should not be considered as acceptable
in the present context (see \cite{GRQ1}, appendix B, for details).
Thus, excluding them from consideration is correct, and not only
for their ``trivial'' nature.\footnote{Note, in passing, that a
similar problem arises when one considers solutions $\psi (x,t)$
to the Schroedinger equation, as these should satisfy $||\psi || =
1 $, with of course $||.||$ the $L_2$ norm.} \EOR

\medskip\noindent
{\bf Remark \nc.4.} In the following we will consider transformations
acting also on the independent variable $t$; it should be recalled
that this will also have an effect on the Wiener processes $w^i
(t)$. This is discussed in very simple terms in \cite{GRQ1} (see Appendix A there).
In particular, if $t \to s = t + \eps \tau (t)$, then
$$ w^i (t) \ \to \ \widetilde{w}^i (s) \ = \ \sqrt{1 + \eps \tau' (s)}
\ w^i (s) \ ; $$ this implies in particular
$$ d \widetilde{w}^i \ = \ (1 + \eps \tau' / 2) \ d w^i \ . \eqno{\odot} $$
More generally, it is known (see e.g. \cite{Oksendal}, Theorem
8.20; or \cite{Varadhan} chap.4) that under a general time change $t \to s = t + \eps
\tau (x,t)$ we have \beq w^i (t) \ \to \ \widetilde{w}^i (s) \ = \
\sqrt{1 + \eps (d \tau / d t)} \ w^i (s) \ , \eeq which of
course implies (for $(d \tau / d t)$ limited, as we always
assume) \beq \label{eq:dwtilde} d \widetilde{w}^i \ = \ [1 + \eps
(1/2) (d \tau / d t)] \ d w^i \ . \eeq This will be of use in
the following. \EOR

\medskip\noindent
{\bf Remark \nc.5.} As well known, the Ito equation \eqref{eq:Ito}
is in a (rather nontrivial, see e.g. \cite {Ikeda,McKean,Stroock})
sense \emph{equivalent} to the Stratonovich equation \beq
\label{eq:strato} d x^i \ = \ b^i (x,t) \, d t \ + \ \s^i_{\ k}
(x,t) \circ d w^k \eeq with \beq \label{eq:itotostrato} b^i (x,t)
\ := \ f^i (x,t) \ - \ \frac12 \left[ \frac{\pa (\s^T)^i_{\ j}
(x,t)}{\pa x^k} \s^{k j} \right] \ . \eeq In the following it will
be convenient to set \beq s^i (x,t) \ := \ \frac12 \left[
\frac{\pa (\s^T)^i_{\ j} (x,t)}{\pa x^k} \s^{k j} \right] \ , \eeq
so that \eqref{eq:itotostrato} reads simply \beq
\label{eq:itotostratosimp} b^i (x,t) \ := \ f^i (x,t) \ - \ s^i
(x,t) \ . \eeq However, one should be careful about this
equivalence, which -- as mentioned above -- goes through some
subtle points (see e.g. \cite{Stroock}, chapter 8).

We will discuss the relation between symmetries of an Ito equation and of
the equivalent Stratonovich one in Sect. \ref{sec:itovsstrat}
below, after discussing also symmetries of equations in
Stratonovich form. \EOR
\bigskip

\medskip\noindent
{\bf Remark \nc.6.} Here again I will \emph{not} consider
variational problems
\cite{CruZam,Gue2,Pavon,Yasue,ZamCV1,ZamCV2,ZamCV3,ZamCV4,ZY} and
their symmetries\footnote{One should also mention that in recent times there has been a surge of interest on variational principles for stochastic fluid dynamics, also in connection with stochastic soliton equations \cite{Holm,HolmTyr}.}. It should however be stressed that stochastic versions of Noether theory exist \cite{PQS2,BaFon,Mis1,TZ,PQS1}. The relation between symmetries and conserved quantities for stochastic non variational systems has been considered by several
authors (see in particular \cite{ADK,AlbFei,ARZ,ARW,Mis2,Mis3,Mis4,Mis5}) and will be discussed in Sect. \ref{sec:scq} below. \EOR

\section{Transformation of an Ito equation, and symmetries}
\label{sec:itotransf}

We will start by providing the (skeleton of the) fundamental
computation concerning transformation of an Ito equation under the
action of a vector field; see also \cite{GS2016} for details.

We consider a fully general vector field in the $(x,t)$ space,
i.e. \beql{eq:compItoVF} X \ = \ \tau (x,t) \, \pa_t \ + \
\xi^i(x,t) \, \pa_i \ . \eeq Under the action of this we have,
recalling also Remark \nc.4 above and in particular
\eqref{eq:dwtilde}, and working always at first order in $\eps$,
\begin{eqnarray*}
x &\to& x^i \ + \ \eps \ \xi^i (x,t) \\
t &\to& t \ + \ \eps \ \tau (x,t) \\
dw^k (t) &\to& dw^k (t) \ + \ \eps \, \frac12 \, (\pa_t \tau) \ d
w^k (t) \ . \end{eqnarray*}

Thus the Ito equation \beql{eq:compItoEQ} d x^i \ - \ f^i (x,t) \,
d t \ - \ \s^i_{\ k} \, d w^k (t) \ = \ 0 \eeq is mapped
into\footnote{Here all functions are to be thought as depending on
$x$ and $t$; that is, $f^i= f^i (x,t)$ and the like.}
\begin{eqnarray} d x^i \ + \ \eps \, d \xi^i &=& [ f^i \ + \ \eps \( \xi^j \pa_j \xi^i + \tau \pa_t \xi^i \) ] \ (d t \, + \, \eps \, d \tau ) \label{eq:compIto1} \\
& & \ + \ [ \s^i_{\ k} \ + \ \eps \( \xi^j \pa_j \s^i_{\ k} + \tau
\pa_t \s^i_{\ k} \) \ [1 + \eps (1/2) (\pa_t \tau)] \ d w^k \ .
\nonumber \end{eqnarray} We now just have to use Ito formula to
evaluate the differentials $d \xi^i$ and $d \tau$, which gives
\begin{eqnarray*}
d \xi^i &=& \( \frac{\pa \xi^i}{\pa t} \) \, d t \ + \ \( \frac{\pa \xi^i}{\pa x^j} \) \, d x^j \ + \ \frac12 \, \( \frac{\pa^2 \xi^i}{\pa x^j \pa x^m} \) \, \s^j_{\ k} \, \s^m_{\ k} \, d t \ , \\
d \tau &=& \( \frac{\pa \tau}{\pa t} \) \, d t \ + \ \( \frac{\pa \tau}{\pa x^j} \) \, d x^j \ + \ \frac12 \, \( \frac{\pa^2 \tau}{\pa x^j \pa x^m} \) \, \s^j_{\ k} \, \s^m_{\ k} \, d t \ . \end{eqnarray*}

Plugging these into \eqref{eq:compIto1}, and restricting to the
flow of \eqref{eq:Ito} -- that is, substituting for $d x^i$
according to it -- we get that of course terms of order zero in
$\eps$ cancel out, while first order terms yield a contribution
\begin{eqnarray}
& & \[ \frac{\pa \xi^i}{\pa t} \ + \ f^j \ \frac{\pa \xi^i}{\pa x^j} \ - \ \xi^j \ \frac{\pa f^i}{\pa x^j} \ - \ \frac{\pa (\tau f^i)}{\pa t} \ - \ f^j \ \frac{\pa \tau}{\pa x^j} \ f^i \right. \nonumber \\
& & \left. \ + \ \frac12 \ \( \frac{\pa^2 \xi^i}{\pa x^j \pa x^m} \, + \, f^i \, \frac{\pa^2 \tau}{\pa x^j \pa x^m} \) \, \s^j_{\ k} \, \s^m_{\ k} \] \ d t \nonumber \\
&+& \[ \s^j_{\ k} \, \frac{\pa \xi^i}{\pa x^j} \ - \ \xi^j \, \frac{\pa \s^i_{\ k}}{\pa x^j} \ - \ \tau \, \frac{\pa \s^i_{\ k}}{\pa t} \ - \ f^i \, \s^j_{\ k} \, \frac{\pa \tau}{\pa x^j} \right. \nonumber \\ & & \left. \ - \ \frac12 \, \s^i_{\ k} \, \frac{\pa \tau}{\pa t} \] \ d w^k \label{eq:hhh} \\
&:=& \a^i (x,t) \, d t \ + \ \b^i_{\ k} (x,t) \, d w^k
\label{eq:hhhshort} \ .
 \end{eqnarray}

In other words, the action of the vector field
\eqref{eq:compItoVF} maps the original equation
\eqref{eq:compItoEQ} into the (generally, different) equation \beq
\label{eq:itotrasformata} d x^i \ - \ [ f^i (x,t) \  + \ \eps \,
\a^i (x,t) ] \, d t \ - \ [ \s^i_{\ k} (x,t) \ + \ \eps \, \b^i_{\
k} (x,t) ] \, d w^k \ , \eeq with $\a^i$ and $\b^i_{\ k}$ defined
above.

Obviously, the transformed equation \eqref{eq:itotrasformata} is
the same as the original one \eqref{eq:compItoEQ} if and only if
$\a^i (x,t) = 0$, $\b^i_{\ k} (x,t) = 0$ for all $i$ and $k$.

Thus, in view of \eqref{eq:hhh} and
\eqref{eq:hhhshort}, we have that:

\medskip\noindent
{\bf Proposition \GS1.} {\it The \ind{determining equations} for general
symmetries (with generator of the form \eqref{eq:compItoVF}) of
the Ito equation \eqref{eq:Ito} read
\begin{eqnarray}
\frac{\pa \xi^i}{\pa t} \ + \ f^j \ \frac{\pa \xi^i}{\pa x^j} \ - \
\xi^j \ \frac{\pa f^i}{\pa x^j} \ - \ \frac{\pa (\tau f^i)}{\pa t} \ - \
f^j \ \frac{\pa \tau}{\pa x^j} \ f^i & & \nonumber \\
\ + \ \frac12 \ \( \frac{\pa^2 \xi^i}{\pa x^j \pa x^m} \, + \,
f^i \, \frac{\pa^2 \tau}{\pa x^j \pa x^m} \) \, \s^j_{\ k} \, \s^m_{\ k} &=& 0 \ ; \nonumber \\
\s^j_{\ k} \, \frac{\pa \xi^i}{\pa x^j} \ - \ \xi^j \, \frac{\pa
\s^i_{\ k}}{\pa x^j} \ - \ \tau \, \frac{\pa \s^i_{\ k}}{\pa t} \
- \ f^i \, \s^j_{\ k} \, \frac{\pa \tau}{\pa x^j} \ - \ \frac12 \,
\s^i_{\ k} \, \frac{\pa \tau}{\pa t} &=& 0 . \label{eq:gdei}
\end{eqnarray} }
\bigskip

These equations are general but also rather involved. It turns out
to be more convenient to consider special classes of
transformations, as we will do below. We stress that limitation to
such simpler types of transformations is not only convenient, but
also justified physically (and mathematically, as we argue in a moment, see Remark \remJCZ), as will be discussed below.

\section{Lie-point (fiber-preserving) symmetries}
\label{sec:liepoint}
\def\cs{\ref{sec:liepoint}}

\subsection{Symmetries of the Ito equation}
\label{ssjcz}

As we have seen above, it is possible to write down explicitly the
\ind{determining equations} for general symmetries of the Ito equation,
but these are rather involved. Moreover, on physical grounds one
wants to consider general transformations in space (i.e. in the
$x$ variables), but the transformation in $t$ should not depend on
the point of space we are at \cite{GKiev,GRQ1}; that is we would
like to consider \emphind{fiber-preserving symmetries} (see
Sect. \ref{sec:symmDS}).

\medskip\noindent
{\bf Remark \nc.7.} It should also be mentioned that for general stochastic processes (in particular, those of non-bounded variation) a time change depending on the spatial coordinates would cause the measure of the transformed process to be not absolutely continuous w.r.t. the original one. This is a feature one would certainly not like to allow\footnote{More generally, dealing with space-dependent time maps opens a number of quite delicate problems \cite{Csink,Freedman, ItoMcKean, McKean, Oksendal} -- which we are not willing to discuss in this context. Thus the reader might think of the discussion to be given in Chapter \ref{chap:RS} and considering time maps which depend on $x$ (which suffices to make this a \ind{random time change}) and/or on the realization $w(t)$ of the Wiener process, apart from $t$ itself, as purely formal.}. See also the discussion in Sect.\ref{sec:RTC} in this respect. \EOR

We will thus consider vector fields of the form \beq
\label{eq:VFP} X_0 \ = \ \tau (t) \, \pa_t \ + \ \xi^i (x,t) \,
\pa_i \ , \eeq where of course $\pa_i = \pa / \pa x^i$. We have
then the following result, which can be proved by direct
computation \cite{GRQ1} or specializing the computations and
result of Sect. \ref{sec:itotransf}.\footnote{These are the first
example of a large list of determining equations for stochastic
equations; these will always be in the form of a couple of
(systems of) equations. To avoid any confusion or awkward phrasing
in referring to them, we will use a notation with (a) and (b) for
the two (sets of) equations; thus e.g. eq.(\ref{eq:deteq_Ito}.a)
and (\ref{eq:deteq_Ito}.b).}

\medskip\noindent
{\bf Proposition \GRQa.} {\it The determining equation for
\ind{fiber-preserving symmetries} of the Ito equation \eqref{eq:Ito} are
\begin{eqnarray} & & (\pa_t \xi^i) \ + \ \[ (f^j \cdot \pa_j) \, \xi^i \ - \
(\xi^j \cdot \pa_j ) \, f^i \] \ - \ \pa_t ( \tau f^i) \ + \
\frac12 \( \s \s^T \)^{jk} \, \pa^2_{jk} \, \xi^i \ = \ 0 \nonumber \\
& & (\s_k^{\ j} \cdot \pa_j ) \, \xi^i \ - \ (\xi^j \cdot \pa_j )
\, \s_k^{\ i} \ - \ \tau \, \pa_t \, \s_k^{\ i} \ - \ \frac12
(\pa_t \tau) \, \s_k^i \ = \ 0 \ .  \label{eq:deteq_Ito}
\end{eqnarray} }

\medskip\noindent
{\bf Remark \nc.8.} One is sometimes willing to further restrict
the set of allowed transformation, and consider only
\emphind{simple symmetries}; these have generator \beq
\label{eq:VFPsimple} X_0 \ = \ \xi^i (x,t) \, \pa_i \ . \eeq The
corresponding \emph{\ind{determining equations} for simple
symmetries} of the Ito equation \eqref{eq:Ito} read
\begin{eqnarray} & & (\pa_t \xi^i) \ + \ \[ (f^j \cdot \pa_j) \, \xi^i \ - \
(\xi^j \cdot \pa_j ) \, f^i \] \ + \
\frac12 \( \s \s^T \)^{jk} \, \pa^2_{jk} \, \xi^i \ = \ 0 \ , \nonumber \\
& & (\s_k^{\ j} \cdot \pa_j ) \, \xi^i \ - \ (\xi^j \cdot \pa_j )
\, \s_k^{\ i}  \ = \ 0 \ ; \label{eq:deteqItosimple}
\end{eqnarray}
they are obtained from \eqref{eq:deteq_Ito} just by setting $\tau
= 0$. \EOR

\subsection{Symmetries of the associated diffusion equation}

A step forward in considering symmetry for SDEs (independently
from a variational origin) was done when symmetries of an Ito
equation were associated and compared to symmetries of the
corresponding \emphind{diffusion equation}.

The idea behind this -- in terms of solutions -- is that a sample
path should be mapped into an \emph{equivalent} one, where
equivalence is meant in statistical sense. We thus have two types
of symmetries for the one-particle process described by a SDE: the
equation can be invariant under the map, or it may be mapped into
a \emph{different} equation which has the \emph{same} associated
diffusion equation. In this way one is to a large extent
considering the symmetries of the associated \index{Fokker-Planck
equation} Fokker-Planck (FP) equation, and this had been studied
in detail in the literature
\cite{CicVit1,CicVit2,Finkel,KMA,Koz1,Koz2,Koz3,Rud,SasDun,ShtSto,SpiSto}.

In the same way as for symmetries of the Ito equation, it is
possible to write down explicitly the \ind{determining equations} for
symmetries of the associated FP equation. Here again we consider
\ind{fiber-preserving symmetries} (see Sect. \ref{sec:symmDS}), i.e.
vector fields of the form \eqref{eq:VFP}. When dealing with the FP
equation we will consider vector fields of the form \beq
\label{eq:VFFP} X \ = \ \tau (t) \, \pa_t \ + \ \xi^i (x,t) \,
\pa_i \ + \ \vphi (x,t,u) \, \pa_u \ . \eeq Note that having
$\tau$ and $\xi^i$ independent of $u$ is needed in order to be
able to project vector fields defined in the $(x,t,u)$ space down
to the $(x,t)$ space; this will be required to compare symmetries
of the Ito and of the associated FP equation (and more generally
vector fields of the form $X$ and those of the form $X_0$). We
have the:

\medskip\noindent
{\bf Proposition \GRQab.} {\it Consider the Ito equation
\eqref{eq:Ito}; the \ind{determining equations} for nontrivial
symmetries of the associated Fokker-Planck equation are
\begin{eqnarray}
\pa_t (\tau A^{ik}) & + & \( \xi^m \, \pa_m A^{ik} \, - \, A^{im}
\, \pa_m \xi^k \, - \,
A^{mk} \, \pa_m \xi^i \) \ = \ 0 \nonumber \\
\pa_t (\tau B^i) & - & \[ \pa_t \xi^i + \( B^m \pa_m \xi^i - \xi^m \pa_m B^i \) \] \nonumber \\
 & & \ \ \ + \ \( A^{ik} \pa_k \b + A^{mi} \pa_m \b \) \ - \ A^{mk} \pa^2_{mk} \xi^i \ = \ 0 \nonumber  \\
\pa_t (\tau C) & + & \pa_t \b \ + \ A^{ik} \pa_{ik}^2 \b \ + \ B^i
\pa_i \b \ + \ \xi^m \pa_m C \ = \ 0 \ . \label{eq:deteq_FP}
\end{eqnarray} }
\bigskip

Note that here ``nontrivial symmetries'' should be understood in
the sense of Remark \nc.2 above. Proposition {\GRQab} was
proved in \cite{GRQ1} by direct computation.

\subsection{Symmetry of Ito versus Fokker-Planck equations}

It follows immediately from comparison of \eqref{eq:deteq_Ito} and
\eqref{eq:deteq_FP} that:

\medskip\noindent
{\bf Proposition \GRQb.} {\it Symmetries of an Ito equation, i.e.
solutions to \eqref{eq:deteq_Ito}, can be extended to
(projectable) symmetries of the associated \ind{Fokker-Planck
equation}, i.e. solutions to \eqref{eq:deteq_FP}, while the
converse is not necessarily true.}
\bigskip

The \ind{determining equations} allow, at least in principle, to
find the symmetries of a given Ito equation; moreover, in view of
Proposition {\GRQb} these can be obtained refining the list of
symmetries of the associated FP equation. As the latter is a
standard (deterministic) PDE, its symmetries are determined by
standard techniques.

Let us look at the statement of Proposition {\GRQb} in more detail.
First of all, we should not consider all symmetries of the FP
equation, but only those which:
\begin{itemize} \item[{\it (i)}]
can be projected to the space in which the Ito equation is
defined, and are projectable\footnote{That is, the change in time
will be the same at all points of space. More general settings are
also possible, see e.g. \cite{ArnImk} and Chapter \ref{chap:RS}
below.} in this; and \item[{\it (ii)}] preserve the normalization
condition for the probability measure.
\end{itemize}

The first requirement leads to consider vector fields
\beq\label{eq:prop8} X \ = \ \tau (t) \, \pa_t \ + \ \xi(x,t) \,
\pa_x \ + \ \phi (x,t,u) \, \pa_u \ := \ X_0 \ + \ \phi (x,t,u) \,
\pa_u \ . \eeq It turns out that the second requirement is
satisfied if and only if \beq \phi (x,t,u) \ = \ \a (x,t) \ + \ \b
(x,t) \, u \eeq with moreover \beq \label{eq:alphabeta} \int \a
(x,t) \ d x \ = \ 0 \ \ (\forall t) \ \ ; \ \ \ \ \b (x,t) \ = \ -
\, \mathrm{div} \, [\xi (x,t)] \ . \eeq This specification gives a
constructive meaning to the statement that the symmetry of
\eqref{eq:Ito} can be extended to a symmetry of the associated FP
equation.

On the other hand \cite{GRQ1}, Proposition {\GRQb} has another simple
consequence (actually a corollary):

\medskip\noindent
{\bf Proposition \GRQbcor.} {\it Consider the Ito equation
\eqref{eq:Ito} and the associated FP equation. A vector field in
the form \eqref{eq:prop8} and which is a symmetry of the
associated FP equation is also a symmetry of the Ito equation
under consideration if and only if $\Gamma = 0$, where \beq \Gamma
\ = \
\[ \s_j^{\ m} \, \pa_m \xi^k \ - \ \xi^m \, \pa_m \s_j^{\ k} \ - \
\tau \, \pa_t \s_j^{\ k} \ - \ \frac12 \, (\pa_t \tau) \, \s_j^{\
k} \] \ . \eeq}

\medskip\noindent
{\bf Remark \nc.9} Most recently the ``diffusive'' approach to
symmetries of SDEs has been reconsidered by F. De Vecchi in his
(M.Sc.) thesis \cite{DVthesis}, making contact with so called
``second order geometry'' \index{second order geometry}
developed by Meyer and Schwartz \cite{Emery,Meyer1,Meyer2,Meyer3,Schwartz}.
This introduces some interesting Geometry
of second order differential operators with no constant part. Note
that his approach is entirely through the associated diffusion
(hence deterministic) equation. The interested reader is referred
to his paper \cite{DVforth} (announced in
\cite{DVMU}).\footnote{For application of the \ind{Meyer-Schwartz
approach}, see also \cite{LCO}.} \EOR

\medskip\noindent
{\bf Example \nc.1.} Let us
consider the Ito equation \beq \label{eq:ex1} d x^i \ = \ \s_0 \ d
w \ , \eeq i.e. a free particle moving in one dimension under the
action of a constant noise (here $\s_0$ is a real constant). In
this case $f=0$ and $s = \s_0$. The corresponding FP equation is
just the heat equation
$$ u_t \ = \ (\s_0 / 2) \ u_{xx} \ . $$ As well known, the
infinitesimal symmetries of the latter are generated by
\begin{eqnarray*}
V_1 &=& \pa_t \ , \ \ V_2 = \pa_x \ , \ \  V_3 \ = \ 2 t \pa_t + x \pa_x \ ; \\
V_4 &=& u \pa_u \ ; \ \ V_5 = \s_0^2 t \pa_x - \s_0 x u \pa_u \ ,
\ \
 V_6 = t^2 \pa_t + x t \pa_x - \frac12 \( t + \frac{x^2}{\s_0^2} \) u \pa_u \ ; \\
 V_\a &=& \a(x,t) \pa_u \ . \end{eqnarray*}
Here $\a (x,t)$ is any solution to the heat equation itself, and
the infinite dimensional algebra spanned by the $V_\a$ corresponds
to the fact the equation is linear, see Sect. \ref{sec:Bluman}.

It is easy to check that $V_1,V_2,V_3$ are also symmetries of the
Ito equation \eqref{eq:ex1}; these vector field do not act on $u$.
Note also that for these $\b = - \pa_x \xi$, which in this
one dimensional case means $\b = - \mathrm{div} (\xi)$, see
\eqref{eq:alphabeta}; the condition is not satisfied by
$V_4,V_5,V_6$. One can actually check that $V_1,V_2,V_3$ span the
symmetry algebra of \eqref{eq:ex1}: in this case the
\eqref{eq:deteq_Ito} read
\begin{eqnarray} \xi_t \ + \ \frac12 \, \s_0^2 \, \xi_{xx} &=& 0 \ , \nonumber \\
\xi_x \ - \ \frac12 \, \tau_t &=& 0 \ . \label{eq:ssbm}
\end{eqnarray} The latter implies (as $\tau_x = 0$) that $\xi_{xx}
= 0$, hence $\xi = a(t) x + b(t)$. Now the first of the above
equations \eqref{eq:ssbm} require that $a$ and $b$ are actually
constant, that is $\xi = c_3 x + c_2$. With this,
(\ref{eq:ssbm}.b) enforces in turn $\tau = 2 c_3 t + c_1$ (the
$c_k$ are real constants). We get exactly the $V_1$ (associated to
$c_1$), $V_2$ (associated to $c_2$) and $V_3$ (associated to
$c_3$). \EOE

\medskip\noindent
{\bf Example \nc.2.}
Consider next a two-dimensional example,
\begin{eqnarray} d x &=& y \, dt \nonumber \\
d y &=& - \, k^2 \ y \ d t \ + \ \sqrt{2 k^2} \ d w (t) \ ;
\label{eq:exKram} \end{eqnarray} the associated FP equation is the
Kramers equation \beq u_t \ = \ k^2 \, u_{yy} \ - \ y \, u_x \ + \
k^2 \, y \, u_x \ + \ k^2 \, u \ . \eeq This is linear, hence we
will have symmetries $V_\a$ with $\a(x,t)$ an arbitrary solution.
Apart from these, the infinitesimal symmetry generators of the
Kramers equation are, as determined in \cite{ShtSto},
\begin{eqnarray*}
V_1 &=& \pa_t \ , \ \ V_2 \ = \ \pa_x \ , \ \ V_3 \ = \ e^{- k^2
t} \
\( k^{-2} \pa_x - \pa_y \) \ ; \\
V_4 &=& u \, \pa_u \ , \ \ V_5 \ = \ t \, \pa_x \ + \ \pa_y \ - \
\frac12 \, \( y + k^2 x \) \ u \, \pa_u \ , \\
V_6 &=& e^{k^2 t} \ \( k^{-2} \pa_x + \pa_y - y u \pa_u \) \ .
\end{eqnarray*} Note that $V_1,V_2,V_3$ (which do not act on $u$,
i.e. are of the form $X_0$) satisfy $\b = - \mathrm{div} (\xi)$,
while $V_4,V_5,V_6$ do violate this condition. One can check
\cite{GRQ1} that $V_1,V_2,V_3$ span the full symmetry algebra for
\eqref{eq:exKram}. \EOE

\medskip\noindent
{\bf Example \nc.3.} In the
two examples above there was full correspondence between
admissible symmetries of the FP equation and symmetries of the Ito
equation. We now consider an example (again two-dimensional, and
again taken from \cite{GRQ1}) in which this is not the case. This
is the equation
\begin{eqnarray}
d x^1 & = & \cos (t) \, d w^1 (t) \ - \ \sin (t) \, d w^2 (t) \nonumber \\
d x^2 & = & \sin (t) \, d w^1 (t) \ + \ \cos (t) \, d w^2 (t) \ ;
\label{eq:ex3} \end{eqnarray} this is also written in vector
notation as \beq \label{eq:ex3b}  d {\bf x} \ = \ R(t) \ d {\bf w}
(t) \ , \eeq where $R(t) = \s (t)$ is the matrix of rotation by an
angle $t$.

The associated FP equation is just the two-dimensional heat
equation, \beq \label{eq:ex3FP}  u_t \ = \ \frac12 \ u_{xx} \ .
\eeq It is immediate to check that $\pa_t$ is (of course) a
symmetry of \eqref{eq:ex3FP} but it does \emph{not} satisfy (the
second of) the \ind{determining equations} \eqref{eq:deteq_Ito} and
hence is \emph{not} a symmetry of \eqref{eq:ex3}.

Note that we would have exactly the same situation for any
orthogonal $\s$ matrix not constant in time. \EOE

\medskip\noindent
{\bf Example \nc.4.} Let us
consider the case of Ito equations with \emph{linear} drift \cite{Alexandrova}, i.e.
\beq d x^i \ = \ M^i_{\ j} \, x^j \, d t \ + \ \s^i_{\ j} \, d w^j
\ ; \eeq we will further simplify the situation by considering the
case of a constant and invertible $\s$ matrix. This system
represents an \ind{Ornstein-Uhlenbeck process} \cite{Nelson,Risken} and,
writing $A = - (1/2) \s \s^T$, the associated FP equation is just
\beq u_t \ + \ M^k_{\ k} \ + \ M^i_{\ k} \, x^k \, u_i \ - \
A^{ij} \, u_{ij} \ = \ 0 \ . \eeq In this case it results
$$ \xi^i (x,t) \ = \ L^i_{\ j} (t) \ x^j \ + \ P^i (t) \ ; $$
the matrix $L$ satisfies $L = (1/2) \tau_t I$.

For $M \not= I$ one gets $\tau_t = 0$, hence $L=0$, and $P(t) =
\exp [M t] P_0$. In the special case $M=I$, one gets $\tau = c_1
e^t + c_2$, hence $L = (c_1/2) e^t I$,  and $P(t) = e^t P_0$.

As for $\b$ in the symmetry of the associated FP equation, we note
that $\mathrm{div} (\xi ) = \mathrm{Tr} (L)$, thus $\b = 0$ for $M
\not= I$, and $\b = (n/2) \tau_t $ for $M=I$. \EOE

\section{W-symmetries}
\label{sec:wsymm}
\def\cs{\ref{sec:wsymm}}

The theory can be extended to consider also transformations acting
on the Wiener processes, also called \emphind{W-symmetries}.
More specifically, we will consider (infinitesimal) maps
\begin{eqnarray}
t & \to & s \ = \ t \ + \ \eps \, \tau (t) \nonumber \\
x^i & \to & y^i \ = \ x^i \ + \ \eps \, \xi^i (x,t) \label{eq:Wmaps} \\
w^k & \to & z^k \ = \ w^k \ + \ \eps \, \mu^k (w,t) \ . \nonumber
\end{eqnarray}
Note that here as well, as in Sect. \ref{sec:liepoint}, we are
just considering \emph{fiber-preserving}
transformations.\footnote{More general ones -- with the cautionary
notes already seen in Remark \remJCZ -- will be considered in
Chapter \ref{chap:RS}.}

The restrictions on $\tau$ and $\xi$ in \eqref{eq:Wmaps} are the
same as before (see Sect. \ref{sec:liepoint}) and do not need
further comments; as for the transformation undergone by the
Wiener processes, i.e. the function $\mu = \mu (w,t)$, we have
allowed this to depend on time and on the processes themselves,
but not on the space variables $x^i$. The rationale for this is
that \emph{we think of the (time-dependent) stochastic processes
$w^i (t)$ as \emph{independent} of the position $x(t)$ reached by
the test particle}, and we would like this property to be still
valid for the transformed processes $z(t)$ (see also Remark
\remJCZ  above).

It should be stressed that $\mu (w,t)$ in \eqref{eq:Wmaps} can not
be an arbitrary function. In fact, if we want the transformed
processes $z^i (t)$ to be still Wiener ones\footnote{Some works in
the literature (about symmetry of SDEs) do only require that the
transformed processes have the same mean and variance of the
original ones, with no check on the fate of higher moments. I am
not understanding the idea behind this choice (which was indeed
questioned by other authors \cite{Unal}), and will not comment on
these.}, it turns out \cite{GRQ2} that the only possibility is to
have \beq z^i (t) \ = \ M^i_{\ j} \ w^j (t) \eeq with $M$ a
\emph{constant} orthogonal matrix, $M M^+ = I$. In terms of
infinitesimal generators, we get \beq \label{eq:muWmaps} \mu^i
(w,t) \ = \ B^i_{\ k} (t) \ w^k \ \ \ \ \ (B^T = - B) \ . \eeq

Having established the class of allowed transformations, one can
compute their effect on the Ito equation \eqref{eq:Ito}. This is
summarized in the following statements \cite{GRQ2}, the second
(Proposition {\GAEb}) being an immediate corollary of the first
(Proposition {\GAEa}).

\medskip\noindent
{\bf Proposition \GAEa.} {\it Under the map \eqref{eq:Wmaps}, with
$\mu$ of the form \eqref{eq:muWmaps}, the Ito equation
\eqref{eq:Ito} is mapped into a new Ito equation
$$ d y^i \ = \ F^i (y,s) \, d s \ + \ S^i_{\ k} (y,s) \ d z^k  $$
with $F^i = f^i + \eps (\de f)^i$, $S^i_{\ k} =  \s^i_{\ k} + \eps
 (\de \s)^i_{\ k}$. The first variations are given
explicitly by
\begin{eqnarray}
(\de f)^i &=& \pa_t \xi^i \ + \ \[ (f^j \pa_j ) \xi^i - (\xi^j
\pa_j) f^i \] \ - \
\pa_t (\tau \, f^i) \ + \ A^{jk} \, \pa^2_{jk} \, \xi^i \ , \label{eq:deteq_W0} \\
(\de \s)^i_{\ k} &=& \[ (\s^j_{\ k} \pa_j ) \xi^i - (\xi^j \pa_j )
\s^i_{\ k} \] \ - \ \tau \, \pa_t \s^i_{\ k} \ - \ \frac12 \,
(\pa_t \tau) \, \s^i_{\ k} \ - \ \s^i_{\ p} \, B^p_{\ k} \ .
\nonumber
\end{eqnarray}}

\medskip\noindent
{\bf Proposition \GAEb.} {\it The map
\eqref{eq:Wmaps}, with $\mu$ as in \eqref{eq:muWmaps}, is a
symmetry of the Ito equation \eqref{eq:Ito} if and only if the
quantities defined in \eqref{eq:deteq_W0} satisfy $(\de f)^i = 0$,
$(\de \s)^i_{\ k} \ = \ 0$ for all $i$ and $k$. In other words,
the \emph{\ind{determining equations} for W-symmetries} of
\eqref{eq:Ito} are
\begin{eqnarray}
\pa_t \xi^i \ + \ \[ (f^j \pa_j ) \xi^i - (\xi^j \pa_j) f^i \] \ -
\
\pa_t (\tau \, f^i) \ + \ A^{jk} \, \pa^2_{jk} \, \xi^i &=& 0 \ , \label{eq:deteqWsymm} \\
\[ (\s^j_{\ k} \pa_j ) \xi^i - (\xi^j \pa_j ) \s^i_{\ k} \] \ - \
\tau \, \pa_t \s^i_{\ k} \ - \ \frac12 \, (\pa_t \tau) \, \s^i_{\
k} \ - \ \s^i_{\ p} \, B^p_{\ k} &=& 0 \ . \nonumber
\end{eqnarray}}

\medskip\noindent
{\bf Remark \nc.10.} Note that the possibility of action on $W$ can
\emph{not} be used to balance the change in $W$ induced by
transformation of $t$, as the latter is \emph{not} an orthogonal
action. \EOR
\bigskip

\medskip\noindent
{\bf Remark \nc.11.} Finally we note that this is definitely
\emph{not} the most general allowable transformation. For example,
in their work on \ind{normal forms}\footnote{As well known, normal forms
- and transformation to normal forms -- is intimately connected
with symmetries \cite{ArnGMDE,CG,Wal331}.} for \ind{stochastic
dynamical systems} \cite{ArnImk}, L. Arnold and P. Imkeller
\index{Arnold-Imkeller approach} considered a more general class. As it was remarked already in
\cite{GRQ2,GRQ1}, albeit the restriction to \eqref{eq:Wmaps} is
well justified physically, from a mathematical point of view (i.e.
to obey just the internal coherence requirements, disregarding the
applications physicists are primarily interested in) one should
extend the theory by including the class of transformation
considered there. We will come back to this point later on in
Chapter \ref{chap:RS}. \EOR
\bigskip

We will now consider some examples. In order to better compare
W-symmetries with standard ones, we will consider some situations
already discussed above in the Examples of Sect.
\ref{sec:liepoint}.

\medskip\noindent
{\bf Example \nc.5.} We start
by considering again Example \nc.2. In this case the determining
equations \eqref{eq:deteqWsymm} require $B=0$, i.e. we have no new
vector fields allowing W-symmetries beside standard ones. \EOE

\medskip\noindent
{\bf Example \nc.6.} Let us
consider the equations
\begin{eqnarray*}
d x &=& (a_1 / x) \, d t \ + \ d w_1 (t) \\
d y &=& a_2 \, d t \ + \ d w_2 (t) \ . \end{eqnarray*} This has
four standard symmetry generators, i.e.
$$ X_1 = \pa_t \ , \ \ X_2 = \pa_x \ , \ \ X_3 = \pa_y \ ; \ \
X_4 \ = \ 2 t \pa_t + x \pa_x + (y + a_2 t) \pa_y \ , $$ and a
proper W-symmetry, generated by $$ X_5 \ = \ (a_2 t - y) \,
\frac{\pa}{\pa x} \ + \ x \, \frac{\pa}{\pa y} \ + \ w_2 \,
\frac{\pa}{\pa w_1} \ - \ w_1 \, \frac{\pa}{\pa w_2} \ . $$
{} \EOE

\medskip\noindent
{\bf Example \nc.7.}
We consider again the equation \eqref{eq:ex3}, i.e. Example \nc.3. In
this case, beside the standard symmetries already obtained in
Sect. \ref{sec:liepoint}, we also have a W-symmetry,
$$ X \ = \ \( y \, \frac{\pa}{\pa x} \ - \ x \, \frac{\pa}{\pa y} \) \ + \
\( w_2 \, \frac{\pa}{\pa w_1} \ - \ w_1 \, \frac{\pa}{\pa w_2} \)
\ . $$ This represents a simultaneous identical rotation in the $(x,y)$ and
the $(w_1,w_2)$ planes. \EOE

\medskip\noindent
{\bf Example \nc.8.} Let us
consider the setting of Example \nc.4. With the notation
introduced above in this Section, the symmetries satisfy \beq
(\tau_t) \ \wt{M} \ + \ [B,\wt{M} ] \ = \ (1/2) \, \tau_{tt} \ I \
, \eeq where we have written $\wt{M} := \s^{-1} M \s$. In
particular, setting $\tau = 0$ we get new W-symmetries
corresponding to matrices $B$ which commute with $\wt{M}$ (e.g. $
B = c \wt{M}$). \EOE

\medskip\noindent
{\bf Example \nc.9.} Finally
let us consider \beq d x^i \ = \ - \, (1 - \la |x|^2 ) \, x^i \, d
t \ + \ d w^i \ , \eeq with $\la \not= 0$. It is easily seen that
the only standard symmetry generator is $V_0 = \pa_t$; allowing
W-symmetries we also have simultaneous rotations in the $x$ and
the $w$ spaces. \EOE

\section{Symmetries of random dynamical systems}
\label{sec:rds}
\def\cs{\ref{sec:rds}}

Up to now we have considered symmetries of an Ito equation
\eqref{eq:Ito} describing a \emphind{one-particle process}.
However in many situations, in particular in Physics, we are
interested in many-particles processes; we also refer to these as
\emphind{random dynamical systems}. One would also like to study
symmetries of these, and in particular of those described by the
same Ito equation, i.e. by an ensemble of non-interacting
identical particles (with different initial positions) undergoing
the stochastic process described by \eqref{eq:Ito}.\footnote{It
should be mentioned that for processes with independent
increments, the two-particle process embodies the full information
needed to determine the $N$ particle process, see \cite{LArnold}
(sect.2.3.9).}

It should be stressed that the resulting equations will be
\emph{covariant} under any \ind{permutation} of the particles, as
follows from considering identical ones.

Not surprisingly, it turns out that -- for a given Ito equation --
any symmetry of the one particle process is also a symmetry of the associated random dynamical system, while the converse is not
true\footnote{Here we are referring to ``symmetries'' in the sense
of infinitesimal generators of continuous symmetries; the
statement is (trivially) true also for the permutation symmetry
mentioned above.}. This can be seen as a corollary to the
following Proposition {\GAEc}; in order to state this, it is
convenient to set some {\it ad hoc} notation.

We will consider $N$ copies ($a = 1,...,N$) of the Ito equation in
$R^n$ (here $i=1,...,n$), \beq \label{eq:Itoy} d y_a^i \ = \
\phi^i (y_a,t) \, d t \ + \ \rho^i_{\ k} (y_a,t) \, d w^k (t) \ ;
\eeq these can be written as a single Ito equation in the form
\eqref{eq:Ito} by setting (in block notation) \beq \label{eq:ItoN}
f \ = \ \pmatrix{ \phi_{(1)} \cr \vdots \cr \phi_{(N)} \cr} \ , \
\ \s \ = \ \pmatrix{ \rho_{(1)} & \ldots & 0 \cr \vdots & \ddots &
0 \cr \rho_{(N)} & \ldots & 0 \cr } \ , \eeq where we have set,
for ease of notation, \beq \phi_{(k)} \ = \ \phi (y_k , t) \ , \ \
\rho_{(k)} = \rho (y_k , t) \ . \eeq We stress that the resulting
$N$ particle equation, which is a Ito equation in dimension $d = N
\, n $, depends only on $n$ Wiener processes; this distinguishes
the situation from the general one, i.e. (in the presently used
language) that of a $d$-dimensional one-particle process.

Note that with the notation \eqref{eq:ItoN} we have
$$ \frac12 \ \s \, \s^T \ = \
\frac12 \ \pmatrix{ \rho_{(1)} \rho_{(1)}^T & \ldots & \rho_{(1)}
\rho_{(N)}^T \cr \vdots & \ddots & \vdots \cr \rho_{(N)}
\rho_{(1)}^T & \ldots & \rho_{(N)} \rho_{(N)}^T \cr} \ := \
\pmatrix{ A (y_1,y_1) & \ldots & A(y_1,y_N) \cr \vdots & \ddots &
\vdots \cr A(y_N,y_1) & \ldots & A(y_N,y_N) \cr} \ . $$

We will set a similar notation for the (symmetry) vector fields.
That is, we will set in general, and again in block notation,
$$ \xi (t; y_1 , ... , y_N) \ = \ \pmatrix{\xi_{(1)} (t;y_1,...,y_N ) \cr
\ldots \cr \xi_{(N)} (t;y_1,...,y_N ) \cr } \ . $$ As recalled
above, the equations should be covariant under $S_N$, the group of
permutation of the $N$ identical particles. This means that
$\xi_{(1)} (t; y_1,...y_N)$ should be invariant under the
permutations which do not affect $y_1$ (thus under a subgroup
$S_{N-1}$), and that one has (considering for simplicity just the
cyclic permutations group $Z_N \ss S_N$)
$$ \xi_{(k)} (t;y_1 , ... , y_N ) \ = \ \xi_{(1)} (t;y_k , y_{k+1} ,..., y_{k-1} ) \ . $$

Finally, we will write for short
$$ \pa_i^{(a)} \ = \ \frac{\pa}{\pa y^i_{(a)} } \ , \ \
\triangle_{(a,b)} \ = \ \sum_{i} \frac{\pa^2}{\pa y^i_{(a)} \, \pa
y^i_{(b)}} \ . $$ With these notations, we have the

\medskip\noindent
{\bf Proposition \GAEc.} {\it The symmetry generators for the $N$
particle process defined by the Ito equation \eqref{eq:Itoy}
satisfy the \ind{determining equations} (no sum on $a$)
\begin{eqnarray}
& & \pa_t \xi^i_{(a)} \ - \ \pa_t \, (\tau \, \phi^i_{(a)} ) \ + \
\[ (\phi_{(a)}^j \pa^{(a)}_j ) \xi_{(a)}^i - (\xi_{(a)}^j \pa^{(a)}_j ) \phi_{(a)}^i \] \nonumber
\\
& & \ \ \ \ + \ A (y_{(a)}, y_{(a)} ) \triangle_{(a,a)}
\xi_{(a)}^i \ + \ \sum_{b \not= a} (\phi_{(b)}^j \pa_j^{(b)} ) \,
\xi^i_{(a)} \nonumber \\
& & \ \ \ \ + \
\sum_{(b,c) \not= (a,a)} A (y_{(b)}, y_{(c)} ) \triangle_{(b,c)} \, \xi^i_{(a)} \ = \ 0 \ , \label{eq:deteqwsymm1} \\
& & \[ (\rho_{(a)})_k^j \, \pa_j^{(a)} \xi_{(a)}^i \ - \
\xi_{(a)}^j \, \pa_j (\rho_{(a)})_k^i \] \ - \ \tau \ (\pa_t
(\rho_{(a)})_k^i ) \ - \ \frac12 \, (\pa_t \tau) \,
(\rho_{(a)})_k^i \nonumber \\
& & \ - \ (\rho_{(a)})_k^i \, M  \ + \ \sum_{b \not= a} \(
(\rho_{(b)})_k^j \pa_j^{(b)} \xi_{(a)}^i \) \ = \ 0 \ .
\label{eq:deteqwsymm2}
\end{eqnarray} }

\medskip\noindent
{\bf Remark \nc.12.} The approach -- and most of the notation --
developed for $N$ non-interacting particles can to a large extent
be applied to the case of interacting particle as well. Albeit
these can be treated (as recalled above) within the frame of a
general stochastic process -- and Ito equation -- in a suitably
large space, one would expect that specializing to the case of $N$
identical (interacting) particles would give some results which
are definitely non-generic\footnote{Such an extension of this
formalism, to the best of my knowledge, has not been developed;
here again one would have an interesting project.}. Similar
extensions would be possible considering sets of different
families of particles; in this case the arguments based on the
covariance under the full $S_N$ permutation group should be
accordingly modified. \EOR

\medskip\noindent
{\bf Example \nc.10.} We start by considering, as in Example \nc.1 above,
the equation \beq dx \ = \ d w \ ; \eeq that is, we have $\phi =
0$, $\rho = 1$. As we are in one dimension, $B=0$. We will
consider the two-particle process associated to this, and write
$y_{(1)} = x$, $y_{(2)} = y$ and $\xi_{(1)} (x,y,t) = \xi_{(2)}
(y,x,t) = \xi (x,y,t)$. The determining equations
\eqref{eq:deteqwsymm1}, \eqref{eq:deteqwsymm2} are now simply
\begin{eqnarray*}
& & 2 \, \eta_t \ + \ \( \xi_{xx} + 2 \xi_{xy} + \xi_{yy} \) \
= \ 0 \\
& & \tau_t \ = \ 2 \, \( \xi_x + \xi_y \) \ . \end{eqnarray*} We
observe that
$$ \xi (x,y,t) \ = \ f (x - y) \ , \ \ \ \tau = 0 $$ is a
solution to these, for any smooth function  $f$. These are
obviously not symmetries (nor meaningful) for the one-particle
process defined by the same Ito equation. \EOE

\medskip\noindent
{\bf Example \nc.11.} Let us now consider again the Example \nc.2, which
we now write as $$ \begin{array}{l} d
x \ = \ y \, d t \\
d y \ = \ - \, k^2 \, d t \ + \ \sqrt{2 k^2} \, d w \ ;
\end{array} $$ we will consider the two-particle process defined by
this system, and write $y_{(k)} = (x_k,y_k)$. In this case we have
$$ \phi \ = \ \pmatrix{x_2 \cr - k^2 x_2 \cr} \ , \ \ \rho \ = \
\pmatrix{0 & 0 \cr 0 & \sqrt{2 k^2} \cr} \ , $$ and will set
$$ \eta \ = \ \pmatrix{\xi_{(1)} \cr \xi_{(2)} \cr } \ . $$
Writing
$$ \zeta \ := \ k^2 \, (x_1 - y_1) \ + \ (x_2 - y_2) \ , $$
we get solutions for any arbitrary smooth function of $\zeta$, in
the form
$$ \eta \ = \ h (\zeta) \ \pmatrix{ x_1 - y_1 \cr 1 \cr} \ . $$
{} \EOE

\medskip\noindent
{\bf Example \nc.12.} Finally we consider (similarly to Examples \nc.4
and \nc.8 above) the Ito equation with constant linear drift
$$ d x^i \ = \ M^i_{\ j} \, x^j \, d t \ + \ \sqrt{2 s^2} \, d w^i
\ . $$ As in previous Examples, we will consider the two-particle
process -- and use the simplified notation set above -- and just
look for solutions with $\tau = 0$, $\xi_{(1)} = \xi_{(2)} = \xi$
and $\pa_t \xi = 0$.

It turns out that a special (and simple) class of solutions is
provided by
$$ \xi^i \ = \ L^i_{\ k} \, (x^k - y^k)  \ + \ B^i_{\ k}
\, x^k $$ for $L$ and $B$ matrices commuting with $M$, i.e.
for $[L,M] = 0 = [B,M]$. \EOE

\chapter{Stratonovich stochastic equations} \label{chap:Strato}
\def\nc{\ref{chap:Strato}}

It is well known that, albeit the proper formulation of Stochastic
Differential Equations corresponds to the Ito formalism, in many
applications it is also convenient to consider the so called
Stratonovich formalism \cite{LArnold,Guerra,Oksendal,Protter,Strato,Stroock}.

The main advantage of this is that a Stratonovich SDE behaves
nicely, i.e. in the ``usual'' way, under coordinate
transformations. As symmetry refers to invariance under
transformations, it is not surprising at all that the first
attempts to consider symmetry in the framework of SDEs
\cite{AlbFei,Mis2,Mis3} focused on Stratonovich equations, and
that the theory of symmetry of Stratonovich SDEs is neatly
formulated -- as we will see below.

We will now analyze symmetries of Stratonovich SDEs. Before doing
this, an important remark is necessary: one would be tempted to
guess that the symmetries of a Stratonovich and of the
``equivalent'' Ito equations should coincide. We will see that in
general\footnote{There are classes of SDEs, e.g. linear ones, for
which there is indeed equivalence. Similarly this holds for
certain classes of symmetries.} this is not the case, even in
quite simple examples; this was already remarked by Unal
\cite{Unal}, and will be discussed in some detail below, see Sect.
\ref{sec:itovsstrat}.

The reason for this non correspondence appears to be rooted in the
non-trivial character of the ``equivalence'' between Ito and
associated Stratonovich equations; this is discussed e.g. in
\cite{Stroock} (see Chapter 8 in there).

\section{Transformation of a Stratonovich equation}
\label{sec:trastrat}

\def\eps{\varepsilon}

Similarly to what we did in Sect. \ref{sec:itotransf} for Ito
equations, we will first derive the explicit expression for the
transformation of a SDE in Stratonovich form, which we write
again (for ease of reference) as \beq\label{eq:mis} d x^i \ =
\ b^i (x,t) \, d t \ + \ \s^i_{\ k} (x,t) \circ d w^k (t) \ , \eeq
under the action of a general vector field \beq \label{eq:A1} X \
= \ \tau (x,t) \, \pa_t \ + \ \vphi^i (x,t) \, \pa_i \ ; \eeq this
induces the map \beq\label{eq:A2} x^i \ \to \ x^i \ + \ \eps \,
\vphi^i (x,t) \ ; \ \ t \ \to \ t \ + \ \eps \, \tau (x,t) \ .
\eeq

We should first of all recall that, as already mentioned in Remark
\ref{chap:Ito}.4, the action on $t$ induces an action on the
increments $d w^k$ of the Wiener processes $w^k (t)$; more
precisely, we have \cite{Oksendal} \beq\label{eq:A3B} d w^k \ \to
\ d w^k \ + \ \eps \, \frac12 \ \frac{d \tau}{d t} \, d w^k \
. \eeq

Thus under \eqref{eq:A2} the equation \eqref{eq:mis} is mapped
into
\begin{eqnarray}
d (x^i + \eps \vphi^i)  &=& b^i (x + \eps \vphi, t + \eps \tau) \, d (t + \eps \tau) \nonumber \\
& & \ + \ \s^i_{\ k} (x + \eps \vphi, t + \eps \tau) \circ d w^k
(t + \eps \tau) \ . \label{eq:A5} \end{eqnarray} (Here it is
understood that $\vphi = \vphi (x,t)$, $\tau = \tau (x,t)$.)

Performing all computations at order $\eps$, and noticing that on
the flow of \eqref{eq:mis} we have
\begin{eqnarray}
d \vphi^i &=& \frac{\pa \vphi^i}{d x^j} \, d x^j \ + \ \frac{\pa \vphi^i}{\pa t} \, d t \nonumber \\
&=& \( \frac{\pa \vphi^i}{d x^j} \) \, \( b^j (x,t) \, d t \ + \ \s^j_{\ k} (x,t) \circ d w^k (t) \) \ + \ \( \frac{\pa \vphi^i}{\pa t} \) \, d t \ ; \label{eq:A6a} \\
d \tau &=& \frac{\pa \tau}{d x^j} \, d x^j \ + \ \frac{\pa \tau}{\pa t} \, d t \nonumber \\
&=& \( \frac{\pa \tau}{d x^j} \) \, \( b^j (x,t) \, d t \ + \
\s^j_{\ k} (x,t) \circ d w^k (t) \) \ + \ \( \frac{\pa \tau}{\pa
t} \) \, d t \label{eq:A6b} \ , \end{eqnarray} eq.\eqref{eq:A5}
reads
\begin{eqnarray}
d x^i &=& b^i (x,t) \, d t \ + \ \s^i_{\ k} (x,t) \circ d w^k \nonumber \\
& & \ - \ \eps \ \[ \( (\pa_t \vphi^i) \, - \, \pa_t (\tau \, b^i) \) \ + \ \( b^j \, (\pa_j \vphi^i) \, - \, \vphi^j \, (\pa_j b^i ) \) \ - \ b^j \, (\pa_j \tau) \, b^i \] \ d t \nonumber \\
& & \ + \ \eps \ \[ \tau \, (\pa_t \s^i_{\ k} ) \ + \ (1/2) (\pa_t \tau) \, \s^i_{\ k} \ - \ \( \s^j_{\ k} \, (\pa_j \vphi^i) \, - \, \vphi^j \, (\pa_j \s^i_{\ k} ) \) \right. \nonumber \\
& & \left. \ \ + \ b^i \, (\pa_j \tau) \, \s^j_{\ k} \] \circ d
w^k (t) \ . \label{eq:A7} \end{eqnarray}

This coincides with \eqref{eq:mis} if and only if both terms in
square brackets vanish; that is, we have \cite{GS2016}:

\medskip\noindent
{\bf Proposition \GSstrat.} {\it The \ind{determining equations}
for deterministic symmetries of the Stratonovich equation
\eqref{eq:mis} are
\begin{eqnarray}
 \( (\pa_t \vphi^i) \, - \, \pa_t (\tau \, b^i) \) \ + \ \( b^j \, (\pa_j \vphi^i) \, - \,
 \vphi^j \, (\pa_j b^i ) \) \ - \ b^j \, (\pa_j \tau) \, b^i  &=& 0 \ , \nonumber \\
 \( \tau \, (\pa_t \s^i_{\ k} ) \ + \ (1/2) (\pa_t \tau) \, \s^i_{\ k} \)
 \ - \ \( \s^j_{\ k} \, (\pa_j \vphi^i) \, - \, \vphi^j \, (\pa_j \s^i_{\ k} ) \) \nonumber \\
 \ + \ b^i \, (\pa_j \tau) \, \s^j_{\ k} &=& 0 \ . \label{eq:A8} \end{eqnarray}}

\medskip\noindent
{\bf Remark \nc.1.} The equations reported in \cite{GS2016} (and
obtained in the same way) are apparently different; but one should
note that in there, based on physical motivation\footnote{Recall also Remark \remJCZ in this respect.}, only the case $\tau = \tau (t)$ was considered. The equations do of course coincide in the case $\tau = \tau (t)$, to be discussed in detail below. \EOR

\section{Strong symmetries}
\label{sec:Misawa}
\def\cs{\ref{sec:Misawa}}

As mentioned above, the first attempts to use symmetry in the
context of SDEs \cite{AlbFei,Mis2,Mis3} involved Stratonovich
equations, and had quite strong requirements for a map to be
considered a symmetry of the SDE. They were based on the idea of a
symmetry as a map taking solutions into solutions; the problem is
that while in the deterministic case it is quite clear what is
meant by ``solution'' (one just has to distinguish between general
and special solutions), and hence by ``mapping a solution to a
solution'' or by ``invariant solution'', in the stochastic case
this can be interpreted in several ways.

Thus, the first approach by Misawa \cite{Mis2,Mis3,Mis4}, then
extended and generalized by Albeverio and Fei \cite{AlbFei} (see
also \cite{MeiXiang,Mis5}), required that for \emph{any given realization}
of the Wiener process \emph{any sample path} satisfying the
equation would be mapped to another such sample path. It is not
surprising that the presence of symmetries was then basically
related to situations where, in suitable coordinates, the
evolution of some of the coordinates is deterministic and not
stochastic. However, uncovering this fact in other (i.e.
non-adapted) coordinates may be not simple; thus this works gave
nontrivial results, and in particular showed that one can have
symmetries -- and, under certain additional conditions, related
conserved quantities -- also in the case of SDEs, pretty much as
in Hojman's work \cite{Hoj} mentioned above, see
Sect. \ref{sec:Hojman}.

It should be stressed that in this case one has quantities which
are \emph{always} conserved under the stochastic flow. This means
that the level manifolds of the conserved quantities are
\emph{always} invariant, i.e. that both the drift and the
stochastic term are zero in directions transversal to these
manifolds. This is by all means a rather strong requirement.

On the other hand, the requirement can be met in practice, and
when this is the case it is surely relevant to be able to detect
the associated conservation laws (note that once again this can be
better seen by passing to adapted coordinates).

\subsection{Time-preserving strong symmetries}

A simple computation -- in practice, a specialization of the
general one presented in Sect. \ref{sec:trastrat} to the case
$\tau = 0$ -- taking advantage of the favorable properties of
stochastic differential equations in Stratonovich form (i.e. the
fact we can just use the chain rule), shows that under the map
\beq\label{eq:cvmis} x^i \ \mapsto \ x^i \ + \ \eps \, \vphi^i
(x,t) \eeq the equation \eqref{eq:mis} is mapped into
\begin{eqnarray} d x^i &=& b^i (x,t)
\, d t \ + \ \s^i_{\ k} (x,t) \circ d w^k (t) \nonumber \\
& & \ + \ \eps \
\[ \( \frac{\pa \vphi^i}{\pa t} \ + \ b^j \frac{\pa \phi^i}{\pa
x^j} \ - \ \vphi^j \frac{\pa b^i}{\pa x^j} \) \ dt \ + \ \right.
\nonumber \\
& & \ \ \left. \ + \ \( \s^j_{\ k} \frac{\pa \vphi^i}{\pa x^j} \ -
\ \vphi^j \frac{\pa \s^i_{\ k}}{\pa x^j} \) \circ d w^k (t) \] \ ;
\end{eqnarray} where arguments are not indicated, it is understood
that $b,\s,\vphi$ should be thought of as functions of $x$ and
$t$. Note that \eqref{eq:cvmis} can be thought as the action of a
vector field \beq \label{eq:missymm} X \ = \ \vphi^i (x,t) \,
\pa_i \ . \eeq

We say that the vector field $X$ is a \emphind{strong
symmetry}\footnote{Needless to say, one should not make confusion
with the notion of strong symmetry in the deterministic context
considered in Sect. \ref{sec:determining}.} for \eqref{eq:mis} if
it leaves the equation invariant. By looking at the discussion
above, we immediately have

\medskip\noindent
{\bf Proposition \MISa.} {\it The vector field $X$ defined in
\eqref{eq:missymm} is a (simple) strong symmetry for
\eqref{eq:mis} if it satisfies
\begin{eqnarray}
\pa_t \vphi^i \ + \ \( b^j \, \pa_j \vphi^i \ - \
\vphi^j \, \pa_j b^i \) &=& 0 \ , \nonumber \\
\s^j_{\ k} \, \pa_j \vphi^i \ - \ \vphi^j \, \pa_j \s^i_{\ k} &=&
0 \ . \label{eq:propmis} \end{eqnarray}}

\medskip\noindent
{\bf Remark \nc.2.} The equations \eqref{eq:propmis} have been
first determined by Misawa \cite{Mis2,Mis3}, and hence are also
known as \emphind{Misawa equations}. \EOR

\medskip\noindent
{\bf Remark \nc.3.} Given a Stratonovich equation \eqref{eq:mis}, one associates to
this $n+1$ vector fields (also called \emphind{Misawa vector fields}): \beq\label{eq:SVF} X_0 \ = \ \pa_t \ + \
b^i (x,t) \, \pa_i  \ ; \ \ X_k \ = \ \s^i_{\ k} (x,t) \, \pa_i \
. \eeq Note these are associated with the deterministic ($X_0$)
and the random part (in the Stratonovich decomposition) of
\eqref{eq:mis}; passing to the Ito formalism -- as mentioned in
Remark \ref{chap:Ito}.5 -- the two would mix.\footnote{It should
be stressed that the Ito formalism is well adapted to the
stochastic framework in that it separates the drift and the
martingale in the stochastic process $x(t)$; the Stratonovich
formalism is convenient from other points of view, but one should
remember that $b (x,t)$ is \emph{not} the drift.} \EOR

\medskip\noindent
{\bf Remark \nc.4.} With the notation introduced above (see
Sect. \ref{sec:symmDS}) the condition stated in Proposition {\MISa} can
also be written as
\begin{eqnarray}
\pa_t \vphi^i \ + \ \{ b , \vphi \}^i  &=& 0 \ , \nonumber \\
\{ \s , \vphi \}^i_{\ k} &=& 0 \ .
\end{eqnarray}
By looking at our discussion in the deterministic case -- and at
\eqref{eq:SVF} -- it is easily seen that, equivalently, $X$ is  a
strong symmetry if and only if \beq \label{eq:misscc} [X,X_0] \ =
\ 0 \ = \ [X,X_k] \ , \eeq i.e. if and only if it commutes with
the Misawa vector fields associated to equation \eqref{eq:mis}.
\EOR

\medskip\noindent
{\bf Example \nc.1.} Consider the three-dimensional
case\footnote{This example was suggested by Misawa in his original
paper \cite{Mis3}, and it thus became a standard example for
checking subsequent results. As the reader will note, we are not
infringing this tradition, and will be repeatedly considering it.}
(depending on a single Wiener process)
\begin{eqnarray*}
d x &=& - (y - z) \, d t \ + \ (z - y) \circ d w \\
d y &=& - (z - x) \, d t \ + \ (x - z) \circ d w \\
d z &=& - (x - y) \, d t \ + \ (y - x) \circ d w \end{eqnarray*}
The vector field
$$ X \ = \ (|{\bf x}|^2 / 2) \ (\pa_x + \pa_y + \pa_z ) $$ is a strong
symmetry for this system. \EOE

\medskip\noindent
{\bf Example \nc.2.} The system considered in Example \nc.1 above
actually admits many symmetries (which now we denote by $Y$, to
avoid confusion with the $X_k$); e.g. for those of the form
$$ Y_k \ = \ \eta_k (x,y,z) \ (\pa_x + \pa_y + \pa_z ) \ , $$
we can choose
$$ \begin{array}{l} \eta_0 \ = \ (x+y+z) \ , \ \
\eta_1 \ = \ (x^2 + y^2 + z^2) \ , \ \ \eta_2 \ = \ (x y + y z + z
x) \ , \\
\eta_3 \ = \ [ x^2 \, (y + z) \, + \, y^2 \, (z + x) \, + \, z^2
\, (x + y) \ + \ 3 \, x y z ] \ , \\
\eta_4 \ = \ [ (x^3 + y^3 + z^3) \ - \ 3 \, x y z ] . \end{array}
$$
{} \EOE

\subsection{Extended symmetries}
\label{sec:albfei}

Following Misawa \cite{Mis2,Mis3} we started by considering
symmetries of the form \eqref{eq:missymm}; the reason to consider
only such transformations is that acting on $t$ means acting on
the Wiener processes as well. One could consider this case as
well, proceeding as seen above in the general case (see Sect.
\ref{sec:trastrat}), but a clever way to circumvent the problem
was devised by Albeverio and Fei \cite{AlbFei}.

Consider a Stratonovich equation \eqref{eq:mis} and a strong
symmetry of it in the form \eqref{eq:missymm}. Then we consider
\emph{symmetries of the symmetry vector field}, i.e. vector fields
\beq S \ = \ \a^0 (x,t) \, \pa_t \ + \ \a^i (x,t) \, \pa_i \eeq
such that \beq S (x^i) \ = \ \varphi^i (x,t) \ = \ \a^i (x,t) \ ;
\eeq note that we are not imposing any condition on $\a^0$, thus
in practice we are just considering extensions of the symmetry
vector field $X$, which acts only on dependent variables, to a
vector field which also acts on the independent variable.

It is immediate from \eqref{eq:misscc} to see that
\begin{eqnarray*} \[ X_0 , S \] &=& \[ X_0 , S_0 \] + \[ X_0 , X \] \ = \ \[ X_0 , S_0 \] \ ; \\
\[ X_k , S \] &=& \[ X_k , S_0 \] + \[ X_k , X \] \ = \ \[ X_k ,
S_0
\] \ .
\end{eqnarray*}

Let us then consider the set $\L$ of vector fields $L$ satisfying
the relations \begin{eqnarray} \[ X_0 , L \] &=& T^0 \, X_0 \ + \
T^m \, X_m \ ; \nonumber \\ \[ X_k , L \] &=&  R^0_k \, X_0 \ + \
R^m_k \, X_m \ . \label{eq:AFeqs} \end{eqnarray}

\medskip\noindent
{\bf Proposition \ALFEI.} {\it The set $\mathcal{L}$ is a
Lie algebra; any symmetry vector field $X$ for the equation
\eqref{eq:mis} satisfies $X \in \mathcal{L}$.}
\bigskip

Thus $\mathcal{L}$ represents an extension of the set of strong
symmetries (in the sense of Misawa) for the Stratonovich equation
under consideration; we will refer to the vector fields $L$ in
$\mathcal{L}$ as \emphind{extended symmetries} for \eqref{eq:mis}.
Proposition \ALFEI is then rephrased saying that the set of
\emph{extended} symmetries of a (Stratonovich) stochastic equation
has the structure of a Lie algebra.

\medskip\noindent
{\bf Remark \nc.5.} Proposition {\ALFEI} follows from a simple
computation. If we have vector fields $L_i$ satisfying
\begin{eqnarray*} \[ X_0 , L_i \] &=& T^0_i \, X_0 \ + \ T^m_i \,
X_m \ , \\ \[ X_k , L_i \] &=&  R^0_{ik} \, X_0 \ + \ R^m_{ik} \,
X_m \ , \end{eqnarray*} and consider $ [ X_0 , [L_i , L_j ] ]$, it
follows from Jacobi identity and some explicit computation that
$$ \[ X_0 \ \[ L_i , L_j \] \] \ = \ \[ \[ X_0 , L_i \] , L_j\] - \[ \[ X_0 , L_j \] , L_i \]
\ = \  \Theta^0_{ij} \, X_0 \ + \
\Theta^m_{ij} \, X_m \ , $$ where
\begin{eqnarray*}
\Theta^0_{ij} &=& \( (T^q_i R^0_{qj} - T^q_j R^0_{qi}) \ + \ (L_i
(T^0_j) - L_j (T^0_i)) \) \ , \\
\Theta^m_{ij} &=& \( (T^0_i T^0_j - T^0_j T^0_i ) \ + \ ( T^q_i
R^m_{qj} - T^q_j R^m_{qi} ) \ + \ (L_i (T^m_j) - L_j (T^m_i) ) \)
\ . \end{eqnarray*} Similarly, and again using the Jacobi
identity,
$$ [X_k \ [L_i , L_j ] \ ] \ = \ [[X_k,L_i],L_j] - [[X_k,L_j],L_i]
\ = \  \Gamma^0_{ij} \, X_0 \ + \ \Gamma^m_{ij} \, X_m \ , $$
where
\begin{eqnarray*}
\Gamma^0_{ij} &=& \( (R^0_i T^0_j - R^0_j T^0_i) \ + \ (R^q_{k i} R^0_{q j} - R^q_{k j} R^0_{q i})
\ + \ (L_i (R^0_j) - L_j (R^0_i) ) \) \ , \\
\Gamma^m_{ij} &=& \( (R^0_i T^m_j - R^0_j T^m_i) \ + \ (R^q_{k i}
R^m_{q j} - R^q_{k j} R^m_{q i} ) \ + \ (L_i (R^m_{k i} - L_j
(R^m_{k i} ) \) \ .
\end{eqnarray*}
These relations express the algebraic structure of the set $\L$.
\EOR

\medskip\noindent
{\bf Example \nc.3.} Consider again Example \nc.1
(see also Example \nc.2). This admits several vector
fields $L$ satisfying the relations $[X_0,L] = 0 = [X_k , L]$;
e.g. we can consider
\begin{eqnarray*}
L_0 &=& (x+y+z) \ (\pa_x + \pa_y + \pa_z) \ , \\
L_1 &=& (x^2 + y^2 + z^2) \ (\pa_x + \pa_y + \pa_z) \, \\
L_2 &=& (x y + y z + z x) \ (\pa_x + \pa_y + \pa_z) \, \\
L_3 &=& [x^2 (y + z) + y^2 (z + x) + z^2 (x + y) + 3 x y z] \ (\pa_x + \pa_y + \pa_z) \ , \\
L_4 &=& [(x^3 + y^3 + z^3) - 3 x y z] \ (\pa_x + \pa_y + \pa_z) \ , \\
... & & ... \end{eqnarray*} These form a Lie algebra; e.g.
considering only the first vector fields we have
$$ \begin{array}{ll}
\[ L_0 , L_1 \] \ = \ - \, L_1 \ + \ 4 \, L_2 , \ \ &
\[ L_0 , L_2 \] \ = \ 2 \, L_1 \ + \ L_2 , \\
\[ L_1 , L_2 \] \ = \ 2 \, L_4 , \ \ &
\[ L_0 , L_3 \] \ = \ 6 \, L_3 \ + \ 2 \, L_4 , \\
\[ L_0 , L_4 \] \ = \ 0 , \ \ &
...  ... \end{array} $$
{} \EOE

\section{W-symmetries of Stratonovich equations}

One could consider, similarly to what was done for Ito equations
(see Sect. \ref{sec:wsymm}), \ind{W-symmetries} for Stratonovich
equations. One considers again maps of the form \eqref{eq:Wmaps},
see the discussion in Sect. \ref{sec:wsymm} for the reasons of this
limitation, and with standard computations it turns out the
Stratonovich equation \eqref{eq:mis} is mapped into an equation
\beq d x^i \ = \ [ b^i (x,t) \ + \ \eps \, (\de b)^i (x,t) ] \, d
t \ + \ [\s^i_{\ k} (x,t) \ + \ \eps \, (\de \s)^i_{\ k} (x,t) ]
\circ d w^k \ , \eeq where the first order variations are given
explicitly by
\begin{eqnarray}
\de b^i &=& \pa_t \vphi^i + b^j \pa_j \vphi^i - \vphi^j \pa_j b^i
- \pa_t (\tau b^i) - b^i b^j (\pa_t \tau ) - \s^i_{\ k} (\pa_t h^k
) \ , \label{eq:wsymmstrat} \\
\de \s^i_{\ k} &=& \s^j_k \pa_j \vphi^i - \vphi^j \pa_j \s^i_k -
b^i \s^j_k \pa_j \tau - \s^i_m (\pa h^k / \pa w^m) - \tau (\pa_t
\s^i_k ) - (1/2) \s^i_k (\pa_t \tau) \ . \nonumber
\end{eqnarray}

The \emph{\ind{determining equations} for W-symmetries of the
Stratonovich equation} \eqref{eq:mis} are therefore given by the
vanishing of $\de b^i$ and $\de \s^i_{\ k}$ as given by
\eqref{eq:wsymmstrat}.

The remarks presented in Sect. \ref{sec:wsymm} do also apply in
this case.

\section{Symmetries of Ito versus Stratonovich equations}
\label{sec:itovsstrat}

As well known, and recalled above (Remark \remcorr), there is a
correspondence between stochastic differential equations in
Stratonovich and in Ito form. In particular, the Stratonovich
equation \eqref{eq:mis} and the Ito equation \eqref{eq:Ito} are
equivalent if and only if the coefficients $b$ and $f$ satisfy the
relation \beql{eq:itostratequiv} f^i (x,t) \ = \ b^i (x,t) \ + \
\frac12 \[ \frac{\pa}{\pa x^k} (\s^T)^i_{\ j} (x,t) \] \, \s^{kj}
\ := \ b^i (x,t) \ + \ \rho^i (x,t) \ . \eeq Note this involve
implicitly the metric (to raise the index in $\s$); as we work in
$\R^n$ we do not need to worry about this. Moreover, for $\s$ (and
hence $\s^T$) a constant matrix, we get $\rho = 0$ i.e. $b^i =
f^i$.

Note also that $\s$ is the same in the Ito and the corresponding
Stratonovich equations, i.e. in \eqref{eq:mis} and in
\eqref{eq:Ito}; thus \eqref{eq:itostratequiv} can be used in both
directions. In particular, we can immediately use it to rewrite
the \ind{determining equations} for symmetries (of different
types) of the Stratonovich equation \eqref{eq:mis} in terms of the
coefficients in the equivalent Ito equation.

One would be tempted to expect that symmetries of an Ito equation
and those of the corresponding Stratonovich one are just the same,
and thus study the former via the latter. This would be
particularly attractive in view of the fact that the
\ind{determining equations} \eqref{eq:A8} for symmetries of
Stratonovich equations are substantially simpler than the
\ind{determining equations} \eqref{eq:gdei} for symmetries of Ito
equations; and similarly for \ind{simple symmetries}, see
\eqref{eq:propmis} and \eqref{eq:deteqItosimple}.\footnote{The
same holds at the level of \ind{determining equations} for
\ind{random symmetries}, as it will be seen in Chapter
\ref{chap:RS} (see in particular Sect. \ref{sec:rndIS}).}

Unfortunately, this way of proceeding would in general give
incorrect results; in order to understand this fact is convenient
to first discuss the case of simple symmetries.
%\footnote{The problem seems to disappear for higher order equations, which we do not consider here \cite{Unal} }

The \ind{determining equations} \eqref{eq:propmis} for simple
symmetries of \eqref{eq:mis} are immediately rewritten in terms of
the coefficients $f^i$ of the equivalent Ito equation as
($i,k=1,...,n$)
\begin{eqnarray} \pa_t \vphi^i \ + \ [ f^j (\pa_j \vphi^i) -
\vphi^j (\pa_j f^i)] \ - \ [ \rho^j (\pa_j \vphi^i) - \vphi^j
(\pa_j \rho^i)] &=& 0 \ , \nonumber \\
\s^j_{\ k} \, (\pa_j \vphi^i) \ - \ \vphi^j \, (\pa_j \s^i_{\ k}
&=& 0 \ ; \label{eq:itosymm0}
\end{eqnarray}
where $\rho^i (x,t)$ is defined in \eqref{eq:itostratequiv}.

It is immediate to check that the equations \eqref{eq:itosymm0} do
\emph{not} coincide with the equations \eqref{eq:deteqItosimple},
which we rewrite here for ease of reference:
\begin{eqnarray} & & (\pa_t \xi^i) \ + \ \[ (f^j \cdot \pa_j) \, \xi^i \ - \
(\xi^j \cdot \pa_j ) \, f^i \] \ + \
\frac12 \( \s \s^T \)^{jk} \, \pa^2_{jk} \, \xi^i \ = \ 0 \ , \nonumber \\
& & (\s_k^{\ j} \cdot \pa_j ) \, \xi^i \ - \ (\xi^j \cdot \pa_j )
\, \s_k^{\ i}  \ = \ 0 \ . \label{eq:disdis}
\end{eqnarray}

More precisely, the equations (\ref{eq:itosymm0}.b) are just the
same as the equations (\ref{eq:disdis}.b), while equations
(\ref{eq:itosymm0}.a) and (\ref{eq:disdis}.a) are different. The
difference corresponds to \beq \delta^i \ := \ \vphi^j \, (\pa_j
\rho^i) \ - \ \rho^j \, (\pa_j \vphi^i) \ - \ \frac12 \, \triangle
\vphi^i \ \not= \ 0 \ . \eeq Note that this inequality generally
holds even in one dimension. In fact, in the one-dimensional case
we get, using the definition \eqref{eq:itostratequiv}, $\rho =
(1/2) (\s \s_x)$, and recalling that now $\triangle \vphi = \s^2
\vphi_{xx}$ (as $\vphi$ is just a function of $x$ and $t$), \beq
\delta \ = \ \frac12 \ \[ \vphi \, \s_x^2 \ - \ \s^2 \, \vphi_{xx}
\ + \ \s \( \s_{xx} \, \vphi \ - \ \s_x \, \vphi_x \)
\] \ . \eeq

Thus a given vector field $X = \vphi^i \pa_i$ is a symmetry for
both the Ito and the corresponding Stratonovich equation if and
only if $\vphi$ satisfies the system made of both
\eqref{eq:itosymm0} and \eqref{eq:disdis}; this is actually a
system of three (sets of) equations, as follows from the identity
of (\ref{eq:itosymm0}.b) and (\ref{eq:disdis}.b):
\begin{eqnarray}
\vphi^i_t \ + \ f^j \, \pa_j \vphi^i \ - \ \vphi^j \, \pa_j f^i &=& - \, \frac12 \ \triangle \vphi^i \label{eq:commonitostrat.a} \\
\vphi^i_t \ + \ f^j \, \pa_j \vphi^i \ - \ \vphi^j \, \pa_j f^i &=& - \, \( \vphi^j \, \pa_j \rho^i \ - \ \rho^j \, \pa_j \vphi^i  \) \label{eq:commonitostrat.b} \\
\s^j_{\ k} \, \pa_j \vphi^i \ - \ \vphi^j \, \pa_j \s^i_{\ k} &=&
0 \ . \label{eq:commonitostrat.c}  \end{eqnarray} We can of course
rearrange the first two equations and use one of them (say
\eqref{eq:commonitostrat.b} to deal with a first order equation)
and their difference, which in the present notation is just $\de^i
= 0$.

Despite the fact \eqref{eq:commonitostrat.a} and
\eqref{eq:commonitostrat.b} are different, they could still admit
the same solutions when restricted to the set of solutions to
\eqref{eq:commonitostrat.c}. The latter is a linear equation, and
can in principle (and sometimes also in practice, see the Examples
below) be solved by the method of characteristics; we will denote
by $\S$ the space of solutions to \eqref{eq:commonitostrat.c}. In
terms of $\delta^i$, this observation means that albeit in general
$\de^i \not= 0$, it may vanish when restricted to $\S$.

It may also happen that $\de^i$ is not zero when restricted to
$\S$, but it is zero when restricted to $\S$ and to the solution
set of \eqref{eq:commonitostrat.b} (or of
\eqref{eq:commonitostrat.a}, equivalently); this will happen in
particular if there are no nontrivial symmetries.

In fact it happens that the identity of symmetries of an Ito and
of the equivalent Stratonovich equations always holds for simple
deterministic symmetries, while for general deterministic
symmetries this is the case only if the function $\tau$ in
\eqref{eq:A1} satisfies a certain condition. (The situation for
random symmetries is quite different, as we will see in
Sect.\ref{sec:rndIS}.)

The question was studied by Unal \cite{Unal}, who gave a complete
answer in the case of deterministic symmetries.\footnote{The proof
of his theorems are based on quite involved computations which are
not reported in his paper; they have recently been checked and
confirmed \cite{Lunini}.}

\medskip\noindent
{\bf Proposition \UNAL.} {\it The simple deterministic symmetries
of an Ito equation and of the associated Stratonovich equation are
always the same. A general deterministic symmetry \eqref{eq:A1} of
an Ito equation is also a symmetry of the associated Stratonovich
equation (or viceversa) if and only if the function $\tau$ in
\eqref{eq:A1} satisfies the condition \beq\label{eq:unal}
\sigma^k_{\ m} \, \sigma^{im} \ \pa_k \[ (\pa_t \tau) \ + \ f^i \,
\pa_i \tau \ + \ \frac12 \, \sigma^p_{\ j} \, \sigma^{qj} \, (
\pa^2_{pq} \tau ) \] \ . \eeq}

\medskip\noindent
{\bf Example \nc.4.} Consider the Ito (and the corresponding
Stratonovich) equation, with $\s = x$ and hence $\rho = x/2$,
$$ \cases{ d x^i \ = \ x \, d t \ + \ x \, d w & (Ito) \cr
d x^i \ = \ (x/2) \, d t \ + \ x \circ d w & (Stratonovich) \cr} \
. $$ In this case the system
\eqref{eq:commonitostrat.a}--\eqref{eq:commonitostrat.c} reads
\begin{eqnarray*}
\vphi_t \ + \ x \, \vphi_x \ - \ \vphi &=& - \, (1/2) \ x^2 \ \vphi_{xx} \\
\vphi_t \ + \ x \, \vphi_x \ - \ \vphi &=& - \, (1/2) \ \( \vphi \ - \ x \, \vphi_x  \) \\
x \, \vphi_x \ - \ \vphi &=& 0 \ ; \end{eqnarray*} obviously the
first two equations do not coincide.

However, the last equation yields
$$ \vphi (x,t) \ = \ \a (t) \ x \ , $$
and now, on the space of these functions (i.e. of solutions to the
last equation) the first two equations do both read
$$  \a' \, x  \ + \ x \, \a \ - \ \a \, x  \ = \ 0 \ , $$
which by the way yields $\a' = 0$ and hence
$$ \vphi (x,t) \ = \ a \, x $$ with $a$ a constant. \EOE

\medskip\noindent
{\bf Example \nc.5.} Consider now the Ito (and the corresponding
Stratonovich) equation
$$ \cases{ d x^i \ = \ x^2 \, d t \ + \ x \, d w & (Ito) \cr
d x^i \ = \ (x^2 - x/2) \, d t \ + \ x \circ d w & (Stratonovich)
\cr} \ . $$ In this case the system \eqref{eq:commonitostrat}
reads
\begin{eqnarray*}
\vphi_t \ + \ x^2 \, \vphi_x \ - \ 2 \, x \, \vphi &=& - \, \frac12 \ x^2 \ \vphi_{xx} \\
\vphi_t \ + \ x^2 \, \vphi_x \ - \ 2 \, x \, \vphi &=& - \, \frac12 \( \vphi \ - \ x \vphi_x  \) \\
x \, \vphi_x \ - \ \vphi &=& 0 \ ; \end{eqnarray*} again the last
equation yields $\vphi = \a (t) x$, and with this {\it ansatz} the
first two equations both read
$$ \a' \, x  \ + \ x^2 \, \a \ - \ 2 \, x^2 \, \a \ = \ 0 \ . $$
However, now the equation enforces $\a (t) = 0$, and hence there
are no nontrivial symmetries. \EOE
\bigskip

The (possible) lack of correspondence between the symmetries of an
Ito equation and of the corresponding Stratonovich equation might
seem rather surprising at first; however, the notion of
correspondence between an Ito and the associated Stratonovich
equation is not so trivial, as discussed e.g. in the last chapter
of the book by Stroock \cite{Stroock} (see in particular Sect.
8.1.2 there), and thus the difference between the symmetries of
the two is not so strange, after all.

\medskip\noindent
{\bf Remark \nc.6.} It should also be stressed that the above
discussion only hints at having possibly different symmetries for
a Stratonovich and the corresponding Ito equations; but it does
not rule out the possibility that symmetries to the two are in
correspondence, albeit not identical\footnote{Once again, we have
here an interesting project to be pursued.}. In the case of
\ind{random symmetries}, to be discussed in Chapter \ref{chap:RS},
we will show explicit examples where symmetries of an Ito and of
the corresponding SDE are not the same, but are in one-to-one
correspondence. \EOR
\bigskip

On the other hand, an Ito equation and the associated Stratonovich
equation do carry the same \emph{statistical} information. In view
of the discussion and results in Sect. \ref{sec:liepoint} (and in
\cite{GRQ1}), we would expect there is a correspondence between
symmetries of the Fokker-Planck equation which are also symmetries
of the Ito equation and symmetries of the equivalent Stratonovich
equation. This is indeed the case, as shown by Spadaro; see
\cite{GS2016} for details of the proof.

\medskip\noindent
{\bf Proposition \SPAD.} {\it Given an Ito equation and the
associated Fokker-Planck equation, the symmetries of the latter
which are also symmetries of the Ito equation, are also symmetries
of the associated Stratonovich equation.}

\medskip\noindent
{\bf Remark \nc.7.} As stressed by Unal \cite{Unal} (and confirmed
by Lunini \cite{Lunini}), in order to really obtain different
\emph{deterministic} symmetries in the Ito and Stratonovich cases,
one should consider rather complex situations. Note that in the
case of \emph{random} symmetries, one can have different
symmetries even in the \index{simple random symmetries} simple
case, as shown in Sect. \ref{sec:rndIS}. \EOR

\medskip\noindent
{\bf Remark \nc.8.} This review focuses on Ito and Stratonovich
equations. However it is also possible to interpolate between
these two classes of equations; in fact, one can have intermediate
type equations depending on a continuous parameter $\alpha \in
[0,1]$ so that for $\a = 0$ we have Stratonovich equations and for
$\a = 1$ Ito equations (see e.g. \cite{LauLub}, where applications
are also considered). It would be interesting to study the
symmetries of these intermediate cases, and ascertain if the whole
family admits -- at least in the case of simple deterministic
symmetries -- the same ones. \EOR

\chapter{Random symmetries}
\label{chap:RS}
\def\nc{\ref{chap:RS}}

\section{Random diffeomorphisms}
\label{sec:randiff}

In the case of deterministic differential equations it is entirely
natural to consider transformations generated by smooth,
deterministic vector fields. But, in the case of \emph{stochastic}
differential equations the restriction to deterministic generators
(as we have considered so far) is not that obvious. In fact, one
could argue that the transformations to be considered should be
stochastic as well.

This point of view was adopted by \index{Arnold-Imkeller approach}
L. Arnold and P. Imkeller in
their seminal work on normal forms for stochastic differential
equations \cite{ArnImk} (see also \cite{LArnold}), and they
considered \emphind{random diffeomorphisms} as generators of the
normalizing transformations. We will follow the same approach for
symmetries of SDEs; i.e. consider, beside the usual (deterministic)
diffeomorphisms, random diffeomorphisms as well.\footnote{This
Chapter will follow a recent paper by the author and F.
Spadaro \cite{GS2016}. I am indebted to C. Lunini for a number
of questions and remarks on these matters.}

\subsection{Random maps}

Arnold and Imkeller \cite{ArnImk} \index{Arnold-Imkeller approach}
define a \emphind{near-identity random map} $h : \Om \times M \to
M$, with $M$ a smooth manifold and $\Om$ a probability space, as a
measurable map such that: (i) $h (\om , . ) \in \mathcal{C}^\infty
(M)$;  (ii) $h(\om, 0) = 0$;
 (iii) $(Dh) (\om , 0) = id$.
%\medskip

Property (i) means that we can consider this as a family of
diffeomorphisms (i.e., passing to generators, of vector fields) on
$M$, depending on elements $\om$ of the probability space $\Om$.
Note that this dependence is rather arbitrary, in particular no
request of smoothness is present.

We will also refer to the generator of such a map, with a slight
abuse of notation, as a \index{random diffeomorphisms}
\emph{random diffeomorphism}. Note that
\ind{random diffeomorphisms} (as well as \ind{random maps}) only
act in the underlying smooth manifold $M$, i.e. they do not act
(but see next Section) on the elements of the probability space
$\Om$.\footnote{In our case, $M = \R \times M_0$, with $\R$
corresponding to the time coordinate, is the phase manifold for
the system, while $\Om$ will be the path space for the
$n$-dimensional Wiener process $W(t) = \{ w^1 (t) , ... , w^n (t)
\}$.}

With local coordinates $x^i$ on $M_0$, we want to consider general
random diffeomorphisms generated by vector fields of the form \beq
X \ = \ \tau (x,t;w) \, \pa_t \ + \ \vphi^i (x,t;w) \, \pa_i \ .
\eeq

A \emph{time-preserving} random diffeomorphism will be
characterized by having $\tau = 0$, while the
\emph{fibration-preserving} ones (with reference to the fibration
$M \to \R$) will be characterized by $\pa_i \tau = 0$ for all $i$,
i.e. $\tau = \tau (t;w)$.

We will start by considering ``simple'' (i.e. time-preserving)
random symmetries in order to tackle the key problem in the
simplest setting \cite{CatLuc,GS2016}; later on (Sect. \ref{sec:general}) we will
consider the general case. Simple random symmetries will be
time-preserving, i.e. have as generator \beq \label{eq:Y} Y \ = \
\vphi^i (x,t;w) \,\pa_i \ . \eeq

In the following it will be convenient to use the notation  \beq
\label{eq:triangle}  \triangle f \ := \ \sum_{k=1}^n \ \frac{\pa^2
f}{\pa w^k \pa w^k} \
 + \ \sum_{j,k=1}^{n}(\s\s^{T})^{jk}
 \frac{\pa^{2} f}{\pa x^j \pa x^k} \ . \eeq
Note that for a function depending only on the $(x,t)$ variables
-- as in the case of deterministic symmetries -- the first term on
the r.h.s. vanishes identically.

We recall once again that, as mentioned above at several points
(or see e.g. Theorem 8.20 in the book by Oksendal
\cite{Oksendal}), a transformation of time will induce a
transformation of the Wiener processes. In particular a map $t \to
t + \eps \tau (x,t)$ will induce the map \beq \label{eq:dwmap} d
w^k \ \to \ d w^k \ + \ \eps \, \frac12 \, \( \frac{d \tau}{d t} \) \ d w^k \ := \ d w^k \ + \ \eps \ \de w^k \ . \eeq

\subsection{Random W-maps}
\label{sec:timechange}

One can also consider general vector fields in the $(x,t;w)$ space
\cite{GS2016}, i.e. \beq \label{eq:Ygen} Y \ = \ \tau (x,t;w) \,
\pa_t \ + \ \vphi^i (x,t;w) \, \pa_i \ + \ h^k (x,t;w) \, \^\pa_k
\ . \eeq Here we started to use the shorthand notation \beq
\^\pa_k \ := \ \pa  / \pa w^k \ , \eeq which will be routinely
used also in the following. We also write $X = \tau \pa_t +
\vphi^i \pa_i$ for the restriction of $Y$ in \eqref{eq:Ygen} to
the $(x,t)$ space.

Note that in \eqref{eq:Ygen} we are considering also the
possibility of direct action on the $w^k$ variables (apart from
the action induced by a change in time), as in the approach to
\ind{W-symmetries} \cite{GRQ2}.

As already pointed out in Sect. \ref{sec:wsymm}, the
requirement that the transformed processes $\^w^k (t) = w^k (t) +
\eps h^k (x,t,w)$ are still Wiener processes, implies that $\^w^k
= M^k_{ \ell} w^\ell$ with $M$ an orthogonal matrix, and hence
that necessarily \beq h^k \ = \ B^k_{\ \ell} (x,t;w) \, w^\ell
\eeq with $B$ a (real) antisymmetric matrix. This will be assumed
from now on. Note moreover that if $B$ does not depend on $w$, then
$\triangle (h^k)$ reduces to its ``deterministic'' part; and that,
as discussed in Sect. \ref{sec:wsymm}, $B$ should not actually
depend on the $x$ nor on the $w$ variables.

\subsection{Random time changes}
\label{sec:RTC}

If we allow a time map which depend on the state of the Wiener
processes $w^k (t)$, or even just on the $x^i (t)$, we are
actually allowing a \emphind{random time change}. This point seems
to have originated some confusion in part of the literature
devoted to symmetry of SDEs, so we will briefly discuss it in the
present subsection.

On physical grounds one would be specially interested in the case
where the change of time does not depend neither on the
realization of the stochastic processes $w^k (t)$ nor on the
spatial coordinates $x^i$; i.e. in the case of
\ind{fiber-preserving maps}. These will be obtained from the
general case by simply setting $\tau = \tau (t)$.\footnote{The
case where time changes depends on the dependent coordinates has
been analyzed by several authors; see e.g.
\cite{FreMaho1,FreMaho2,SMS3,SMS1,SMS2}. A further complication
comes from the fact that in some of these papers the authors
considered transformations which could map the Wiener processes
driving the SDE into a process of different nature. As far as I
know this point was first raised by Unal \cite{Unal} and his
remarks originated a debate which I will not report here: we want
transformations leaving the SDE, and {\it a fortiori} the nature
of the processes driving it, unchanged.}

It should also be recalled that, as already mentioned in Remark
\remJCZ, considering a time change which depends on $x$ and/or $w$
(and is therefore a \ind{random time change}) will in general
destroy the absolute continuity of the measure of the transformed
process w.r.t. that of the original one, and will give raise to a
number of quite delicate questions
\cite{ChuZam,Csink,Freedman,Ikeda,ItoMcKean,Kunita,McKean,Oksendal,Varadhan};
thus the discussion concerning such transformations should to a
large extent be considered, from the mathematical point of view,
as a formal one.

It is maybe worth providing some further detail on this point
(also to explain to which extent our discussion will \emph{not} be
just formal, and what are the underlying problems). When
considering a time change depending on $x$ one should bear in mind
that $x(t)$ follows a SDE and is therefore a random process; this
also holds, of course, for $w(t)$. Such time changes are thus in
general not acceptable. To make a long story short and
non-rigorous (see e.g. \cite{Freedman,Ikeda,ItoMcKean,McKean,
Oksendal,Stroock} for a precise discussion), one should consider
time changes described by \emph{integrals} of functions of $x$
and/or $w$; the integration has of course a regularizing effect.
Thus one should consider time changes of the form \beql{eq:RTC1} t
\ \to \ \wt{t} \ = \ \b (x,t,w) \ = \ \int_0^t \gamma (x,s,w) \, d
s \ ; \eeq in our case the function $\gamma$ should be of the form
\beql{eq:RTC2} \gamma (x,s,w) \ = \ 1 \ + \ \eps \, \vartheta
(x,s,w) \ , \eeq and the relation between $\vartheta$ and $\tau$
is \beql{eq:RTC3} \vartheta \ = \ d \tau / d t \ . \eeq

Under the map \eqref{eq:RTC1}, the standard Wiener process $w(t)$ with increments $ d w $ is mapped into a Wiener process $z(t)$ with increments
$$ d z \ = \ \int_0^t \sqrt{\ga (x,s,w)} \, d w (s) \ . $$
A brief self-contained discussion is provided e.g. in Chapter 8 of \cite{Oksendal}.

\section{Ito equations}

\subsection{Simple random symmetries of Ito equations}

We will now consider the case of \ind{simple random symmetries}, i.e.
vector fields of the form \eqref{eq:Y}; under this the Ito
equation
\beql{eq:ItoC5} d x^i \ = \ f^i (x,t) \, d t \ + \ \s^i_{\ k} (x,t) \, d w^k \eeq
is in general mapped into a new, generally
different, Ito equation. The equation remains invariant if and
only if the component of the vector field \eqref{eq:Y} satisfy
appropriate relations, i.e. the \ind{determining equations}. More
precisely, as shown in \cite{GS2016},

\medskip\noindent
{\bf Proposition \GSaa.} {\it The \ind{determining equations} for
\ind{simple random symmetries} for the Ito equation \eqref{eq:ItoC5} are
\begin{eqnarray}
(\pa_t \vphi^i) \ + \ f^j (\pa_j \vphi^i) \ - \ \vphi^j (\pa_j f^i)
&=& - \ \frac12 \, (\triangle \vphi^i) \ , \nonumber \\
(\^\pa_k \vphi^i) \ + \ \s^j_{\ k} (\pa_j \vphi^i) \ - \ \vphi^j
(\pa_j \s^i_{\ k} ) &=& 0  \ . \label{eq:Itosymm} \end{eqnarray} }

\medskip\noindent
{\bf Remark \nc.1.} Apparently the only difference w.r.t. the
\ind{determining equations} for (simple) deterministic symmetries
\eqref{eq:deteqItosimple} is the presence of the $\^\pa_k \vphi^i$
term in the second equation. But one should however recall that --
despite the formal analogy -- the term $\triangle \vphi^i$ does
now also include derivatives w.r.t. the $w^k$ variables, which are
of course absent in \eqref{eq:deteqItosimple}. \EOR

\medskip\noindent
{\bf Remark \nc.2.} The solutions to the \ind{determining equations}
should then be evaluated on the flow of the evolution equation
(the Ito SDE); this can lead some function to get less general, or
even trivial; see in this respect Examples \nc.1 and \nc.3 below.
\EOR

\medskip\noindent
{\bf Example \nc.1.} We start by considering a rather trivial
example, i.e. the equation \eqref{eq:ex1} of Example
\ref{chap:Ito}.1. In this case ($f=0$, $\s = \s_0$) we just have a system of two
equations for the single function $\vphi = \vphi (x,t;w)$, and
\eqref{eq:Itosymm} read
$$ \pa_t \vphi \ = \ - \, \frac12 \ \triangle \vphi \ , \ \ \
\pa_w \vphi  \ = \ - \, \s_0 \ (\pa_x \vphi) \ . $$ The solution
to the second of these is $ \vphi (x,t,w) = \psi (z , t)$, where
$F$ is an arbitrary (smooth) function of $z := x - \s_0 w$ and
$t$. Now the first equation reads
$$ \psi_t \ + \ \s_0^2 \, \psi_{zz} \ = \ 0 \ , $$ and Fourier transforming we write $\psi = \rho_k (t) \, \exp[ i k z]$ and obtain
$ \rho_k (t) = r_k \, \exp[ \s_0^2 k^2 t]$, with $r_k$ a constant.
It should also be noted that $dz = 0$ on solutions to our
equation \eqref{eq:ex1} (see Remark \nc.2). \EOE

\medskip\noindent
{\bf Example \nc.2.} We consider another one-dimensional example,
i.e. \beq \label{eq:example2} d x \ = \ d t \ + \ x \ d w \ ; \eeq
this was considered in \cite{GRQ1}, where it was shown it admits
no deterministic symmetries. The determining equations
\eqref{eq:Itosymm} read now
$$ \vphi_t \ + \ \vphi_x \ = \ - \, \frac12 \ \( \vphi_{ww} \, + \, x^2 \, \vphi_{xx} \) \ , \ \ \
\vphi_w \ + \ x \, \vphi_x \ = \ \vphi \ ; $$
the second equation implies
$$ \vphi (x,t,w) \ = \ x \ \psi (z,t) \ , \ \ \ \ z \ := \ x \, e^{- w} \ . $$
Plugging this into the first equation we get
$$ \psi \ + \ z \, \psi_z \ + \ x \, \( \psi_t \, + \, \frac32 \, z \, \psi_z \, + \, z^2 \, \psi_{zz} \) \ , $$
which yields
$$ \psi (z,t) \ = \ c \ \frac{e^{- t/2}}{z} \ , $$ with $c$ a constant; we hence have a simple random symmetry identified by
$$ \vphi \ = \ e^{w - t/2} \ . $$
{} \EOE

\medskip\noindent
{\bf Example \nc.3.} We pass to consider examples in dimension
two\footnote{We will write, here and below, the explicit vector indices (in $x$, $w$, $\vphi$) as lower ones in order to avoid any possible misunderstanding.}.
The first case we consider is a system related to work by Finkel
\cite{Finkel}, i.e.
\begin{eqnarray}
d x_1 &=& (a_1 / x_1) \, d t \ + \ d w_1 \nonumber \\
d x_2 &=& a_2 \, d t \ + \ d w_2 \ ; \label{eq:example3} \end{eqnarray}
here $a_1,a_2$ are two non-zero real constants.

In this case the second set of determining equations
\eqref{eq:Itosymm} imply that, setting $z_k := w_k - x_k$,
$$ \vphi_1 (x_1,x_2,t;w_1,w_2) \ = \ \eta_1 (t,z_1,z_2) \ , \ \ \
\vphi_2 (x_1,x_2,t;w_1,w_2) \ = \ \eta_2 (t,z_1,z_2) \ . $$
Plugging these into the first set of equations in
\eqref{eq:Itosymm}, and recalling that $\eta_i = \eta_i
(t,z_1,z_2)$, and hence the coefficient of different powers of
$x_1$ must vanish separately, we have
$$ \eta_1 (t,z_1,z_2) \ = \ 0 \ , \ \ \eta_2 (t,z_1,z_2) \ = \ \xi (t,z_2) \ ; $$
finally plugging this into the equation for $\eta_2$ we obtain
that $\xi$ is an arbitrary function of $\zeta = a_2 t + z_2$. In
conclusion, we got
$$ \vphi_1 \ = \ 0 \ , \ \ \ \vphi_2 \ = \ \xi (a_2 t - x_2 + w_2 ) \ . $$
Note that, again, $\zeta$ (and hence $\xi (\zeta )$) is trivially constant, i.e. $d \zeta = 0$, on solutions to \eqref{eq:example3} (and again see Remark
\nc.2 in this respect). \EOE

\medskip\noindent
{\bf Example \nc.4.} We will consider another
two-dimensional example, which is an Ornstein-Uhlenbeck type
process related to the Kramers equation: \begin{eqnarray*} d x
&=& y \, dt \\
dy &=& - \, k^2 \, y \, d t \ + \ \sqrt{2 k^2} \, d w (t)  \ ;
\end{eqnarray*} this is the system considered in Example \ref{chap:Ito}.2.
Note we have a single Wiener process $w(t)$, and correspondingly
we will look for solutions $\vphi^i = \vphi^i (x_1,x_2,t;w)$.

As in the previous example, we will start from the second set of
equations in \eqref{eq:Itosymm}; for our system these read
$$
\frac{\pa \vphi_1}{\pa w} = 0 \ , \ \ \frac{\pa \vphi_1}{\pa x_2}
=  0 \ , \ \ \frac{\pa \vphi_2}{\pa w} = 0 \ ,  \ \ \frac{\pa
\vphi_2}{\pa x_2}  =   0 \ . $$ These of course rule out any
possible dependence on $w$, i.e. show that there is no simple
random symmetry. \EOE

\medskip\noindent
{\bf Example \nc.5.} In applications one is often faced with
$n$-dimensional system, depending on $n$ Wiener processes,
$$ d x^i \ = \ (M^i_{\ j} \, x^j) \, d t \ + \ \s^i_{\ j} \ d w^j (t) \ , $$
with $M$ and $\s$ \emph{constant} matrices. It is also frequent that $\s$ is diagonal.

In this case (for general, i.e. non necessarily diagonal, $\s$)
the determining equations for simple random symmetries read
\begin{eqnarray}
(\pa_t \vphi^i) \ + \ M^j_{\ q} \, x^q \ (\pa_j \vphi^i ) \ - \
M^i_{\ j} \, \vphi^j \ + \ \frac12 \, \triangle \vphi^i &=& 0 \nonumber \\
(\^\pa_k \vphi^i) \ + \ \s^j_{\ k} \, (\pa_j \vphi^i ) &=& 0 \ .
\end{eqnarray} We start by considering the second set of
equations; these yield
$$ \vphi^i (x^1,...,x^n, t , w^1 , ... , w^n ) \ = \ \psi^i (z^1,....,z^n;t) \ , $$
where we have defined $z^k := x^k - \s^k_{\ j} w^j$.
For functions of this form we get immediately that $$ \triangle \vphi^i \ = \ 2 \ \( \s^\ell_{\ k} \, \s^m_{\ k} \ \frac{\pa^2 \psi^i}{\pa z^\ell \pa z^m} \) \ , $$ and hence the
first set of determining equations read simply \beq \frac{\pa
\psi^i}{\pa t} \ + \ \( M^j_{\ k} \, x^k \) \, \( \frac{\pa
\psi^i}{\pa z^j} \) \ - \ M^i_{\ j} \, \psi^j \ + \ \s^\ell_{\ k} \, \s^m_{\ k} \ \frac{\pa^2 \psi^i}{\pa z^\ell \pa z^m} \ . \eeq This requires in particular
$$ \( M^T \)_p^{\ j} \, \( \frac{\pa \psi^i}{\pa z^j} \) \ = \ 0 \ , $$ which means $\nabla \psi^i$ (for all $i=1,...,n$) is in the kernel of $M^T$. If $M$ is non singular, necessarily $\nabla \psi^i = 0$, hence $\psi^i (z^1,...,z^n;t) = \eta^i (t)$; moreover, substituting this into the equation again, we get the $\eta^i (t)$ satisfy $ \pa_t \eta^i = M^i_{\ j} \eta^j$, i.e.
$$ \eta^i (t) \ = \ \( \exp[ M t] \)^i_{\ j} \ \eta^j (0) \ . $$
{} \EOE

\subsection{General random symmetries of Ito equations}
\label{sec:general}

We can now pass to consider general symmetries (note these could
possibly also be \ind{W-symmetries}) of Ito equations. The vector
field \eqref{eq:Ygen} induces -- taking into account also the
discussion of the previous Section \ref{sec:timechange} and in
particular eq.\eqref{eq:dwmap} -- the infinitesimal map
\begin{eqnarray} x^i &\to& x^i \ + \ \eps \ \vphi^i (x,t;w) \ , \nonumber \\
t &\to& t \ + \ \eps \ \tau (x,t;w) \ , \label{eq:genrndmaps} \\
w^k &\to& w^k \ + \ \eps \ h^k (x,t;w) \ + \ \eps \, \de w^k \ , \nonumber \end{eqnarray}

With this, the Ito equation \eqref{eq:ItoC5} will again be mapped
into a new, generally different, equation. It is convenient to
introduce some further notation; we define the \ind{Misawa vector
fields}\footnote{Note we are extending the definition of Misawa
vector fields given previously, to include the $\^\pa_k$ terms.} $Y_\mu$ and the second order
operator $L$ by \beq \label{eq:MVFExt} Y_0 := \pa_t + f^j \pa_j \ , \ \ Y_k :=
\^\pa_k + \s^j_{\ k} \pa_j \ ; \ \ L := Y_0 + \frac12 \triangle \
. \eeq

With these, the condition for the Ito equation \eqref{eq:ItoC5} to
be invariant are readily determined \cite{GS2016}.

\medskip\noindent
{\bf Proposition \GSab.} {\it The determining equation for
general random \ind{W-symmetries} of \eqref{eq:ItoC5} are
\begin{eqnarray}
X(f^i) \ - \ L (\vphi^i) \ + \ f^i \, L(\tau ) \ + \ \s^i_k \, L(h^k) &=& 0
\ , \nonumber \\
X (\s^i_k ) \ - \ Y_k (\vphi^i ) \ + \ f^i \, Y_k (\tau) \ + \ \s^i_m  \, Y_k (h^m )
&=& 0 \ . \label{eq:deteqsito}
\end{eqnarray}}
\bigskip

The equations \eqref{eq:deteqsito} can also be rewritten, using
the explicit form of $L$ and $\psi$, as\footnote{It appears there
is no simple way to write these in terms of commutation with
\ind{Misawa vector fields}. This should not be surprising, as that
possibility -- even in the deterministic case -- is peculiar to
Stratonovich equations.}
\begin{eqnarray}
X(f^i) & - & Y_0 (\vphi^i) \ + \ f^i \, Y_0 (\tau) ) \ + \ \s^i_k \, Y_0 (h^k ) \ = \nonumber \\
& & \ = \ \frac12 \, \[ \triangle (\vphi^i) \ + \ f^i \, \triangle (\tau) \ + \ \s^i_k \, \triangle (h^k ) \]  \ , \nonumber \\
X(\s^i_k) & - & Y_k (\vphi^i) \ + \ f^i \,  Y_k (\tau) \ + \
\s^i_m \, Y_k (h^m ) \ = \  - \, \frac12 \, (\pa_t \tau ) \,
\s^i_k \ . \label{eq:deteqsitoexp}
\end{eqnarray}

In the case of general symmetries \emph{not} acting directly on
the $w$ variables, i.e. on the Wiener processes (and thus
excluding the case of \ind{W-symmetries}) the equations
\eqref{eq:deteqsito} reduce to
\begin{eqnarray}
X(f^i) \ - \ L (\vphi^i) \ + \ f^i \, L(\tau ) &=& 0
\ , \nonumber \\
X (\s^i_k ) \ - \ Y_k (\vphi^i ) \ + \ f^i \, Y_k (\tau)
&=& 0 \ , \label{eq:deteqsito2}
\end{eqnarray}
i.e. using the explicit form of $L$ and $\psi$, to
\begin{eqnarray}
X(f^i) & - & Y_0 (\vphi^i) \ + \ f^i \, Y_0 (\tau) ) \ = \
\frac12 \, \[ \triangle (\vphi^i) \ + \ f^i \, \triangle (\tau) \]  \ , \nonumber \\
X(\s^i_k) & - & Y_k (\vphi^i) \ + \ f^i \,  Y_k (\tau) \ = \  - \, \frac12 \, (\pa_t \tau ) \,
\s^i_k \ . \label{eq:deteqsito2exp}
\end{eqnarray}

\medskip\noindent
{\bf Remark \nc.3.} This system is over-determined for all $n >
1$, and in general we will have no symmetries; even in the case
there are symmetries, the equations are not always easy to deal
with, despite being linear, due to the high dimension. For $n = 1$
the counting of equations and unknown functions would suggest we
always have symmetries, but the solutions could be only local in
some of the variables. \EOR

\medskip\noindent
{\bf Remark \nc.4.} It is easily checked that in the case of
deterministic vector fields, i.e. $\vphi = \vphi (x,t)$, $\tau =
\tau (x,t)$, $h = 0$, the equations \eqref{eq:deteqsito} reduce to
the \ind{determining equations} for deterministic symmetries seen
in previous Chapters. Similarly, in the case of \ind{simple random
symmetries}, i.e. $\vphi = \vphi (x,t;w)$, $\tau = h = 0$, we get
the equations \eqref{eq:Itosymm} derived above, and for $\vphi$
not depending on $h$ these further reduce to the \ind{determining
equations} for simple (deterministic) symmetries. \EOR

\medskip\noindent
{\bf Remark \nc.5.}
It is also worth considering random \ind{fiber-preserving symmetries}, i.e. require $\tau =\tau (t)$ in \eqref{eq:deteqsito2exp} above. In this case we  write the \ind{determining equations} as
\begin{eqnarray}
\pa_t \vphi^i \ + \ f^j \, \pa_j \vphi^i \ - \ \vphi^j \, \pa_j f^i \ - \ \tau \, \pa_t f^i \ - \ f^i \, \pa_t \tau \ + \ \frac12 \ \triangle \vphi^i &=& 0 \ , \nonumber \\
\^\pa_k \vphi^i \ + \ \s^j_{\ k} \, \pa_j \vphi^i \ - \ \vphi^j \, \pa_j \s^i{\ k} \ - \ \tau \, \pa_t \s^i_{\ k} \ - \ \frac12 \ (\pa_t \tau) \ \s^i_{\ k} &=& 0 \ , \label{eq:itorandomprojectable} \end{eqnarray}
having reverted to a more explicit notation. \EOR

\medskip\noindent
{\bf Example \nc.6.} We will consider again the equations of
Example \nc.2, i.e. \beq \label{eq:example5} d x \ = \ d t \ + \ x
\, d w \ ; \eeq we have seen this does not admit any deterministic
symmetry, and it admits a simple random symmetry, identified by $\vphi = \exp[w -t/2]$. We will now check this admits some more general random symmetry; in order to keep computations simple, we will restrict to the time-independent case $\tau=0$ and $\vphi_t = h_t = 0$. (Note $\vphi_t = 0$ rules out the simple random symmetry obtained above.)

In this case the equations \eqref{eq:deteqsito} read
\begin{eqnarray*}
 x h_x + x^2 h_{xx} - \vphi_x  - \frac12 \( \vphi_{ww} + x^2 \vphi_{xx} + x h_{ww} \) &=& 0 \\
 \vphi - \vphi_w - x \vphi_x + x h_w + x^2 h_x &=& 0 \ . \end{eqnarray*}

The second equation requires
$$ \vphi (x,w) \ = \ x \ \( h(x,w)  + \eta (z) \) \ , \ \ \ \ \ \ z := w - \log (|x|) \ ; $$
plugging this into the first one we get
$$ - \eta (z) \ + \ \eta' (z) \ + \ \frac12 \, \eta' (z) \ - \ x \ , \eta''(z) \ = \ h (x,w) \ - \ x^2 \, h_x (x,w) \ . $$
Solutions to these are provided by
$$ h (x,w) \ = \ e^{1/x} \, \b (w) \ + \ k \ , \ \ \eta (z) = - k \ , $$
with $k$ an arbitrary constant and $\b$ an arbitrary smooth function.

The random symmetries we obtained in this way are
$$ Y \ = \ \[ x \, e^{1/x} \, \b (w) \] \, \pa_x \ + \ \[ e^{1/x} \, \b (w) \ + \ k \] \, \pa_w \ . $$
{} \EOE

\medskip\noindent
{\bf Example \nc.7.} We consider the system
\begin{eqnarray}
d x_1 &=& [1 - (x_1^2 +x_2^2)] \, x_1 \, d t \ + \ d w_1 \nonumber \\
d x_2 &=& [1 - (x_1^2 +x_2^2)] \, x_2 \, d t \ + \ d w_2 \ ;
\label{eq:example6} \end{eqnarray} this is manifestly covariant
under simultaneous rotations in the $(x_1,x_2)$ and the
$(w_1,w_2)$ planes \cite{GRQ2}.

In order to simplify (slightly) the computations, we will look for
symmetries which are time-preserving and time-independent; that
is, we assume again $\tau = 0$, $(\pa_t \vphi^i ) = 0 = (\pa_t h^k)$.
Setting $z_k :=  x_k  -  w_k$, the first set of
\eqref{eq:deteqsito} provides
\begin{eqnarray*}
h_1 (x_1,x_2,w_1,w_2) &=& \vphi_1 (x_1,x_2,w_1,w_2) \ + \ \rho_1 (z_1 , z_2 ) \\
h_2 (x_1,x_2,w_1,w_2) &=& \vphi_2 (x_1,x_2,w_1,w_2) \ + \ \rho_2 (z_1 , z_2 ) \ ,  \end{eqnarray*}
where the $\rho_i$ are arbitrary smooth functions of $(z_1,z_2)$.

Plugging these into the first set of \eqref{eq:deteqsito} we
obtain two equations involving $\vphi^i$ and derivatives of the
$\rho^i$. These equations can then be solved for the $\vphi^i$ in
terms of the $\rho^i$, yielding some complicate expression we do
not report.

This shows we have random symmetries in correspondence with arbitrary functions $\rho_i (z_1,z_2)$.
When these are linear,
$$ \rho_1 \ = \ r_{10} \ + \ r_{11} \, z_1 \ + \ r_{12} \, z_2 \ ; \ \
\rho_2 \ = \ r_{20} \ + \ r_{21} \, z_1 \ + \ r_{22} \, z_2 \ , $$
and writing $ \chi :=  [-  1  +  3  (x_1^2 +x_2^2)]$, the
resulting random symmetries can be explicitly identified via
standard simple computations. With the choice
$$ r_{10} = 0 , \ r_{20} = 0 \ ; \ \ r_{11} = 0 , \ r_{12} = 1 , \ r_{21} = - 1 , \ r_{22} = 0 $$
we get just simultaneous rotations (by the same angle) in the $(x_1,x_2)$ and $(w_1,w_2)$ planes \cite{GRQ2}. \EOE

\section{Stratonovich equations}

As in the case of Ito equations, we have first (in Chapter
\ref{chap:Strato}) considered deterministic transformations for
Stratonovich SDEs. But again, as in the Ito case, one would like
to consider \ind{random diffeomorphisms} as well. This will be
done in the present Section.

\subsection{Simple random symmetries
of Stratonovich equations} \label{sec:stratosimple}

We will start by considering \ind{simple random symmetries}, i.e.
generators of the form \eqref{eq:Y}, for the Stratonovich equation
\beql{eq:stratoC5} d x^i \ = \ b^i (x,t) \, d t \ + \ \s^i_{\ k} (x,t) \circ d w^k \ . \eeq

Under the action of $Y$, the equation \eqref{eq:stratoC5} is mapped
into another equation of the same type. More precisely, the latter
is \beq d x^i \ + \ \eps \, d \vphi^i \ = \ (b^i + \eps \phi^j
\pa_j b^i ) \, d t \ + \ (\s^i_{\ k} + \eps \vphi^j \pa_j \s^i_{\
k} ) \circ d w^k \ ; \eeq taking into account \eqref{eq:stratoC5}
and expanding the term $d \vphi$, we have that terms of first
order in $\eps$ cancel out if and only if \beq (\pa_t \vphi^i) \,
d t \ + \ (\pa_j \vphi^i) \, d x^j \ + \ (\^\pa_k \vphi^i) \circ d
w^k \ = \ (\vphi^j \pa_j b^i) \, d t \ + \ (\vphi^j \pa_j \s^i_{\
k} ) \circ d w^k \ . \eeq Note that the last term in the l.h.s. is
the only difference with respect to the computation in the
deterministic case.

Considering now $x$ on the solutions to \eqref{eq:stratoC5}, we
get with a simple rearrangement \cite{GS2016} $$ \begin{array}{l}
\[ \pa_t \vphi^i \ + \ b^j (\pa_j \vphi^i ) \ - \ \vphi^j (\pa_j
b^i)
\] \, d t \\
\ + \ \[ (\^\pa_k \vphi^i ) \ + \ \s^j_{\ k} (\pa_j \vphi^i) \ - \ \vphi^j
(\pa_j \s^i_{\ k} )
\] \circ d w^k \ = \ 0 \ ; \end{array} $$
it follows immediately that

\medskip\noindent
{\bf Proposition \SimStrat.} {\it The \ind{determining equations}
for \ind{simple random symmetries} (of the form \eqref{eq:Y}) of
the Stratonovich SDE \eqref{eq:stratoC5} are
\begin{eqnarray}
\pa_t \vphi^i \ + \ b^j (\pa_j \vphi^i ) \ - \ \vphi^j
(\pa_j b^i)  & = & 0 \ \ \ (i=1,...,n) \nonumber \\
\^\pa_k \vphi^i \ + \ \s^j_{\ k} (\pa_j \vphi^i) \ - \ \vphi^j (\pa_j
\s^i_{\ k} ) & = & 0 \ \ \ (i,k=1,...,n) \ .
\label{eq:deteqrndstr}
\end{eqnarray}}

\medskip\noindent
{\bf Remark \nc.6.} In order to express these \ind{determining
equations} in compact terms, it is convenient to consider the
\ind{Misawa vector fields} \eqref{eq:MVFExt} associated with the
SDE; in terms of these, the \ind{determining equations}
\eqref{eq:deteqrndstr} read simply \beql{eq:symmYk} [ Y , Y_\mu ]
\ = \ 0  \ \ \ \ (\mu = 0,1,...,n ) \ . \eeq These can be compared
with \eqref{eq:misscc} for deterministic symmetries. \EOR

\medskip\noindent
{\bf Remark \nc.7.} The equations \eqref{eq:deteqrndstr} should be compared with the corresponding \ind{determining equations} for \ind{simple random symmetries} of Ito equations, \eqref{eq:Itosymm}. Here too -- as in the deterministic case -- the second set of equations is just the same in the two cases. \EOR

\medskip\noindent
{\bf Remark \nc.8.} More precisely, the \ind{determining equations}
\eqref{eq:deteqrndstr} for simple random symmetries of \eqref{eq:stratoC5}
are immediately rewritten in terms of the coefficients $f^i$ of
the equivalent Ito equation as ($i,k=1,...,n$)
\begin{eqnarray} \pa_t \vphi^i \ + \ [ f^j (\pa_j \vphi^i) -
\vphi^j (\pa_j f^i)] \ - \ [ \rho^j (\pa_j \vphi^i) - \vphi^j
(\pa_j \rho^i)] &=& 0 \ , \nonumber \\
\^\pa_k \vphi^i \ + \ [ \s^j_{\ k} \, (\pa_j \vphi^i) \ - \
\vphi^j \, (\pa_j \s^i_{\ k} ] &=& 0 \ ; \label{eq:itosymm0R}
\end{eqnarray}
where $\rho^i (x,t)$ is defined in \eqref{eq:itostratequiv}.

Note that the equations \eqref{eq:itosymm0R} can be expressed in
the compact form \eqref{eq:symmYk} of commutation with the vector
fields $Y_\mu$ defined in \eqref{eq:MVFExt}, except that now the
same vector field $Y_0$ should now better (but equivalently) be
defined as \beq Y_0 \ = \ \pa_t \ + \ [f^i (x,t) + \rho^i (x,t)]
\, \pa_i \ . \eeq

It is immediate to check that the equations \eqref{eq:itosymm0R}
do \emph{not} coincide with the equations
\eqref{eq:Itosymm} determined above. The difference is due
to\footnote{This expression is formally equal to the one seen in Sect. \ref{sec:itovsstrat}, but one should recall that now $\triangle$ also contains derivativs w.r.t. the $w^k$ variables.} \beq \delta^i \ := \ \vphi^j \, (\pa_j \rho^i) \ - \ \rho^j \,
(\pa_j \vphi^i) \ - \ \frac12 \, \triangle \vphi^i \ \not= \ 0 \ .
\eeq This inequality generally holds (for $\s_x \not= 0$) even in
one dimension. Actually, as we have seen above (Sect.
\ref{sec:itovsstrat}), this is true even for \emph{deterministic}
vector fields, i.e. for $\^\pa_k \vphi^i \equiv 0$.
But, as in the deterministic case, one should evaluate $\delta^i$ on the space $\S$ of solutions to the (common) second set of determining equations. \footnote{Actually, for random symmetries one could argue that the random transformations should also be set in Stratonovich form, and thus be inherently different from those considered for
Ito equations.} \EOR

\subsection{Time-changing random symmetries of Stratonovich equations}
\label{sec:stratotime}

The computations presented in Section \ref{sec:stratosimple} above
can be extended to cover the case where the considered
transformations act on time as well\footnote{We stress that here
we are \emph{not} considering maps acting directly on the Wiener
processes; that is, here we are \emph{not} considering
\ind{W-symmetries}; these will be considered later on in Sect.
\ref{sec:Wrndstrat} (where we also give general formulas for
possibly non fiber-preserving maps).}; in this case one should, as
usual, take into account the map induced on the Wiener processes.
Here we will consider only \ind{fiber-preserving maps}, i.e. $\tau =
\tau (t,w)$. Proceeding in the standard way \cite{GS2016}, we have
that

\medskip\noindent
{\bf Proposition \GSac.} {\it The \ind{determining equations} for
\ind{random fiber-preserving symmetries} of the Stratonovich equation
\eqref{eq:stratoC5} are
\begin{eqnarray}
(\pa_t \vphi^i) + b^j (\pa_j \vphi^i)  - \vphi^j (\pa_j b^i)
- (\pa_t \tau b^i)  & = & 0 \ , \nonumber \\
\^\pa_k \vphi^i + \s^j_{\ k} (\pa_j \vphi^i) - \vphi^j (\pa_j
\s^i_{\ k}) - \tau (\pa_t \s^i_{\ k} ) - (1/2) (\pa_t \tau) \s^i_{\ k}
&=& 0 \ . \label{eq:deteqSgen}
\end{eqnarray}
}

\medskip\noindent
{\bf Remark \nc.9.} If we want to express these in terms of
commutation properties, introducing the vector fields $$ Z_0 = \pa_t
+ b^i (x,t;w) \pa_i \ , \ \ Z_k  = \^\pa_k + \s^i_{\ k} (x,t;w) \pa_i \ , $$
the \ind{determining equations} \eqref{eq:deteqSgen} are rewritten
-- recalling we assume $\tau$ does not depend on the $x$ variables
-- as $$ \[ Z_0 , Z \] \ = \ (\pa_t \tau) \, Z_0 \ ; \ \ \
\[ Z_k , Z \] \ = \ (\^\pa_k \tau ) \, \pa_t \ + \ \frac12 \, \s^i_{\ k} \, (\pa_t
\tau) \, \pa_i \ . $$ Thus, in this case we do \emph{not} obtain
that the determining equations amount to a simple condition in
terms of commutators (contrary to the case of simple symmetries,
see Remark \nc.6). \EOR

\medskip\noindent
{\bf Remark \nc.10.} The \ind{determining equations}
\eqref{eq:deteqSgen} for \ind{random fiber-preserving symmetries} of \eqref{eq:stratoC5} can be rewritten in terms of the coefficients $f^i$ of
the equivalent Ito equation as
\begin{eqnarray} \pa_t \vphi^i \ + \ [ f^j (\pa_j \vphi^i) -
\vphi^j (\pa_j f^i)] &=& [ \rho^j (\pa_j \vphi^i) - \vphi^j
(\pa_j \rho^i)] \ + \ (f^i + \rho^i) \( \pa_t \tau \)  \ , \nonumber \\
\^\pa_k \vphi^i \ + \ [ \s^j_{\ k} \, (\pa_j \vphi^i) \ - \
\vphi^j \, (\pa_j \s^i_{\ k} ] &=& \tau \, \pa_t \s^i_{\ k} \ + \ \frac12 \ (\pa_t \tau) \ \s^i_{\ k} \ ; \label{eq:itosymmPR}
\end{eqnarray}
here $i.k=1,...,n$ and $\rho^i (x,t)$ is defined in \eqref{eq:itostratequiv}.

These equations should be compared with the corresponding \ind{determining equations} for \ind{simple random symmetries} of the equivalent Ito equation, see \eqref{eq:itorandomprojectable}. Once again the second set of equations coincide in the two cases.

It is immediate to check that the first set of equations \eqref{eq:itosymmPR}
do \emph{not} coincide with the corresponding equations in
\eqref{eq:itorandomprojectable} determined above. However, as already remarked for deterministic and simple random symmetries, one should consider these equations restricted to the space $\S$ of solutions to the (common) second set of equations.  \EOR

\medskip\noindent
{\bf Example \nc.8.} Let us consider the equation
$$ d x \ = \ - \, x \, dt \ + \ x \circ d w \ ; $$
in this case the Misawa vector fields are
$$ Y_0 \ = \ \pa_t \ - \ x \, \pa_x \ ; \ \ Y_1 \ = \ \pa_w \ + \ x \, \pa_x \ . $$
The requirement that $X := \vphi (x,t,w) \pa_x$ commutes with both $Y_0$ and $Y_1$ yields
$$ \vphi (x,t,w) \ = \ e^{- t} \ \eta (z) \ , \ \ \ z := (e^w / x ) \ . $$
{} \EOE

\medskip\noindent
{\bf Example \nc.9.} Let us consider the system
\begin{eqnarray*}
d x_1 &=& - \, x_2 \, dt \ + \ \a \, x_1 \circ d w_1 \\
d x_2 &=& - \, x_1 \, dt \ + \ \a \, x_2 \circ d w_2 \ . \end{eqnarray*}

The Misawa vector fields are now
$$ Y_0 \ = \ \pa_t \ - \ x_2 \, \pa_1 \ + \ x_1 \, \pa_2 \ ; \ \
Y_1 \ = \ \^\pa_1 \ + \ \a r \, \pa_1 \ , \ \ Y_2 \ = \ \^\pa_2 \ + \ \a r \, \pa_2 \ . $$
Requiring the vector field
$$ X \ = \ \vphi^1 (x_1,x_2,t,w_1,w_2) \ \pa_1 \ + \ \vphi^2 (x_1,x_2,t,w_1,w_2) \ \pa_2 $$
to commute with $Y_1$ and $Y_2$ enforces
$$ \vphi^1 \ = \ x_1 \, \eta^1 (z_1,z_2,t) \ , \ \ \vphi^2 \ = \ x_2 \, \eta^2 (z_1,z_2,t) \ , $$
where we have defined
$ z_k := [(a w_k - \log|x_k|)/a]$. Requiring now that $X$ also
commutes with $Y_0$, we get that actually it must be $\eta^1 =
\eta^2 = c$; thus in the end the only simple random symmetry of
the system under consideration is
$$ X \ = \ \pa_1 \ + \ \pa_2 \ ; $$
this is actually, obviously, a simple \emph{deterministic}
symmetry. \EOE

\medskip\noindent
{\bf Example \nc.10.} We consider again the equation
$$ d x \ = \ d t \ + \ x \, d w \ , $$ as in Example \nc.2 above.
The corresponding Stratonovich equation is
$$ d x \ = \ \[ 1 \, - \, \frac{x}{2} \] \, d t \ + \ x \circ d w \ ; $$
the determining equations \eqref{eq:deteqrndstr} for simple random
symmetries of this Stratonovich equation read
\begin{eqnarray*}
\pa_t \vphi \ + \ [1 - (x/2)] \, (\pa_x \vphi) \ + \ (1/2) \, \vphi &=& 0 \\
\pa_w \vphi \ + \ x \, (\pa_x \vphi) \ - \ \vphi &=& 0 \ .
\end{eqnarray*} It is immediate to check these, or more precisely
the first of these, do \emph{not} correspond to the equations
obtained in Example \nc.2.
However this set of equations
does admit a solution, which is just \emph{the same} as that obtained in Example \nc.2:
$$ \vphi (x,t,w) \ = \ c_0 \ \exp[w - t/2] \ . $$
This fact will be discussed below, see Sect.\ref{sec:rndIS}. \EOE

\medskip\noindent
{\bf Example \nc.11.} When dealing with symmetries of Stratonovich
equations, it is customary to consider the Misawa system
\cite{Mis1}
\begin{eqnarray*}
d x_1 &=& (x_3 - x_2) \, d t \ + \ (x_3 - x_2) \, \circ \, d w \\
d x_2 &=& (x_1 - x_3) \, d t \ + \ (x_1 - x_3) \, \circ \, d w \\
d x_3 &=& (x_2 - x_1) \, d t \ + \ (x_2 - x_1) \, \circ \, d w \ ;
\end{eqnarray*}
it is well known -- and immediately apparent -- that this admits
the simple symmetry generated by
$$ X = (1/2) (x_1^2 + x_2^2 + x_3^2) \ (\pa_1 \, + \, \pa_2 \, + \, \pa_3 )  $$
(and many others, as discussed by Albeverio and Fei \cite{AlbFei};
see Sect. \ref{sec:albfei}). Note that this involves only one
Wiener process, which will induce a non-symmetric expression for
the equivalent Ito system.

Using \eqref{eq:itostratequiv}, the equivalent system of Ito
equations turns out to be
\begin{eqnarray*}
d x_1 &=& (1/2) \, (3 x_3 - x_2 - 2 x_1) \, d t \ + \ (x_3 - x_2) \, d w  \\
d x_2 &=& (x_1-x_3) \, d t \ + \ (x_1 - x_3) \, d w \\
d x_3 &=& (x_2 - x_1) \, d t \ + \ (x_2 - x_1) \, d w \ .
\end{eqnarray*}

It is immediate to check that the determining equations
\eqref{eq:Itosymm} are \emph{not} satisfied by $X$; more
precisely, the second set of \eqref{eq:Itosymm} are (of course)
satisfied, while the first set is not: in fact, we get (for all
$i=1,2,3$)
$$ \pa_t \vphi^i \ + \ f^j \, (\pa_j \vphi^i) \ - \ \vphi^j \, (\pa_j f^i)
\ + \ \frac12 \, (\triangle \vphi^i) \ = \  F (x) \ , $$
where $$ F(x) \ = \ 2 \, x_1^2 \ + \ 3 \, x_2^2 \ + \ 3 \, x_3^2
\ - \ \( \frac52 \, x_1 x_2 \ + \ 3 \, x_2 x_3 \ + \ \frac52 \, x_1 x_3 \) \ . $$
{} \EOE

\section{Random W-symmetries for Stratonovich equations}
\label{sec:Wrndstrat}

In the previous Sect. \ref{sec:stratotime} we have considered
transformations acting on time, and through this on the Wiener
processes, but \emph{not} directly on the $w$ variables. In order
to complete our discussion, we should allow also for this
possibility, i.e. consider (random) \ind{W-symmetries} as well;
this is precisely the subject of the present Section.

We will thus consider again a map of the general form
\eqref{eq:genrndmaps}. Under this, the Stratonovich equation
\eqref{eq:stratoC5} is mapped into (all functions should be thought
as functions of $(x,t)$ or $(x,t;w)$ as appropriate)
\begin{eqnarray*}
d x^i +\eps d \vphi^i &=& \[ b^i + \eps \( \vphi^j \pa_j b^i + \tau \pa_t b^i + h^k \^\pa_k b^i \) \] \ ( d t + \eps d \tau ) \\
&+&\[ \s^i_{\ k} + \eps \( \vphi^j \pa_j \s^i_{\ k} + \tau \pa_t \s^i_{\ k} \) \] \ \[ d w^k + \eps \( (1/2) (\pa_t \tau) d w^k + d h^k \) \] \ . \end{eqnarray*}
The terms of order $\eps$ provide the equation
\begin{eqnarray*}
d \vphi^i &=& \( \vphi^j \pa_j b^i + \tau \pa_t b^i \) \, d t \ + \ b^i \, d \tau \ + \ (1/2) \, (\pa_t \tau) \, \s^i_{\ k} \, d w^k \\
& & \ + \ \s^i_{\ k} \, d h^k \ + \ \( \vphi^j \pa_j \s^i_{\ k} +
\tau \pa_t \s^i_{\ k} \) \, d w^k \ . \end{eqnarray*} We should
now expand the differentials $d \tau$, $d \vphi^i$, $d h^k$, which
gives $$ d \tau \ = \ (\pa_t \tau) \, d t \ + \ (\pa_j \tau) \, d
x^j \ + \ (\^\pa_k \tau) \, d w^k \ , $$ and the like for $d
\vphi^i$ and $d h^k$ (note we are \emph{not} yet considering the
restrictions of $h^k$ which result from the discussion of Sect.
\ref{sec:wsymm}; these will be introduced in a moment). Doing
this, the equation results in the vanishing of the expression
\begin{eqnarray*}
& & \[ \pa_t \vphi^i - \tau \pa_t b^i - \vphi^j \pa_j b^i - b^i \pa_t \tau - \s^i_{\ k} \pa_t h^k \] \ d t \\
& + & \[\^\pa_k \vphi^i - b^i \^\pa_k \tau - (1/2) (\pa_t \tau) \s^i_{\ k} - \s^i_{\ m} \^\pa_k h^m - \vphi^j \pa_j \s^i_{\ k} - \tau \pa_t \s^i_{\ k} \] \ d w^k \\
& + & \[ (\pa_j \vphi^i) - b^i (\pa_j \tau) - \s^i_{\ k}  (\pa_j h^k) \] \ d x^j \ . \end{eqnarray*}

We should now use \eqref{eq:stratoC5} itself, i.e. restrict our
expression to the flow of the Stratonovich equation. Doing this,
our previous expression reduces to
\begin{eqnarray}
& & \[ \pa_t \vphi^i \ + \ \( b^j \pa_j \vphi^i \, - \, \vphi^j \pa_j b^i \) \ - \ \pa_t \( \tau b^i \) \ - \ b^i \, b^j \, (\pa_j \tau) \right. \nonumber \\
& & \left. \ - \ \s^i_{\ k} \, (\pa_t h^k) \ - \ \s^i_{\ k} \, (\pa_j h^k) \, b^j \] \ d t \nonumber \\
&+& \[\^\pa_k \vphi^i \ + \ \( \s^j_{\ k} \pa_j \vphi^i \, - \, \vphi^j \pa_j \s^i_{\ k} \) \ - \ b^i \, \( \^\pa_k \tau \, + \, \s^j_{\ k} \, (\pa_j \tau) \) \right. \label{eq:rndstr0} \\
& & \left. \ - \ \( \tau \, \pa_t \s^i_{\ k} \, + \, (1/2) \,
\s^i_{\ k} \pa_t \tau \) \ - \ \s^i_{\ m} \, (\^\pa_k h^m ) \ - \
\s^i_{\ m} \, (\pa_j h^m) \, \s^j_{\ k} \] \ d w^k \nonumber
 \end{eqnarray}

It is now time to recall the discussion of Sect. \ref{sec:wsymm}
and its conclusions, i.e. that the $h^k$ should \emph{not} depend
on the $x$ (and actually be linear in the $w$). Taking this into
account, the equations \eqref{eq:rndstr0} get slightly simplified
and we end up with
\begin{eqnarray}
& & \[ \pa_t \vphi^i \ + \ \( b^j \pa_j \vphi^i \, - \, \vphi^j \pa_j b^i \) \ - \ \pa_t \( \tau b^i \) \ - \ b^i \, b^j \, (\pa_j \tau) \ - \ \s^i_{\ k} \, (\pa_t h^k) \] \ d t \nonumber \\
&+& \[\^\pa_k \vphi^i \ + \ \( \s^j_{\ k} \pa_j \vphi^i \, - \, \vphi^j \pa_j \s^i_{\ k} \) \ - \ b^i \, \( \^\pa_k \tau \, + \, \s^j_{\ k} \, (\pa_j \tau) \) \right. \nonumber \\
& & \left. \ - \ \s^i_{\ m} \, (\^\pa_k h^m ) \ - \ \( \tau \,
\pa_t \s^i_{\ k} \, + \, (1/2) \, \s^i_{\ k} \pa_t \tau \) \] \ d
w^k \label{eq:rndstr1} \ ; \end{eqnarray}
hence we have the

\medskip\noindent
{\bf Proposition \WSTRAT.} {\it The \ind{determining equations}
for general random symmetries (including possibly
\ind{W-symmetries}) of the Stratonovich equation \eqref{eq:stratoC5}
are
\begin{eqnarray}
\pa_t \vphi^i &+& \( b^j \pa_j \vphi^i \, - \, \vphi^j \pa_j b^i \) \ - \ \pa_t \( \tau b^i \) \ - \ b^i \, b^j \, (\pa_j \tau) \ - \ \s^i_{\ k} \, (\pa_t h^k) \ = \ 0 \ , \nonumber \\
\^\pa_k \vphi^i &+& \( \s^j_{\ k} \pa_j \vphi^i \, - \, \vphi^j \pa_j \s^i_{\ k} \) \ - \ b^i \, \( \^\pa_k \tau \, + \, \s^j_{\ k} \, (\pa_j \tau) \) \nonumber \\
& & \ - \ \s^i_{\ m} \, (\^\pa_k h^m ) \ - \ \( \tau \, \pa_t
\s^i_{\ k} \, + \, (1/2) \, \s^i_{\ k} \pa_t \tau \) \ = \ 0 \ .
\label{eq:rndstr2} \end{eqnarray} }

\medskip\noindent
{\bf Remark \nc.11.} Not surprisingly, it appears that in this case
too (see Remark \nc.9) there is no simple way to express the
\ind{determining equations} \eqref{eq:rndstr2} in terms of
commutation properties with the \ind{Misawa vector fields}. \EOR

\medskip\noindent
{\bf Example \nc.12.} Let us consider again, as in Example \nc.8,
the equation
$$ d x \ = \ - \, x \ d t \ + \ x \circ d w \ . $$ Now the
determining equations \eqref{eq:rndstr2}  read simply
\begin{eqnarray*}
\vphi_t - x \vphi_x + \vphi + x \tau_t - x^2 \tau_x - x h_t &=& 0
\\
\vphi_w + x \vphi_x - \vphi + x ( \tau_w + x \tau_x) - (1/2) x
\tau_t - x h_w &=& 0 \ . \end{eqnarray*} This is an
under-determined system of two equations for the three unknown
functions $\vphi,\tau,h$; we will look for solutions with $\tau
\equiv 0$, which yields the simplified equations
\begin{eqnarray*}
\vphi_t \ - \ x \, \vphi_x \ + \ \vphi &=& x \, h_t \\
\vphi_w \ + \ x \, \vphi_x \ - \ \vphi &=& x \, h_w \ .
\end{eqnarray*}
Setting
$$ \vphi (x,t,w) \ = \ \psi (t,w) \ x \ , \ \ \ h(x,t,w) \ = \
\psi (t,w) $$ we get a family of solutions. Note that, as
discussed above, $h$ -- and therefore $\psi$ -- should actually be
linear in $w$; thus we will in fact have $\psi (t,w) = \b (t) w$,
with $\b$ an arbitrary function. \EOE

\medskip\noindent
{\bf Example \nc.13.} Let us consider again the equation studied
in Example \nc.10, i.e.
$$ d x \ = \ (1 - x/2) \, dt \ + \ x \circ d w \ ; $$
thus we have $b=(1- x/2)$, $\s = x$. The determining equations
will again be under-determined, and again we will look for
solutions with $\tau \equiv 0$; moreover, we set $h = \eta (t) \,
w$. In this frame, \eqref{eq:rndstr2} read
\begin{eqnarray*}
\vphi_t \ + \ \( 1 - \frac{x}{2} \) \, \vphi_x \ + \ \frac12 \, \vphi &=&
x  \, w \, \eta' \ , \\
\vphi_w \ + \ x \, \vphi_x \ - \ \vphi &=& x \, \eta \ .
\end{eqnarray*}
The second equation yields immediately
$$ \vphi (x,t,w) \ = \ x \ \[ \eta(t) \, \log (|x|) \ + \ \psi(t, z) \] \ ,
\ \ \ z \ = \ x \, e^{- w} \ . $$ Plugging this into the first
equation we obtain that necessarily $\eta (t) = 0$ (as seen by
considering the coefficients of terms with $\log (|x|)$), thus
in this case we get no new symmetries with respect to those found
in Example \nc.10. \EOE

\section{Random symmetries of Ito versus Stratonovich equations}
\label{sec:rndIS}

In Sect.\ref{sec:itovsstrat} we have compared the \ind{determining
equations} for deterministic symmetries of an Ito and of the
corresponding Stratonovich equation. Here we aim at doing the same
for random symmetries\footnote{This section is based on ongoing
work with C. Lunini \cite{GaeLun}.}. We will actually focus on
\index{simple random symmetries} \emph{simple} ones, and just
consider \emph{scalar} equations\footnote{A more exhaustive
analysis, considering more general random symmetries, is also
possible along the same lines; but the simple case will already
present the different situations which can occur in the general
ones, see below.}; we will see by concrete examples that one can
indeed have \emph{different} symmetries (as well as just
coinciding ones) in the two cases.

We first of all rewrite, for ease of reference, the
\ind{determining equations} for \ind{simple random symmetries} of
scalar Ito and Stratonovich equations; for the Ito case we have
\begin{eqnarray}
\vphi_t \ + \ f \, \vphi_x \ - \ \vphi \, f_x &=& - \frac12 \[ \vphi_{ww} \ + \ \s^2 \,\vphi_{xx} \] \nonumber \\
\vphi_w \ + \ \s \, \vphi_x \ - \ \vphi \, \s_x &=& 0
\label{eq:IS1} \end{eqnarray} while in the Stratonovich case
(using the expression of $\rho$ in terms of $\s$) the equations
are
\begin{eqnarray}
\vphi_t \ + \ f \, \vphi_x \ - \ \vphi \, f_x &=& - \frac12 \[ \vphi \, \s_x^2 \ + \ \vphi \, \s \, \s_{xx} \ - \ \s \, \s_x \, \vphi_x \] \nonumber \\
\vphi_w \ + \ \s \, \vphi_x \ - \ \vphi \, \s_x &=& 0  \ .
\label{eq:IS2} \end{eqnarray}

The first equations in the two sets are obviously different in
general, as stressed by Unal \cite{Unal}. The second equation in
both sets is just the same; note also it is a linear equation,
which makes it solvable, and that its solutions are a linear
space. We will denote by $\S$ the space of solutions to this
equation; needless to say, this depends on the function $\s = \s
(x,t)$.

We can then compare the first equations in the systems
\eqref{eq:IS1} and \eqref{eq:IS2} \emph{when restricted to $\S$}.
This allows to identify situations in which the equations (and
hence their solutions) coincide, situations in which the equations
are different but they both admit only the trivial solution $\vphi
(x,t,w)  = 0$, and situations in which the two equations are
different -- and indeed they do not admit any common solution.

The condition for the two equations to coincide is immediately
apparent from \eqref{eq:IS1} and \eqref{eq:IS2}, and is just
\beql{eq:IS3} \[ \vphi_{ww} \ + \ \s^2 \,\vphi_{xx} \]_\S \ = \ \[
\vphi \, \s_x^2 \ + \ \vphi \, \s \, \s_{xx} \ - \ \s \, \s_x \,
\vphi_x \]_\S \ . \eeq This is a kind of ``compatibility
equation''.

If we consider a given $\s (x,t)$, the second equation in
\eqref{eq:IS1} and \eqref{eq:IS2} can be solved, providing a
concrete expression for functions in $\S$. One can then proceed to
solve the compatibility condition \eqref{eq:IS3}. Note that this
provides a set of functions $\vphi (x,t,w)$. By considering either
one of the (first equations in) \eqref{eq:IS1} or \eqref{eq:IS2}
considered as equations for $f(x,t)$ one can determine the cases
in which the Ito and the equivalent Stratonovich equation have the
same symmetry.

\medskip\noindent
{\bf Remark \nc.12.} We note that we are guaranteed to have only
common solutions if \eqref{eq:IS3} is satisfied. In fact, the
condition to have a common simple random symmetry to an Ito and
the associated Stratonovich equation is that \emph{both}
\eqref{eq:IS1} and \eqref{eq:IS2} are satisfied, i.e. that $\vphi$
satisfies the system
\begin{eqnarray}
\vphi_t \ + \ f \, \vphi_x \ - \ \vphi \, f_x &=& - \frac12 \[ \vphi_{ww} \ + \ \s^2 \,\vphi_{xx} \] \label{eq:ccs1} \\
\vphi_t \ + \ f \, \vphi_x \ - \ \vphi \, f_x &=& - \frac12 \[ \vphi \, \s_x^2 \ + \ \vphi \, \s \, \s_{xx} \ - \ \s \, \s_x \, \vphi_x \] \label{eq:ccs2} \\
\vphi_w \ + \ \s \, \vphi_x \ - \ \vphi \, \s_x &=& 0 \ .
\label{eq:ccs3}
\end{eqnarray} By considering the difference of \eqref{eq:ccs1} and \eqref{eq:ccs2}, we reduce to a system which is just made of
\eqref{eq:IS1} (or equivalently \eqref{eq:IS2}) and the
\ind{compatibility equation} \eqref{eq:IS3}. We conclude that
common solutions are possible only if \eqref{eq:IS3} is satisfied
\emph{on solutions to the other equations} in the system.

Note that this contains $\vphi$ itself; thus it just provides a
check (to know if a symmetry of, say, the Ito equation is also a
symmetry for the associated Stratonovich one) once solutions have
been determined. On the other hand, as we show in the Examples
below, it allows to provide a general discussion and in particular
to ascertain when (that is, for which $f(x,t)$, given that a
certain $\s (x,t)$ has been assigned) there can be nontrivial
common symmetries.  \EOR

\medskip\noindent
{\bf Remark \nc.13.} As already mentioned, the scalar
one-dimensional situation is rich enough to present the different
possibilities, as shown by the following examples; one should
nevertheless remark that (at least in principle) in higher
dimensions one could have a situation where some of the symmetries
are common, and some are different. Once again, we have a problem
worth exploring. \EOR

\medskip\noindent
{\bf Example \nc.14.} Let us consider the case $\s = 1$. Now the
equation (\ref{eq:IS1}.b), and the identical equation
(\ref{eq:IS2}.b), reads
$$ \vphi_w \ + \ \vphi_x \ = \ 0 \ , $$ and its general solution is
\beql{eq:ISvphiex14} \vphi (x,t,w) \ = \ \psi(z,t) \ , \ \ \ \ z
:= x - w \ . \eeq The compatibility equation \eqref{eq:IS3} is
therefore
$$ \psi_{zz} \ = \ 0 \ , $$ with general solution
$$ \psi (z,t) \ = \ \a(t) \ + \ \b(t) \ z \ . $$
By considering the determining equations as equations for
$f(x,t)$, with $\s = 1$ and $\vphi$ as determined above, it turns
out that the equations admitting such a symmetry are characterized
by
$$ f(x,t) \ = \ F(t) \ + \ G(t) \, x \ ; $$
the functions $F$ and $G$ are related to the $\a$ and $\b$
characterizing $\psi$ and hence $\vphi$ by several equations,
which we do not write explicitly. Note that this discussion shows
that when $f$ is not of the above (linear) form, there will be no
nontrivial common symmetries for the Ito and the corresponding
Stratonovich equations.\footnote{The solution sets may coincide in
that they reduce to the trivial one.} \EOE

\medskip\noindent
{\bf Example \nc.15.} A similar discussion can be conducted for
the case $\s = x$. Now the second equation in \eqref{eq:IS1} and
\eqref{eq:IS2} reads
$$ \vphi_w \ + \ x \, \vphi_x \ - \ \vphi \ = \ 0 \ , $$ and its general solution is
\beql{eq:ISvphi} \vphi (x,t,w) \ = \ x \ \psi(z,t) \ , \ \ \ \ z
:= x \ e^{- w} \ . \eeq The compatibility equation \eqref{eq:IS3}
is therefore
$$ \[ x \, \vphi_x \ + \ x^2 \, \vphi_{xx} \ + \ \vphi_{ww} \ - \ \vphi \]_\S \ = \ 0 \, $$ which using our functional form for $\vphi$ reads explicitly
$$ z \, \psi_z \ + \  z^2 \, \psi_{zz} \ = \ 0 \ , $$ with general solution
$ \psi (z,t)  =  \a(t) + (1/z) \b(t) $ and hence \beql{eq:IS5}
\vphi (x,t,w) \ = \ x \, \a(t) \ + \ e^{- w} \, \b (t) \ . \eeq
Correspondingly, the $f(x,t)$ admitting a symmetry identified by
\eqref{eq:IS5} are identified by plugging the latter into
\eqref{eq:IS1} or \eqref{eq:IS2}, which yields $f_{xx} = 0$, i.e.
$$ f(x,t) \ = \ g(t) \ + \ h(t) \ x \ . $$
Again the functions $g,h$ are linked to the $\a,\b$ by certain
simple relations which we do not write explicitly.

This discussion shows immediately what are the $f(x,t)$ which --
for the given $\s (x,t)$ -- can give common solutions to the
determining equations \eqref{eq:IS1} and \eqref{eq:IS2}; and again
we just get linear functions. \EOE

\medskip\noindent
{\bf Example \nc.16.} In Example \nc.2 and Example \nc.10 we have
considered the Ito equation
$$ d x \ = \ d t \ + \ x \, d w $$
and the associated Stratonovich one; we obtained that albeit the
determining equations are not the same, their solutions do
coincide and are given by $$ \vphi \ = \ c_0 \, \exp[ w - t/2 ] \
. $$ The discussion of the previous Example \nc.15 explains why
and how this happens. \EOE

\medskip\noindent
{\bf Example \nc.17.} Motivated by Example \nc.14, let us consider
the linear Ito equations \beq d x \ = \ \( \a (t) \ + \ \b (t) \ x
\) \, d t \ + \ d w \ . \eeq The second of \eqref{eq:IS1} will
give \eqref{eq:ISvphiex14}. With this, the other determining
equation reads simply \beql{eq:e16a} \psi_t \ + \ (\a + \b x) \,
\psi_z \ - \ \b \, \psi \ + \ \psi_{zz} \ = \ 0 \ . \eeq This
implies $\psi_t = 0$, i.e. $\psi (x,t) = \theta (t)$, and the
equation is thus reduced to \beql{eq:e16b} \theta' (t) \ = \ \b
(t) \ \theta (t) \ , \eeq with solution
$$ \theta (t) \ = \ \theta (0) \ \exp \[ \int_0^t \b (s) \ d s \] \ . $$
We thus have as solution to the determining equations
$$ \vphi (x,t,w) \ = \ \vphi (t) \ = \ c_0 \ \exp \[ \int_0^t \b (s) \ d s \] \ . $$

Let us now consider the corresponding Stratonovich equations and
the determining equations \eqref{eq:IS2} for their simple
symmetries. The first equation reads \beql{eq:e16c} \psi_t \ + \
(\a + \b x) \, \psi_z \ - \ \b \, \psi \ = \ 0 \ . \eeq This is of
course still different from \eqref{eq:e16a}, but it still requires
$\psi_z = 0$ and hence $\psi (z,t) = \theta (t)$; the equation for
$\theta$ is again \eqref{eq:e16b}, and we thus have that simple
symmetries are common to the Ito and the corresponding
Stratonovich equations. \EOE

\medskip\noindent
{\bf Example \nc.18.} Consider now the Ito equation
$$ d x \ = \ x^2 \, d t \ + \ d w \ ; $$
according to the discussion in Example \nc.14, this should not
admit common symmetries for this and the corresponding
Stratonovich equation.

In facts, the first of \eqref{eq:IS1} reads now, assuming again
\eqref{eq:ISvphiex14},
$$ \psi_t \ + \ x^2 \, \psi_z \ + \ \psi_{zz} \ - \ 2 \, x \, \psi \ = \ 0 \ ; $$
this enforces $\psi (z,t) = 0$.
If we consider the first of \eqref{eq:IS2} we get
$$ \psi_t \ + \ x^2 \, \psi_z \ - \ 2 \, x \, \psi \ = \ 0 \ , $$
which also enforces $\psi (z,t) = 0$; again the equations are
different and again they admit the same solutions -- but now they
are just the trivial one, i.e. provide no symmetry. Note that this
discussion is easily extended to the case where $x^2$ is replaced
by a general polynomial of order $n > 1$. \EOE

\medskip\noindent
{\bf Example \nc.19.} Finally, let us consider
$$ d x \ = \ e^{- k x} \, d t \ + \ d w \ . $$
Now the first of \eqref{eq:IS1} reads
$$ \psi_t \ + \ e^{- k x} \( \psi_z \, + \, k \, \psi \) \ + \ \psi_{zz} \ = \ 0 \ . $$
This requires
$$ \psi (z,t) \ = \ \ga (t) \ e^{- k z} \ , $$ which should moreover satisfy
$$ \[ \ga' (t) \ + \ k^2 \, \ga (t) \] \ e^{- k z} \ = \ 0 \ ; $$
this in turn enforces $\ga (t) = c_0 \exp [ - k^2 t]$. We thus
have
$$ \vphi_{ito} (x,t,w) \ = \ c_0 \ \exp [ - k (x - w) - k^2 t ] \ . $$

Let us now consider the corresponding Stratonovich equation; the
first of \eqref{eq:IS2} reads
$$ \psi_z \ + \ e^{- k x} \, \psi_z \ + \ k \, e^{- k x} \, \psi \ = \ 0 \ . $$
This again requires $ \psi_z = -  k \psi$ and hence $\psi (z,t) =
\ga (t) \exp (- k z) $, but this should now moreover satisfy
$$ \ga' (t) \ e^{- k z} \ = \ 0 \ , $$
which of course provides $\ga (t) = c_0 $. Thus we get
$$ \vphi_{strat} (x,t,w) \ = \ c_0 \ \exp [ - k(x -w) ] \ . $$

That is, in this case the Ito and the Stratonovich equations do
admit nontrivial symmetries, but they are not the same. On the
other hand, there is a (simple) one to one correspondence between
the two sets. \EOE

\medskip\noindent
{\bf Remark \nc.14.} As mentioned in Remark \remLL, one could
consider intermediate cases between Ito and Stratonovich equations
depending on a continuous parameter $\a$ \cite{LauLub}. This opens
the possibility to study how the symmetries of these intermediate
equations depend on such a parameter, and possibly if there is a
correspondence between different symmetries in the extreme -- i.e.
the Ito and the Stratonovich -- cases we are here considering, as
was the case in Example \nc.19 above. \EOR

\chapter{Use of symmetries for studying SDEs}
\label{chap:Use}
\def\nc{\ref{chap:Use}}

So far, our discussion -- like most of the literature devoted to
symmetries of stochastic differential equations -- focused on
determining what is the ``right definition'' for symmetries of a
SDE rather than on the \emph{use} of these symmetries.

It may look surprising that applications of symmetries -- except
for extension of the Noether theorem to the stochastic framework
\cite{PQS2,Mis1,TZ,PQS1}, which is of course an extremely
important application! -- were not more actively pursued neither
at the time of the early works of Misawa and Albeverio \& Fei
\cite{AlbFei,Mis2,Mis3,Mis4}, nor more recently.

In my opinion one of the reason for this is quite simply that the
way to proceed for ``applications'' was (at least in principle)
clear: if we have a symmetry, we should use (as in the
deterministic case, and as done in \ind{stochastic Noether
theory}) \emphind{symmetry-adapted coordinates}.

In this chapter we will consider -- rather briefly, which also
keeps us in line with the general literature -- several
applications of this general idea. There is no need to stress that
a lot of work remains to be done in this direction, both in terms
of general theory and in terms of concrete applications to
specific problems, but as we will see below
(Sect. \ref{sec:kozlov}) there is a close analogue of the symmetry
reduction holding for deterministic ODEs, see Propositions {\KOZg}
and {\KOZh}. These appear to be the gateway for much of this future work.

We will actually follow, for once, the time development of the
subject. We start from considering Stratonovich type equation, as
in the early works of Misawa and Albeverio and Fei
\cite{AlbFei,Mis2,Mis3,Mis4} mentioned above (Sect. \ref{sec:scq});
we will then also consider Ito equations (Sect. \ref{sec:adapted})
and the problem of linearization of SDEs (Sect.
\ref{sec:linearization}), which is also a special case of the more
general problem of reducing a SDE to a simpler (i.e. more
convenient) form, see Sect. \ref{sec:kozlov}. The
\emphind{reduction} of SDEs will also be considered in this
context.

\medskip\noindent
{\bf Remark \nc.1.} In the previous chapters, see in particular Chap. \ref{chap:Ito}, we have also considered the relations between symmetries of a SDE and those of the associated \ind{Fokker-Planck equation}; symmetries of the Fokker-Planck equation can of course be used to determine explicit solutions to it (as discussed in Chap. \ref{chap:det}). However this falls within the realm of deterministic equations and we will thus not discuss this aspect, referring the reader to the literature \cite{CicVit1,CicVit2,Finkel,KMA,Koz1,Koz2,Koz3,Rud,SasDun,ShtSto,SpiSto}.

\section{Stratonovich equations.
Strong symmetries and strongly conserved quantities}
\label{sec:scq}
\def\cs{\ref{sec:scq}}

As already mentioned, the early work by Misawa, and then by
Albeverio and Fei, on symmetries of non-variational stochastic
equations were prompted by the attempt to extend Hojman's results
(see Sect. \ref{sec:Hojman}) to the realm of SDE.

In fact, under an additional condition, a \ind{strong
symmetry}\footnote{These were defined in Sect. \ref{sec:Misawa}.}
of a SDE is related to a \emphind{strongly conserved quantity} for
the flow of the SDE under study. This is a stochastic counterpart
to the Hojman theorem (Proposition {\HOJ} above, see
Sect.\ref{sec:Hojman}).

A \emphind{strongly conserved quantity} $J$ for the Stratonovich
equation \eqref{eq:mis}, which we rewrite here once again as
\beq\label{eq:strat6} d x^i \ = \ b^i (x,t) \, d t \ + \ \s^i_{\
k} \circ d w^k  \eeq for ease of reference, is a smooth function
$J (x,t)$ which is invariant under \emph{both} $X_0$ and all of
the $X_k$ defined in \eqref{eq:SVF}, thus under all the Misawa
vector field associated to \eqref{eq:strat6}, i.e. such that \beq
\label{eq:miscq} X_\mu (J) = 0 \ \ \ \ \forall \mu = 0,...,n \ .
\eeq

\medskip\noindent
{\bf Remark \nc.2.} The name is justified by the fact that if $J$
is a strongly conserved quantity for \eqref{eq:strat6} then $d J
\equiv 0$ on solutions to \eqref{eq:strat6}. Indeed, under
\eqref{eq:strat6} we have in general \beq d J \ = \ [X_0 (J)] \, d
t \ + \ [X_k (J)] \circ d w^k (t) \ , \eeq hence the above
condition \eqref{eq:miscq} guarantees that $d J \equiv 0$. \EOR

\subsection{Time-preserving strong symmetries}

We will first consider time-preserving strong symmetries; the
following result was shown by Misawa \cite{Mis3}.

\medskip\noindent
{\bf Proposition {\MISb}.} {\it Let the vector field \beq
\label{eq:symaf0} X \ = \ \vphi^i (x,t) \, \pa_i \eeq be a
\ind{strong symmetry} for the equation \eqref{eq:strat6}; assume
\emph{moreover} there exists a function $\la (x,t)$ such that
\begin{eqnarray} \frac{\pa b^i}{\pa x^i} &=& \div
(b) \ = \ - \ X_0 (\la) \ ; \nonumber \\ \frac{\pa \s^i_{\ k}}{\pa
x^i} &=& \div (\s_k ) \ = \ - \ X_k (\la ) \ \ \ (k=1,...,n) \ .
\label{eq:misla} \end{eqnarray} Then the quantity \beq J_\la \ :=
\ \frac{\pa \vphi^i}{\pa x^i} \ + \ X (\la ) \ = \ \div (\vphi ) \
+ \ X (\la ) \eeq is conserved under the flow of
\eqref{eq:strat6}.}

\medskip\noindent
{\bf Remark \nc.3.} Note that for $\la$ a constant (or more generally
a strongly conserved quantity), and hence $\mathrm{div} (b) = 0 =
\div (\s)$, the conserved quantity is just $J_0
= \mathrm{div} (\vphi)$. \EOR

\medskip\noindent
{\bf Proposition {\MISc}.} {\it If $J$ is a conserved quantity for
\eqref{eq:strat6} and $X = \vphi^i (x,t) \pa_i$ a symmetry for it,
then $X(J)$ is also conserved \cite{AlbFei}.}

\medskip\noindent
{\bf Remark \nc.4.} In fact, if $J$ satisfies \eqref{eq:miscq} we
immediately have, using \eqref{eq:misscc}, $X_k [ X (J)] = X [X_k
(J)] = 0$, for all $k = 0,1,...,n$. \EOR

\medskip\noindent
{\bf Remark \nc.5.} On the other hand, as already seen above (see
Proposition {\ALFEI}) it is apparent from \eqref{eq:misscc} (see
Remark \ref{chap:Strato}.1) that strong symmetries of a given
equation \eqref{eq:strat6} form a Lie algebra \cite{AlbFei}. This,
together with the previous Proposition \MISc, means that we have a
Lie algebraic structure for the set of conserved quantities under
\eqref{eq:strat6}, see again \cite{AlbFei} for a discussion. \EOR

\medskip\noindent
{\bf Remark \nc.6.} In the same way, we observe that multiplying a
symmetry vector field by a (nontrivial) constant of motion we
obtain a (new) symmetry vector field. In fact, let $X$ satisfy
\eqref{eq:misscc}, $J$ be a conserved quantity, and consider $Y =
J X$. Then \beq [X_k , Y] \ = \ (X_k (J)) \ X \ + \ J \ [X_k,X] \
= \ 0 \ . \eeq Thus, exactly as in the case of deterministic
dynamical systems \cite{CG}, the set of symmetries of a given
stochastic differential equation \eqref{eq:strat6} has also the
structure of a \emphind{Lie module} over the algebra of constants
of motion, or conserved quantities (it seems this fact went
unnoticed so far). \EOR

\medskip\noindent
{\bf Example \nc.1.} Let us consider again, as in Example
\ref{chap:Strato}.1, the Misawa example; i.e. the system with \beq\label{eq:ex1scq} b \ = \
- \ \pmatrix{y - z \cr z - x \cr x - y\cr} \ , \ \ \s \ = \ - \
\pmatrix{y - z & 0 & 0 \cr z - x & 0 & 0 \cr x - y & 0 & 0 \cr} \
. \eeq Equation \eqref{eq:misla} is satisfied with $\la$ any
constant. As already observed, the vector field $ X = (|{\bf x}|^2
/ 2) (\pa_x + \pa_y + \pa_z ) $ is a strong symmetry; the
associated conserved quantity according to Proposition {\MISb} is
just
$$ J_0 \ = \ |{\bf x}|^2 \ . $$
That is, the stochastic process lives on a sphere of constant
radius, set by initial conditions. \EOE

\medskip\noindent
{\bf Example \nc.2.} Let us consider again Example \nc.1 above,
but from the point of view of Proposition {\MISc}  \cite{AlbFei}.
As observed earlier on, this admits many symmetries, e.g. those of
the form
$$ Y_\eta \ = \ [\eta (x,y,z) ] \ (\pa_x + \pa_y + \pa_z ) \ , $$
where we can e.g. choose
\begin{eqnarray*} \eta_0 &=& (x+y+z) \ , \\
\eta_1 &=& (x^2 + y^2 + z^2) \ , \\ \eta_2 &=& (x y + y z + z
x) \ , \\
\eta_3 &=& [ x^2 \, (y + z) \, + \, y^2 \, (z + x) \, + \, z^2 \,
(x + y) \ + \ 3 \, x y z ] \ . \end{eqnarray*} The associated
conserved quantities $J_k$ are $J_0 = 1$ (that is, a trivial one),
$J_1 = J_2 = |{\bf x}|^2 = (x^2 + y^2 + z^2) = \eta_1$ (as in
Example \nc.1), and
$$ J_3 \ = \ 2 \ ( x^2 + y^2 + z^2 ) \ + \ 7 \ (x y + y z + z x) \ = \
2 \, \eta_1 \ + \ 7 \, \eta_2 \ ; $$
note that in view of $J_1$, this means that $\wt{J}_3 = (x y + y z
+ z x) = \eta_2$ is conserved.

Actually the $\eta_k$ are also themselves conserved quantities on
the flow of the system, as easily checked by direct computation.

We want now to check that symmetry maps conserved quantities into
conserved quantities. In fact, writing $Y_k = Y_{\eta_k}$ for
short, we have (beside, of course, $Y_k (J_0) = 0$) at first steps
$$ \begin{array}{lcl}
Y_0 (J_1) \ = \ 2 \ \( \eta_0 \ + \ 2 \, \eta_2 \) & , & \
Y_1 (J_1) \ = \ 2 \, \eta_0 \, \eta_1 \ , \\
Y_2 (J_1) \ = \ 2 \, \eta_0 \, \eta_2 & , & \
Y_3 (J_1) \ = \ 2 \, \eta_0 \, \eta_3 \ ; \\
Y_0 (J_2) \ = \ 2 \ \( \eta_0 \ + \ 2 \, \eta_2 \) & , & \
Y_1 (J_2) \ = \ 2 \, \eta_0 \, \eta_1 \ , \\
Y_2 (J_2) \ = \ 2 \, \eta_0 \, \eta_2 & , & \
Y_3 (J_2) \ = \ 2 \, \eta_0 \, \eta_3 \ ; \\
Y_0 (J_3) \ = \ 2 \ \( \eta_0 \ + \ 2 \, \eta_2 \) & , & \
Y_1 (J_3) \ = \ 2 \, \eta_0 \, \eta_1 \ , \\
Y_2 (J_3) \ = \ 2 \, \eta_0 \, \eta_2 & , & \
Y_3 (J_3) \ = \ 2 \, \eta_0 \, \eta_3 \ . \end{array}
$$
As the $\eta_k$ are conserved, this verifies indeed Proposition {\MISc}. \EOE

\subsection{Extended strong symmetries}

In Proposition {\MISb}, the symmetry vector field was bound to be
of the form \eqref{eq:symaf0}. When considering vector fields of
the form \beq \label{eq:symaf} X \ = \ \tau (x,t) \, \pa_t \ + \
\vphi^i (x,t) \, \pa_i \ , \eeq this result extends
\cite{AlbFei} to

\medskip\noindent
{\bf Proposition {\ALFb}.} {\it Let $X$ be a vector field of the
form \eqref{eq:symaf} and let it be a strong symmetry for the
equation \eqref{eq:strat6}; assume moreover there exists a
function $\la (x,t)$ such that \beq \label{eq:afla} \frac{\pa
b^i}{\pa x^i} \ = \ - \ X_0 (\la) \ ; \ \ \frac{\pa \s^i_{\
k}}{\pa x^i} \ = \ - \ X_k (\la ) \ \ \ (k=1,...,n) \ . \eeq Then
the quantity \beq K_\la \ := \ \frac{\pa \tau}{\pa t} \ + \
\frac{\pa \vphi^i}{\pa x^i} \ + \ X (\la ) \ - \ X_0 (\tau ) \eeq
is conserved under the flow of \eqref{eq:strat6}.}

\medskip\noindent
{\bf Remark \nc.7.} The constructions of this Section are based on
conserved quantities. We have seen in sect.\ref{sec:orbital} (see
in particular Proposition {\WALa} there) that in the case of
deterministic dynamical systems there is an essential relation
between conserved quantities and \ind{orbital symmetries}
\cite{Wal332,Wal333}. I am not aware of any study of this question
in the context of stochastic dynamical systems; quite clearly this
would have some interest. \EOR

\medskip\noindent
{\bf Example \nc.3.}\wfoot{{\blu questo e' ex 3 di AF}} Let us
consider, as suggested in \cite{AlbFei}, the system (defined for
$t > 0$, say $t \ge 1$) given by
\begin{eqnarray*}
dx &=& (x/t) \, d t \ + \ (x/t) \, (z - y) \circ d w \\
dy &=& (y/t) \, d t \ + \ (y/t) \, (x - z) \circ d w \\
dz &=& (z/t) \, d t \ + \ (z/t) \, (y - x) \circ d w \ ;
\end{eqnarray*} this is again involving a single Wiener process. We
consider the vector field
$$ X \ = \ \tau \, \pa_t \ := \ (x+y+z) \, \pa_t \ ; $$
its relation with the Misawa vector fields is given by
$$ [X_0,X] \ = \ \frac{1}{t} \ (x+y+z) \ X_0 \ := \ \rho_0 \, X_0  \
; \ \ [X_1,X] \ = \ \rho_0 \, X_1 \ . $$ In this case, eq. \eqref{eq:afla} is
satisfied with $\la = -  3 \log (t)$. The associated conserved
quantity is
$$ J \ = \ - \, \frac{4}{t} \ (x+y+z) \ = \ - \, \frac{4 \, \tau}{t} \ . $$
In fact, we have
\begin{eqnarray*}
d J &=& - 4 \frac{d \tau}{t} \ + \ 4 \frac{\tau}{t^2} d t \ = \
- \frac{4}{t} (d x + d y + d z) \ + \ 4 \frac{\tau}{t^2} d t \\
&=& - \frac{4}{t} \ \[ \frac{\tau}{t} \, d t \ + \ \frac{1}{t} [
(xz - xy) + (yx - yz) + (zy - zx) ] \circ d w \] \ + \ \frac{4 \,
\tau}{t^2} \, d t \\
& & \ = \ 0 \ . \end{eqnarray*}
{} \EOE

\section{Ito equations. Adapted coordinates}
\label{sec:adapted}
\def\cs{\ref{sec:adapted}}

The experience built with deterministic differential equation --
as well as common sense -- suggests that in the presence of a
symmetry it is convenient to reformulate the equation under study in terms of
\emphind{symmetry-adapted coordinates}. It is quite natural to
expect this also holds when dealing with stochastic differential
equations, albeit in this case one does not (yet?) have as
detailed results and theorems as in the deterministic case.

A substantial advance was provided by R. Kozlov \cite{Koz3} who
noted that the same kind of results relating symmetry and
reduction for ODEs also hold for SDEs, albeit with one (not so
weak) extra condition: that is, only symmetries \emph{not} acting
on the time can actually be used. The precise results in this
direction will be discussed in Sect. \ref{sec:kozlov}, where we
review Kozlov's work.

In this section we will instead briefly reconsider some of the
examples discussed above from the point of view of symmetry
adapted coordinates, obtaining other kind of results; as mentioned
above, in Sect. \ref{sec:linearization} and even more in
Sect. \ref{sec:kozlov} we will see other applications -- again
based on the idea of \ind{symmetry-adapted coordinates} -- of the
symmetry analysis of Ito equations.

\medskip\noindent
{\bf Example \nc.4.} Let us consider again, as in Example
\nc.1, the system \eqref{eq:ex1scq}

Passing to spherical coordinates $(r,\vartheta, \varphi)$ via
$$
x = r  \cos (\vartheta ) \cos (\varphi ) \ , \ \ y = r  \sin
(\vartheta ) \cos (\varphi ) \ , \ \ z = r  \sin (\varphi )
$$ (where of course $\vartheta \in [0,2 \pi]$, $\varphi \in [-
\pi/2 , \pi/2 ]$), and writing
$$ \a (\vartheta, \varphi ) \ := \ 1 \ - \ \( \sin \vartheta \, + \, \cos
\vartheta \) \ \tan \varphi \ , \ \ \b (\vartheta , \varphi ) \ :=
\ \sin \vartheta \, - \, \cos \vartheta \ , $$ the system
considered in this example reads simply
\begin{eqnarray*}
d r &=& 0 \\
d \vartheta &=& \a (\vartheta, \varphi ) \, d t \ + \ \a (\vartheta, \varphi ) \circ d w  \\
d \varphi &=& \b (\vartheta , \varphi ) \,  d t \ + \ \b
(\vartheta , \varphi ) \circ d w \ ; \end{eqnarray*} thus the
conservation of $r$ is completely explicit. \EOE

\medskip\noindent
{\bf Example \nc.5.} Let us consider again the system
\eqref{eq:exKram} seen in Example \ref{chap:Ito}.2 and related to
the Kramers equation, i.e.
\begin{eqnarray*} d x &=& y \, dt \\
d y &=& - \, k^2 \ y \ d t \ + \ \sqrt{2 k^2} \ d w (t) \ .
\end{eqnarray*}
By focusing on the vector field $$ V_3 \ = \ e^{- k^2 t} \ \(
k^{-2} \pa_x - \pa_y \) $$ in its symmetry algebra (see Example
\ref{chap:Ito}.2), we pass to coordinates
$$ p \ = \ x \ , \ \ q \ = \ y \ + \ k^2 \, x \ . $$
With these, we also have $$ d x \ = \ d p \ , \ \  d y
\ = \ d q \ - \ k^2 \, d p  \ ; $$ thus the stochastic systems reads
\begin{eqnarray*} dp &=& (q - k^2 p) \ dt \\  d q &=& \sqrt{2 k^2} \ d w (t) \ . \end{eqnarray*}
In other words, using symmetry-adapted coordinates led us to
decompose the system into a stochastic and a deterministic part. \EOE

\medskip\noindent
{\bf Example \nc.6.} We consider again the system
\begin{eqnarray*}
d x_1 &=& [1 - (x_1^2 +x_2^2)] \, x_1 \, d t \ + \ d w_1 \\
d x_2 &=& [1 - (x_1^2 +x_2^2)] \, x_2 \, d t \ + \ d w_2 \ ; \end{eqnarray*}
seen in Example \ref{chap:RS}.6.

We change coordinates as suggested by the symmetries determined in
there; we will thus set
$$ x_1  \ = \ \rho \ \cos (\vartheta) \ , \ \ x_2 \ = \ \rho \ \sin (\vartheta ) \ ; \ \ w_1 \ = \ \chi \ \cos (\la ) \ , \ \ w_2 \ = \ \chi \, \sin (\la ) \ . $$
With these coordinates, the system reads simply
\begin{eqnarray*}
d \rho &=& (1 - \rho^2) \, \rho \, d t \ + \ \cos (\la - \vartheta) \, d \chi \ - \ \chi \, \sin (\la - \vartheta) \, d \la \ , \\
d \vartheta &=& (1/\rho) \ [ \sin (\la - \vartheta) \, d \chi \ +
\ \chi \, \cos (\la - \vartheta) \, d \la] \ . \end{eqnarray*} The
invariance under simultaneous rotations in the $(x_1,x_2)$ and
$(w_1,w_2)$ planes (i.e. simultaneous shifts in the angles
$\vartheta$ and $\la$) is now completely explicit. \EOE

\section{Ito equations. Linearization}
\label{sec:linearization}
\def\cs{\ref{sec:linearization}}

%\subsection{Exact linearization}

Transforming an equation to linear form is a convenient way to solve it, and thus  the problem of linearizing a given SDE was considered by several
authors. I am not aware of a direct analogue of the Bluman-Kumei
theorem mentioned in Sect. \ref{sec:Bluman} above; but I will be
mentioning here some results on this topic. I will actually only consider the case of a one-dimensional SDE.

In particular\footnote{For other approaches to (symmetry-related) linearization of SDEs, see \cite{Koz2,Koz3,MWS}.}, Grigoriev, Ibragimov, Meleshko and  Kovalev
\cite{Meleshko} considered the problem of linearizing a SDE via a
smooth deterministic change of the dependent variable -- they
called this \emphind{strong linearization} -- i.e. via a map \beq
y \ = \ \phi (x,t) \ ; \eeq we assume $\phi_x \not= 0$, so the map
is invertible, and we denote the inverse as $x = \psi (y,t)$.
Under this map, the Ito equation \beq \label{eq:mel} d x \ = \ f
(x,t) \ d t \ + \ g (x,t) \ d w \eeq is mapped into a new Ito
equation for $y$, \beq d y \ = \ F (y,t) \ dt \ + \ G (y,t) \ d w
\ , \eeq with \beq \label{eq:melmap} F(y,t) \ = \ \phi_t \ + \ f
\, \phi_x \ + \ \frac12 \, g^2 \, \phi_{xx} \ , \ \ G(y,t) \ = \ g
\ \phi_x \ , \eeq and the functions on the r.h.s. of these
formulas should be understood to be expressed in terms of $y$ via
$x = \psi (y,t)$.

One would like to get a \emph{linear} equation for $y$, i.e. to
have \beq \label{eq:mellin} F(y,t) \ = \ \a_0 (t) \ + \ \a_1 (t)
\, y \ ; \ \ G(y,t) \ = \ \b_0 (t) \ + \ \b_i (t) \, y \ . \eeq
This is, of course, not always possible, so the problem lies in
identifying the equations (that is, the functions $f$ and $g$)
which are linearizable, and the linearizing map $\phi$.

Comparing \eqref{eq:melmap} and \eqref{eq:mellin} it is clear that
this amounts to studying conditions for the existence of solutions
-- and if these are satisfied, one would also like to find such
solutions -- to the system
\begin{eqnarray}
\phi_t \ + \ f \, \phi_x \ + \ \frac12 \, g^2 \, \phi_{xx} &=&
\a_0 \ + \ \a_1 \, \phi \ , \nonumber \\
g \ \phi_x &=& \b_0 \ + \ \b_1 \ \phi \ . \end{eqnarray}

These can be considered as identifying the partial derivatives
$\phi_t$ and $\phi_x$; the request that the mixed partial
derivatives determination through the two definitions should
coincide (that is, $\pa_x \phi_t = \pa_t \phi_x$) yields the
solvability condition \beq \a_0 \, \b_1 \ + \ (\b_0)_t \ - \ \b_0
\, (\a_1 + \mu ) \ + \ \phi \, [(\b_1)_t - \b_1 \mu] \ = \ 0 \ ,
\eeq where we have defined $\mu = \mu (x,t)$ by \beq
\label{eq:melmu} \mu \ := \ \frac{1}{g} \ \( g_t \ + \ f \, g_x \
+ \ \frac12 \, g^2 \, g_{xx} \ - \ g \, f_x \) \ . \eeq Note that
$\mu$ is defined entirely in terms of the coefficients of the
original SDE.

The situation is simpler when $\mu_x = 0$; in this case one has
\cite{Meleshko}

\medskip\noindent
{\bf Proposition {\MELa}.} {\it Let the function $f$ and $g$ in
equation \eqref{eq:mel} be such that $\mu$ defined in
\eqref{eq:melmu} satisfies $\mu_x = 0$. Then the Fokker-Planck
equation associated to \eqref{eq:mel} is equivalent to the heat
equation, and \eqref{eq:mel} is reduced to the linear SDE \beq d y
\ = \ h(t) \ d w \ , \ \  \ \ h(t) = \exp \[ \int \mu (t) \, d t
\] \ ; \eeq the linearizing map $y = \phi (x,t)$ is the solution
to the compatible system of PDEs \beq \phi_t \ = \ h \, \(
\frac{g_x}{2} \ - \ \frac{f}{g} \) \ , \ \ \phi_x  \ = \
\frac{h}{g} \ . \eeq}
\bigskip

The general case, i.e. $\mu_x \not=0$, corresponds to a more
involved situation. One can show \cite{Meleshko} that a different
condition also leads to similar results; the condition is now
expressed by two equations:
\begin{eqnarray} \mu_{xxx} & - & \frac{1}{g \, \mu_x} \ \( g \,
\mu_{xx}^2 \ - \ g_{xx} \, \mu_x^2
\ - \ g_x \, \mu_x \, \mu_{xx} \) \ = \ 0 \ , \label{eq:melmunot0a} \\
\mu_{xxt} & - & \frac{1}{g \, \mu_x} \ \[ \( g \, \mu_{xt} \, + \,
g \, \mu \, \mu_x \, - \, g_t \, \mu_x \) \, \mu_{xx} \right. \nonumber \\
& & \left. \ \ \ \ - \ \( g_{xt} \, - \, g_x \, \mu \, + \, g \,
\mu_x \) \, \mu_x^2 \] \ = \ 0 \ . \label{eq:melmunot0b}
\end{eqnarray}
With this, one has \cite{Meleshko}:

\medskip\noindent
{\bf Proposition {\MELb}.} {\it Let the function $f$ and $g$ in
equation \eqref{eq:mel} be such that $\mu$ defined in
\eqref{eq:melmu} satisfies \eqref{eq:melmunot0a},
\eqref{eq:melmunot0b}. Then \eqref{eq:mel} is reduced to the
linear SDE \beq d y \ = \ \a (x,t) \, d t \ + \ \b(x,t) \, y \  d
w \ , \eeq where
\begin{eqnarray*} \a (x,t) &=& \exp \[ \int \( \b \, \frac{g_x -
\b}{2} \ + \ \frac{g_t - f \b}{g} \ + \ \frac{\mu_{xt}}{\mu_x} \ -
\ 2 \, \frac{g \mu_x}{\b} \ - \ \mu \) \ dt \] \ , \\
\b (x,t) &=& - \ \frac{g_x \, \mu_x \ + \ g \, \mu_{xx}}{\mu_x} \
; \end{eqnarray*} the linearizing map $y = \phi (x,t)$ is given by
\beq \phi \ = \ \frac{\a \, \b}{\b_t \, - \, \b \, \mu} \ . \eeq}

\medskip\noindent
{\bf Remark \nc.8.} Second order SDE have been analyzed along the
same lines in a number of publications; see e.g. \cite{Meleshko,Koz1,Koz3,SMS3,Sri,SMS1,SMS2}. We
will just refer the interested reader to these. \EOR

%\subsection{Perturbative linearization}

\medskip\noindent
{\bf Remark \nc.9.} Linearization of SDE, and more generally
reduction of SDE to an analogue of \index{Poincar\'e}
\index{Dulac} Poincar\'e-Dulac \index{normal forms} \emph{normal
form} \cite{ArnGMDE,Elp,Iooss,Wal330}, has been considered by
Arnold and Imkeller \cite{LArnold,ArnImk}; they provided a
complete solution to the problem, and hence I will just refer the
reader to their work.

In this context, it may be noted that they mainly focused on the
formal aspects of the theory and the determination of a (formal)
Taylor series for the normalizing transformation, while as for the
existence of an actual random diffeomorphism (see below) whose
Taylor series coincides with the one they determine, they refer to
a theorem by \ind{Borel} as used and proved by Vanderbauwhede (see
\cite{VdB}, page 142); it is possible that further progress can be
obtained by looking at the problem of normal form convergence in a more
specific way. \EOR

\medskip\noindent
{\bf Remark \nc.10.} It should be stressed that Arnold and
Imkeller did consider maps more general than those considered by
Grigoriev {\it et al.}, i.e. \emphind{random diffeomorphisms} (see
Chapter \ref{chap:RS}); this puts on equal footing the dynamical
SDE and the change of variables required by the normalizing
procedure, which can thus be seen as the flow of an equation of
the same type, i.e. a SDE \cite{GS2016}.\footnote{This parallels
the approach to Poincar\'e normalization through \index{Lie
transform} Lie transforms, i.e. through the time-one flow of a
dynamical system \cite{BGG,BroTh,BroLNM}.} \EOR

\medskip\noindent
{\bf Remark \nc.11.} Perturbative linearization is also related to \emphind{approximate symmetries}  (normalization corresponds to having the linear part of the system as a symmetry for the full system); in this respect
one should consider \cite{IUJ} (see also \cite{UJ}). \EOR

\section{Transformation of an Ito equation to simpler form. Reduction}
\label{sec:kozlov}
\def\cs{\ref{sec:kozlov}}

The approach considered in the previous Section
\ref{sec:linearization} aimed at linearization of a SDE.
More generally, one can aim at transforming a SDE into a
convenient form. This is in general a simpler (than the original)
form, but it may also be an equation which has already been
studied and which is thus just more convenient, albeit possibly not simpler
than the original one.

As symmetries are present -- or absent -- independently of the
coordinates used, they are a natural tool to investigate if two
equations might be transformed one into the other\footnote{This
remark goes back to \ind{Moser}, who used it in the framework of
normalization theory to advocate that a \index{Poincar\'e}
\index{Birkhoff} (Poincar\'e-Birkhoff) normal form could be
actually (and not just formally) conjugated to the original form
of the system under study \emph{only} if the latter has a symmetry
(this since the normal form admits its linear -- or quadratic in
the Hamiltonian case -- part as a symmetry).}. This point of view
has been advocated by several authors, and here we are specially
interested in the results obtained by R. Kozlov in a series of
papers \cite{Koz1,Koz2,Koz3}, which we follow quite
closely.\footnote{One should mention that the reduction (and
reconstruction) problem was also considered by Lazaro-Cam\'i and
Ortega \cite{LCO3} through a more geometric approach (they were
also giving substantial attention to the Hamiltonian -- thus
variational --setting); we will not discuss their work here.}

The idea was to classify possible symmetry groups of SDEs of
different orders (in particular, low order ones); apart from the
interest of the classification in itself, as a byproduct one would
like to detect the possibility of transforming the equation into a
simpler form.\footnote{It is maybe worth recalling that a similar
approach was pursued by \ind{Lie} in dealing with ordinary differential
equations \cite{LieScheffers} (see also \cite{Gonzalez}); in most
cases a maximal dimensional (for the order of the equation)
symmetry group is a guarantee that the equation can be
linearized.}

\medskip\noindent
{\bf Remark \nc.12.} It should be stressed that in this section --
and actually in the whole Chapter -- the symmetries are always
meant to be \emph{deterministic} ones; as far as I know there is
no study along these lines considering \emph{random} symmetries as
well. This is a topic which is surely worth investigating. \EOR

\subsection{First order scalar SDEs}
\label{sec:koz1}

In this Section, at difference with the other parts of this work,
we will also consider higher order SDEs.

The case of a single first order equation is of course specially
relevant, and we will hence start by considering it; the reader
should be warned it is also in a way a degenerate one, as will be
clear from comparing the results obtained here with those in Sect.
\ref{sec:kozho} and Sect. \ref{sec:kozsyst} below.

For an equation of first order, i.e. a standard Ito equation
\beq\label{eq:kozeq} d x \ = \ f (x,t) \, d x \ + \ g (x,t) \, d w
(t) \ , \eeq we have the following results\footnote{Unfortunately,
the proofs of these are not always given in full detail. After the
completion of this paper, C. Lunini has provided detailed proofs
of Kozlov's theorem; she has also shown that the condition in
Proposition \KOZa is not only sufficient but also necessary for
the possibility to map and equation of the type \eqref{eq:kozeq}
into one of the type \eqref{eq:kozprop27} \cite{Lunini}.}
\cite{Koz1}.

\medskip\noindent
{\bf Proposition \KOZa.} {\it If a scalar SDE \eqref{eq:kozeq}
admits a fiber-preserving symmetry, with generator $X = \tau (t)
\pa_t + \xi (x,t) \pa_x$, then there exists a fiber-preserving
change of coordinates $s = s (t)$, $y = y (x,t)$ which maps $w
(t)$ into the Wiener process $\om (s)$ and the equation
into\beq\label{eq:kozprop27} d y \ = \ \phi (y) \, d s \ + \ \ga
(y) \, d \om (s) \ . \eeq}

\medskip\noindent
{\bf Proposition \KOZb.} {\it A scalar SDE \eqref{eq:kozeq} can
have a symmetry algebra of dimension $r = 0,1,2,3$. In the case
$r=3$, the equation can be transformed into a Brownian motion
equation, $d y = d \om (s)$, via a change of variables with non-random time
transformation.}

\medskip\noindent
{\bf Proposition \KOZc.} {\it A scalar SDE \eqref{eq:kozeq} can
have a symmetry algebra of dimension $r=3$ if and only if it
admits a symmetry generator of the form \beq \label{eq:X*}
X_* \ = \ \xi (x,t) \ \pa_x \ . \eeq
This in turn is the case if and only if the functions appearing in
\eqref{eq:kozeq} satisfy the relation \beq \label{eq:kozsymmcond}
\[ \frac{g_t}{g} \ - \ g \, \( \frac{f}{g} \)_x \ + \ \frac12 \, g
\, g_{xx} \]_x \ = \ 0 \ . \eeq}
\bigskip

It is interesting to note that the presence of a symmetry of the
form \eqref{eq:X*} leads to integrability
of the equation (see also Sect. \ref{sec:kozsyst} in this respect).
In fact, while in the case of a nonzero $\tau$ we have to take
into account the effect on the Wiener process, in the case $\tau =
0$ we do not have to worry about this. Passing to
\emphind{symmetry-adapted coordinates} $(y,t)$ means the symmetry
vector field will be expressed in these as \beq X_* \ = \ \pa_y  \
, \eeq while the equation will be written as \beql{eq:kozsac} d y\
= \ \phi (y,t) \, d t \ + \ \ga (y,t) \, d w(t) \ . \eeq But in
this case the \ind{determining equations} for symmetries of the equation
(which is by hypothesis satisfied by $X_* = \eta(y,t) \pa_y$)
read simply
\begin{eqnarray}
\frac{\pa \eta}{\pa t} \ + \ \phi \ \frac{\pa \eta}{\pa y} \ - \
\eta \ \frac{\pa \phi}{\pa y} &=& 0 \nonumber \\
\ga \ \frac{\pa \eta}{\pa y} \ - \ \eta \ \frac{\pa \ga}{\pa y}
&=& 0 \ ; \end{eqnarray} as in the \ind{symmetry-adapted
coordinates} we have $\eta = 1$, these actually read
$$ \frac{\pa \phi}{\pa y} \ = \ 0 \ \ , \ \ \ \ \frac{\pa \ga}{\pa y} \ = \ 0 \ . $$
In other words, the equation \eqref{eq:kozsac} is actually
\beql{eq:kozsac2} d y \ = \ \phi (t) \, d t \ + \ \ga (t) \, d
w(t) \ , \eeq and is therefore promptly integrated to give \beq
y(t) \ = \ y(t_0) \ + \ \int_{t_0}^t \phi (s) \ d s \ + \ \ \[ w(t)
- w(t_0) \] \ . \eeq

\medskip\noindent
{\bf Example \nc.7.} The equation \cite{Koz1}
\beq\label{eq:koz1} dx \ = \ f(t) \, d t \ + \ g(t) \, d w(t) \eeq
admits a three-dimensional symmetry algebra, with generators
\begin{eqnarray*}
X_1 &=& (1/g^2) \ [ \pa_t \ + \ f \, \pa_x ] \ , \\
X_2 &=& \pa_x \ , \\
X_3 &=& (2 G / g^2) \ [ \pa_t \ + \ f \, \pa_x ] \ + \ (x \, - \,
F) \, \pa_x \ , \end{eqnarray*} where we have written
$$ F(t) \ := \ \int f(t) \, dt \ , \ \ G(t) \ := \ \int g(t) \, dt \ . $$
These vector fields satisfy the commutation relations
$$ [X_1,X_2] = 0 \ , \ \ [X_1 , X_3] = 2 X_1 \ , \ \ [X_2 , X_3] = X_2 \ . $$

With the change of variables \beq\label{eq:cvkoz1} s \ = \ \int
g^2 (t) \, d t \ , \ \ y \ = \ x \ - \ \int f(t) \, d t \ , \eeq
the equation is mapped into the Brownian motion equation
\beq\label{eq:koz1bm} d y \ = \ d \om (s) \ ; \eeq here $\om (s )$
is the Wiener process obtained from $w(t)$ via the above change of
variables. This equation admits the symmetries
$$ X_1 \ = \ \pa_s \ , \ \ X_2 \ = \ \pa_y \ , \ \ X_3 \ = \ 2 s \pa_s + y \pa_y \ . $$
Solution to the original equation \eqref{eq:koz1} are obtained by
solutions to \eqref{eq:koz1bm}, which read \beq\label{eq:koz1bm2}
y (s ) \ = \ y (s_0) \ + \ \[ \om (s) - \om (s_0) \] \ , \eeq by
inverting the change of variables \eqref{eq:cvkoz1}; with this the
function \eqref{eq:koz1bm2} is mapped into \beq x(t) \ = \ x (t_0)
\ + \ \int_{t_0}^t f(s) \, d s \ + \ \int_{t_0}^t g(s) \, d w (s)
\ , \eeq which provides a solution to the original equation
\eqref{eq:koz1bm}. Needless to say, this solution could be derived
immediately from \eqref{eq:koz1}, but the example shows how to use
the Kozlov result and procedure. \EOE

\medskip\noindent
{\bf Example \nc.8.} Let us consider the vector field $X = \pa_t +
x \pa_x$, which of course is \emph{not} of the form required by
Proposition \KOZc.\footnote{We note that this vector field
generates the one-parameter group (here $\la$ is the group
parameter) $t \to t + \la$, $x \to e^{\la} x$, so that we have a
special behavior in $x=0$. Thus we should consider separately the
domains $x > 0$ and $x < 0$; we will consider just the former.}

In view of \eqref{eq:deteq_Ito}, the more
general Ito equation which admits $X$ as a symmetry is
\beql{eq:ex8geneq} d x \ = \ x \, \phi (z) \ d t \ + \ x \, \ga
(z) \ d w(t) \ , \eeq where we have written as $z$ the
characteristic function for $X$, i.e.
$$ z \ := \ x \ e^{- t} \ . $$
Passing  to symmetry adapted coordinates means passing from
$(x,t)$ to $(y,z)$, where $y = y (x,t)$ is the solution to
$$ X (y) \ = \ \pa_t y \ + \ x \, \pa_x y \ = \ 1 \ . $$
The latter is given by
$$ y \ = \ \log (x) \ + \ \b (z) \ , $$
with $\b$ an arbitrary smooth function; we set $\beta (z) \equiv
0$. Thus our change of coordinates and the inverse one are given
by
$$ y = \log (x) \ , \ \ z = x \, e^{- t} \ ; \ \ x = e^y \ , \ \ t = y - \log (z) \ . $$
In these coordinates, the equation \eqref{eq:ex8geneq} reads
$$ d y \ = \ - \, \frac{\phi (z) }{[1 - \phi(z)] \ z} \ d z \ + \
\frac{\ga (z)}{[1 - \phi(z)]} \ d w (y - \log (z) ) \ . $$

Actually, it is more convenient to use coordinates $(y,t)$, so
that the Wiener process depends on these variables in a simple
way. In such coordinates, $z = \exp(y-t)$ and our equation
\eqref{eq:ex8geneq} reads
$$ d y \ = \ \phi (z) \, d t \ + \ \ga (z) \, d w(t) \ = \
\phi [\exp(y-t)] \, d t \ + \ \ga [\exp(y-t)] \, d w (t)  \ . $$
In general, this equation is \emph{not} integrable.

Thus this example shows concretely that a symmetry which is
\emph{not} of the form \eqref{eq:X*} (as stipulated by Kozlov)
does not, in general, imply the integrability of the equation. \EOE

\subsection{Higher order scalar SDEs}
\label{sec:kozho}

The results discussed above for a first order scalar SDE can be
generalized to the case of scalar higher order SDEs\footnote{A
single equation of order $n$ could of course also be mapped into a
system of $n$ first order equations, and analyzed as a system; see
Sect. \ref{sec:kozsyst}.}, i.e. equations of the form
\beql{eq:kozeqho} d x^{n-1} \ = \ f (x,x^{(1)}, ... , x^{(n-1)},
t) \, d t \ + \ g (x,x^{(1)}, ... , x^{(n-1)}, t) \, d w (t) \ .
\eeq

Albeit we have not considered this kind of equations in our discussion\footnote{Symmetries of higher order SDEs have been considered by several authors; see e.g. \cite{SMS3,Sri,MWS}.}, it is worth mentioning the results obtained by Kozlov \cite{Koz3}.

\medskip\noindent
{\bf Proposition \KOZd.} {\it The scalar SDE \eqref{eq:kozeqho} of
order $n$ admits a symmetry algebra of dimension at most $(n+2)$.
If \eqref{eq:kozeqho} admits a symmetry algebra of dimension $r =
n+2$, then it can be mapped into the higher order Brownian motion
equation $$ d y^{(n-1)} \ = \ \ga_0 \, d \om (t) \ . $$

The symmetry subalgebra for the scalar SDE \eqref{eq:kozeqho}
generated by vector fields of the form $X = \xi (x,t) \pa_x$ has
dimension $r_0 \le n$.}

\medskip\noindent
{\bf Remark \nc.13.} A complete classification of symmetry algebras
for equations \eqref{eq:kozeqho} of order $n=2$ and $n=3$ is
provided in \cite{Koz3}. \EOR

\subsection{Systems of SDEs}
\label{sec:kozsyst}

The same kind of approach can be pursued for \emph{systems} of
SDEs. The case of a degenerate diffusion matrix $\s^i_j$ would
introduce several degenerations (and subcases to be considered) in
the problem, so one likes to consider the case where the diffusion
matrix has full rank \cite{Koz2}.

More precisely, one considers the system \beql{eq:kozsyst} d x^i \
= \ f^i (x^1,...,x^n;t) \, d t \ + \ \s^i_{\ k} (x^1,...,x^n;t) \,
d w^k (t) \ \eeq with $i = 1,...n$, $k = 1,...,m$ and assumes $n
\le m $ with $\s$ of rank $n$. In this case symmetries are
fiber-preserving \cite{Koz2}, i.e. of the form $X = \tau (t) \pa_t
+ \xi^i (x,t) \pa_i$ (where as usual $\pa_i = \pa/ \pa x^i$); this
guarantees we will have a non-random change of time.

\medskip\noindent
{\bf Proposition \KOZf.} {\it The $n$-dimensional system of SDEs
\eqref{eq:kozsyst} admits a symmetry algebra of dimension $r \le
(n+2)$. The symmetry subalgebra generated by vector fields of the
form $X = \xi^i (x,t) \pa_i$ has dimension $r_0 \le n$.}
\bigskip

The full classification of symmetry properties for two dimensional
systems of SDEs ($n=2$, hence $r \le 4$) is provided in
\cite{Koz2}.

In the case of systems one can also use symmetry properties to
reduce (as for ordinary differential equations) the dimension of
the system; in the case of a sufficiently large algebra with a
suitable structure (again as for ODEs) one can infer integrability
of the equation. It should be stressed that in this case (at
difference with ODEs) only symmetries acting in the space of
dependent variables can actually be used.\footnote{Looking back at
the examples considered above, we note that in Example \nc.7 we
had symmetries of this form (and the equation could be
integrated), while in Example \nc.8 we had a symmetry but it was
not of the required form, and correspondingly the equation could not,
in general, be integrated.}

\medskip\noindent
{\bf Proposition \KOZg.} {\it If the $n$-dimensional system
\eqref{eq:kozsyst} admits a symmetry generated by a vector field
of the form $$ X_* \ = \ \xi^i (x,t) \ \pa_i \ , $$ then there
exists a regular change of variables $y = y (x,t)$ which maps the
system into a system \beql{eq:kozsystred} d y^i \ = \ \phi^i
(y^1,...y^{n-1};t) \, d t \ + \ \rho^i_{\ k} (y^1,...,y^{n-1};t)
\, d w^k (t) \eeq of dimension $(n-1$) plus a \index{reconstruction}
``reconstruction equation'' \beq y^n (t) \ = \ y^n (t_0) \ +
\ \int_{t_0}^t \phi^n (y^1,...,y^{n-1};s) \, d s \ + \
\int_{t_0}^t \rho^i_{\ k} (y^1,...,y^{n-1};s) \, d w^k (s) \ .
\eeq}

\medskip\noindent
{\bf Proposition \KOZh.} {\it Let the $n$-dimensional system
\eqref{eq:kozsyst} admits a solvable symmetry group of dimension
$r$, acting regularly\footnote{This requires the group orbits to
be $r$-dimensional manifolds with regular embedding in the ambient
space; in other words, points which are on the same orbit and
nearby in the space should also be nearby along the orbit. See
e.g. \cite{DuiKolk,Olver1} for \ind{regular group action}.}. Then the system can
be reduced to a system of dimension $q = (n - r)$. The solutions
of the original system are in correspondence with solutions to the
reduced system and are obtained by the latter via quadratures. If
$r=n$, the original system is integrable and its general solution
is obtained by quadratures.}

\medskip\noindent
{\bf Example \nc.9.} Consider the system \eqref{eq:kozsyst} for $m=n=2$ and $\s$ constant,
$$ \s \ = \ \pmatrix{S_{11} & S_{12} \cr S_{21} & S_{22} \cr} \ , $$
with of course $\mathrm{det} (\s) \not= 0$, and $f$
linear\footnote{We write all indices as lower ones in order to
avoid any confusion.}, $$ f_i \ = \ a_i \ + \ b_i \, x_1 \ + \ c_i
\, x_2 \ . $$

This turns out \cite{Koz2} to admit two symmetries of the required form, which are actually of the type
$$ X_i \ = \ A_{ij} (t) \pa_j \ \ \ \ \[ \mathrm{det} (A) \not= 0 \] \ . $$ With the change of coordinates
$$ y_1 \ = \ \frac{A_{22} x_1 - A_{21} x_2}{A_{11} A_{22} - A_{12} A_{21}} \ , \ \ y_2 \ = \ \frac{A_{11} x_2 - A_{12} x_1 }{A_{11} A_{22} - A_{12} A_{21}} $$
and writing \begin{eqnarray*} \a_1 &=& \frac{A_{22} a_1 - A_{12}
a_2}{A_{11} A_{22} - A_{12} A_{21}} \ , \ \ \a_2 \ = \
\frac{A_{11} a_2 - A_{21} a_1}{A_{11} A_{22} - A_{12} A_{21}} \ ;
\\
\b_{11} &=& \frac{A_{22} S_{11} - A_{12} S_{21}}{A_{11} A_{22} -
A_{12} A_{21}} \ , \ \ \b_{12} \ = \ \frac{A_{22} S_{12} - A_{12}
S_{22}}{A_{11} A_{22} - A_{12} A_{21}} \ , \\
\b_{21} &=& \frac{A_{11} S_{21} - A_{21} S_{11}}{A_{11} A_{22} -
A_{12} A_{21}} \ , \ \ \b_{22} \ = \ \frac{A_{11} S_{22} - A_{21}
S_{12}}{A_{11} A_{22} - A_{12} A_{21}} \ , \end{eqnarray*} the
system is mapped into
\begin{eqnarray*}
d y_1 &=& \a_1 \ dt \ + \ \b_{11} \ d w_1 (t) \ + \ \b_{12} \ d w_2 (t) \\
d y_2 &=& \a_2 \ dt \ + \ \b_{21} \ d w_1 (t) \ + \ \b_{22} \ d
w_2 (t) \ ,
\end{eqnarray*} which is readily integrated. \EOE

\chapter{Conclusions}
\label{chap:conclusions}

The symmetry approach is a general way to attack deterministic differential equations (and actually one can set in terms of it all the different
solution methods for differential equations, provided suitable generalizations of the concept of symmetry are considered); in the deterministic
framework it proved invaluable both for the theoretical study of
differential equations and for obtaining concrete solutions.

The theory is comparatively much less advanced in the case of
stochastic differential equations. There is now some general
agreement on what the ``right'' -- that is, \emph{useful} --
definition of symmetry for stochastic differential equations is,
but only few applications have been considered, most of these
concerning ``integrable'' equations. There is ample space for
considering new applications, first and foremost considering ``non
integrable'' equations.

Correspondingly, there is ample space for concrete applications,
i.e. applying the approaches already existing or to be developed
to new concrete stochastic systems.

Albeit here we have mentioned it only in passing, symmetry theory flourished and expanded its role by considering generalization of the ``standard'' (i.e. Lie-point) symmetries in
several directions, some of them classical and some developed only in recent years. As far as I know, there is no attempt in this direction for stochastic systems yet; any work in this direction is very likely to collect success and relevant results.

It should also be stressed that actually in dealing with SDEs one
could legitimately consider more general classes of
transformations, as already done in normal forms theory for
stochastic \ind{dynamical systems} \index{Arnold-Imkeller
approach}  \cite{LArnold,ArnImk}. A first attempt in this
direction was proposed only very recently \cite{GS2016}, and here
again there is ample space for new work and interesting results.

Summarizing in a single sentence: {\it the first attempts to use
symmetry in the analysis of stochastic equations were promising;
time is now ripe for extending fully fledged symmetry theory to
stochastic systems.}

I hope that the present text, aiming at providing an overview of
the state of the art, can be of help in promoting work on this
fascinating subject.

%\vfill\eject

\printindex

\end{document}

\title{Erratum for \\ ``Symmetry of stochastic non-variational
differential equations'' \\ ({\it Physics Report} {\bf 686}
(2017), 1-62)}

\author{G. Gaeta\thanks{ORCID 0000-0003-3310-3455; e-mail: {\tt giuseppe.gaeta@unimi.it}} \\
{\it Dipartimento di Matematica, Universit\`a degli Studi di
Milano} \\ {\it via Saldini 50, 20133 Milano (Italy)} }

\maketitle

In my recent paper \cite{GGPR}, due to a regrettable and rather trivial
mistake, a mixed derivatives term is missing in the expression
(5.3) for the Ito Laplacian -- which is essentially a Taylor
expansion. The correct formula is, of course
$$ \Delta f \ := \ \sum_{k=1}^n \ \frac{\pa^2 f}{\pa w^k \pa w^k} \ + \
\sum_{j,k=1}^n \ \left( \s \, \s^T \right) \ \frac{\pa^2 f}{\pa
x^j \, \pa x^k} \ + \ 2 \ \sum_{j,k=1}^n \s^{ik} \ \frac{\pa^2
f}{\pa x^j \, \pa w^k} \ . \eqno(5.3) $$ (The reader is alerted
that the same mistake found its origin in a previous paper of mine
\cite{GS}, on which some of this review was based.)

This error has no consequence on our general discussion --
conducted in terms of the $\Delta$ operator -- except for what is
said below; but it does affect the specific computations occurring
in most of the concrete examples of Section 5.

%\bigskip

The error in (5.3) has some more substantial consequence in Remark
5.8 and Section 5.5.

The part of Remark 5.8 following eq.(5.32) is simply wrong: once
the correct formula for $\Delta (\varphi)$ is used, the quantity
$\delta^i$ defined in eq.(5.33) is exactly zero, in any dimension,
as proved in \cite{GL}.

All the discussion in Sect.5.5 should be revised in the light of
this fact; in particular, $\delta^i = 0$ means that for
\emph{simple} (deterministic or random) \emph{symmetries}, there
is a full equivalence between an Ito and the corresponding
Stratonovich equation. (In the deterministic case, this was proved
by Unal, see ref.200 in \cite{GGPR}.)

Note this holds in \emph{any} dimension (while in Section 5.5 we
only considered the scalar case); the correct statement concerning
the matter considered in this Section is therefore as follows:
\par\noindent ``The simple (deterministic or random) symmetries of
an Ito equation and those of the corresponding Stratonovich one do
coincide''.

I apologize to the readers, and thank the anonymous Referee of \cite{GL} for pointing out the mistake.